\documentclass[aps,twocolumn,superscriptaddress,preprintnumbers,floatfix]{revtex4}
\usepackage[colorlinks, citecolor=blue,anchorcolor=red,menucolor=red, linkcolor=red,filecolor=red,urlcolor=blue,frenchlinks=red]{hyperref}
\usepackage{amsfonts}
\usepackage{amsmath}
\usepackage{amssymb}
\usepackage{CJKutf8}
\usepackage{color}
\usepackage{comment}
\usepackage{epsfig}
\usepackage{epstopdf}
\usepackage{float}
\usepackage{graphicx,booktabs}
\usepackage{indentfirst}
\usepackage{longtable,lscape}
\usepackage{mathrsfs}
\usepackage{mathtools}
\usepackage{morefloats}
\usepackage{pifont}
\usepackage{txfonts}
\usepackage{multirow}
\usepackage{slashed}
\usepackage{braket}
\usepackage{bm}
\usepackage{microtype}

\begin{document}

\title{Unified study of hyperon semileptonic decays in a relativistic three-quark model}

\author{Ru-Hui Ni}
\affiliation{School of Physical Sciences, University of Chinese Academy of Sciences, Beijing 100049, China}
\author{Zhen-Yang Wang}
\affiliation{Physics Department, Ningbo University, Zhejiang 315211, China}
\author{Jia-Jun Wu}
\email[Corresponding author: ]{wujiajun@ucas.ac.cn}
\affiliation{School of Physical Sciences, University of Chinese Academy of Sciences, Beijing 100049, China}
\affiliation{Southern Center for Nuclear-Science Theory (SCNT), Institute of Modern Physics, Chinese Academy of Sciences, Huizhou 516000, China}
\author{Bing-Song Zou}
\email[Corresponding author: ]{zoubs@mail.tsinghua.edu.cn}
\affiliation{Department of Physics and Center for High Energy Physics, Tsinghua University, Beijing 100084, China}

\begin{abstract}
We present a unified theoretical study of semileptonic decays of ground state octet hyperons using the relativistic three-quark model (R3QM).
A key innovation of our approach is that all baryon wave functions are determined by fitting the baryon mass spectrum with a semirelativistic potential model, leading to predictions for weak transition amplitudes without free parameters.
Adopting these wave functions, we calculate the decay widths, branching fractions, lepton flavor universality ratios, as well as the angular correlation and spin asymmetry parameters for the octet channels.
The calculated values agree with the available experimental data and give predictions for channels with limited experimental information.
We further compute the complete set of octet transition form factors without any additional free parameters, so that the weak current can be examined beyond the rate observables.
In the well measured $\Lambda \to p \ell^-\bar{\nu}_\ell$ channel, the calculated leading vector and axial-vector form factors, $f_1(0)$ and $g_1(0)$, agree well with recent lattice QCD results, and the $g_1/f_1$ ratio is consistent with recent BESIII measurements.
Beyond the leading vector and axial-vector terms, the complete form factor set separates the weak magnetism, second class, and the pole contribution associated with the partially conserved axial current (PCAC) relation.
The weak magnetism term $f_2$ shows the clearest channel dependence compared with lattice QCD results, and its smaller values in some channels may point to transverse current strength not fully saturated by pure $qqq$ valence components.
This work provides a framework for connecting octet hyperon weak form factors to the spin--flavor and spatial structure of baryons at the quark level, and gives testable weak current observables for future hyperon semileptonic decay measurements.
\end{abstract}

\pacs{}

\maketitle

\section{Introduction}
\label{sec:intro}

Hyperon semileptonic decays proceed through the charged current weak interaction.
At tree level, the amplitude separates into two vertices.
The leptonic vertex $W\ell\nu_\ell$ is fixed by the Standard Model, while the effective hadronic vertex $WHH$ contains the nonperturbative baryon transition dynamics.
With the leptonic vertex fixed, the baryon matrix element can be used to examine the spin--flavor and spatial structure of hyperons.
In this sense, the virtual $W$ boson plays a role analogous to that of the virtual photon in deep inelastic scattering.
The separation of the fixed leptonic vertex from the hadronic matrix element makes semileptonic decays cleaner than purely hadronic decays, where final-state interactions mix the weak transition with subsequent hadronic rescattering.

Theoretically, all descriptions of these decays rely on Lorentz invariant hadronic form factors.
For ground state spin-$1/2 \to 1/2$ octet transitions, the matrix element is usually described by six form factors, including three vector form factors $f_{1,2,3}$ and three axial-vector form factors $g_{1,2,3}$~\cite{Gaillard:1984ny,Garcia:1985xz,Cabibbo:2003cu,Mateu:2005wi}.
The leading form factors $f_1$ and $g_1$ determine the total decay rate, encoding the vector charge and axial-vector strength, respectively. 
The subleading form factors carry more detailed structural information, with $f_2$ describing weak magnetism, $f_3$ and $g_2$ corresponding to second-class currents induced by flavor symmetry breaking, and $g_3$ representing the longitudinal axial-vector component.
Despite decades of theoretical effort, our understanding of these form factors remains incomplete.

These form factors have been studied using various approaches at both the hadron level and the quark--gluon level. 
At the hadron level, flavor $\mathrm{SU}(3)$ analyses~\cite{Gaillard:1984ny, Garcia:1985xz, Donoghue:1986th, Flores-Mendieta:1998tfv,Flores-Mendieta:2004cyh, Ratcliffe:1998wn, Cabibbo:2003ea, Cabibbo:2003cu, Mateu:2005wi, Pham:2012db, Wang:2019alu} adopt the Cabibbo $F$ and $D$ couplings, introduce flavor $\mathrm{SU}(3)$ breaking through fitted parameters, and estimate $f_2$ from the baryon magnetic moments with the conserved vector current (CVC) hypothesis. 
At the same level, baryon chiral perturbation theory (B$\chi$PT)~\cite{Bijnens:1985kj,Krause:1990xc,Anderson:1993as,Villadoro:2006nj,Lacour:2007wm,Geng:2009ik,Luty:1993gi,Jiang:2008aqa,Jiang:2009sf,Ledwig:2014rfa,Sauerwein:2021jxb,Jenkins:1990jv,Jenkins:1991es,Zhu:2000zf,Kaiser:2001yc} treats the weak current in a chiral expansion, where meson loops and low-energy constants generate the recoil corrections, induced terms, and flavor breaking effects. 
All approaches at the hadron level rely on phenomenological parameterizations, such as the widely used dipole form factor, to describe the $q^2$ dependence of form factors. 
These parametrizations remain purely descriptive, because the form factors are fitted to experimental decay data and the same fitted parameters are then used to account for those data. 
This procedure leads to a circular interpretation and gives little insight into the hadronic structure behind the weak transition.

At the quark--gluon level, hyperon semileptonic matrix elements can be computed from first principles in lattice QCD (LQCD)~\cite{Guadagnoli:2004qw,Becirevic:2004bb,Guadagnoli:2006gj,Sasaki:2006jp,Sasaki:2008ha,Sasaki:2011hu,Sasaki:2012ne,Shanahan:2015dka,Sasaki:2017jue,Bacchio:2025auj,Alexandrou:2026noh}. 
These matrix elements give the hadronic part of the decay amplitudes, from which the decay rates and rate ratios are obtained.
The $q^2$-dependent form factors can also be obtained from correlation functions in QCD sum rule (QCDSR) studies~\cite{Wang:2006yz,Zhang:2024ick,Ahmadi:2025oal} and used to calculate decay widths and branching fractions.
However, the matrix elements and form factors obtained in this way are usually given as complete numerical quantities.
Their decomposition into intuitive quark--gluon level mechanisms, or their connection to specific features of the baryon wave function, is still difficult.
Furthermore, none of these approaches provide a unified, dynamically consistent description of all ground state hyperon semileptonic decays.

Such a unified description can be naturally carried out in quark models, since the weak matrix elements for all ground state hyperon channels can be calculated with the baryon wave functions and the appropriate quark weak current.
In early quark model studies~\cite{Donoghue:1981uk,Barik:1985in,Beyer:1986cf}, the relevant axial-vector, weak magnetism, and weak electricity form factors in baryon semileptonic decays were calculated.
Recoil corrections, relativistic spin effects, and chiral $\mathrm{SU}(3)$ breaking contributions were later incorporated in light-front, covariant, chiral-quark, and chiral-soliton calculations~\cite{Schlumpf:1994fb,Ohlsson:1998bk,Kim:1999uf,Ledwig:2008ku,Sharma:2009hg,Dahiya:2024ekj,Faessler:2008ix,Ramalho:2015jem}.
These calculations showed that explicit internal dynamics beyond the static $\mathrm{SU}(6)$ limit are needed for a realistic description of the hyperon weak form factors.
However, most existing quark model calculations are still affected by the same circularity problem, because the wave function parameters are typically tuned to reproduce weak decay data rather than fixed independently, and the calculations are often restricted to only a few channels.

In the past decade, two critical experimental developments have made a new approach both timely and feasible.
First, experimental information on hyperon semileptonic decays has become much more precise and complete.
Early fixed target experiments established the branching fractions and leading form factor ratios $g_1/f_1$ for the main $\Lambda$, $\Sigma$, and $\Xi$ decay channels~\cite{Bristol-Geneva-Heidelberg-Orsay-Rutherford-Strasbourg:1981uuv,Bristol-Geneva-Heidelberg-Orsay-Rutherford-Strasbourg:1983jzt,Bristol-Geneva-Heidelberg-Orsay-Rutherford-Strasbourg:1983jpz,Bristol-Geneva-Heidelberg-Orsay-Rutherford-Strasbourg:1983rna,Wise:1980xx,Dworkin:1990dd,Hsueh:1988ar}.
The KTeV and NA48/1 Collaborations later measured the rare $\Xi^0 \to \Sigma^+ e^- \bar{\nu}_e$ decay~\cite{KTeVE832E799:1999tte,KTeV:2001djr,NA48I:2006yat,NA481:2012dtx}.
Most recently, the BESIII Collaboration reported a precise measurement of the $\Lambda \to p \mu^- \bar{\nu}_\mu$ branching fraction~\cite{BESIII:2021ynj}, and the LHCb Collaboration reported the lepton flavor universality ratio $R_{\mu e}$~\cite{LHCb:2025wld}.
In a landmark 2025 study, the BESIII Collaboration further reported the first experimental determinations of the subleading weak magnetism and weak electricity ratios in $\Lambda \to p e^- \bar{\nu}_e$~\cite{BESIII:2025hgj}. These determinations marked the first experimental sensitivity to the full structure of the hadronic weak current.
Second, with high-precision measurements of the ground state baryon mass spectrum~\cite{ParticleDataGroup:2024cfk}, the spatial and spin--flavor structure of baryon wave functions can now be constrained much more tightly, making it possible to determine these wave functions with much higher accuracy.

Against this background, we present a unified study of all ground state octet hyperon semileptonic decays within the relativistic three-quark model (R3QM), with no additional free parameters in the weak transition amplitudes. 
Compared with our previous nonrelativistic work~\cite{Wu:2013kla}, the present R3QM calculation differs in both the input wave functions and the treatment of recoil dynamics. 
The constituent quark masses and baryon wave functions are fixed a priori by fitting the baryon mass spectrum, and the transition amplitudes are calculated in the instant-form Bakamjian--Thomas relativistic formalism with full Dirac spinors, finite recoil, and Wigner rotations.
These amplitudes are used to obtain the decay widths and branching fractions and to project out the full set of form factors, with $g_3$ reconstructed through the partially conserved axial current (PCAC) relation.

This approach addresses all limitations of previous methods: it breaks the circular logic of parameter fitting, provides a unified treatment of all channels, enables explicit tracing of form factors to wave function structure at the quark level, and establishes a clean valence three-quark reference against which future studies can quantify nonvalence contributions, such as meson cloud effects and other higher Fock components.
With the branching fractions, rate ratios, angular correlation and spin asymmetry parameters, and form factors obtained in the same calculation, we compare them with available experimental data and theoretical calculations.
We also discuss how these observables reflect the internal structure and weak transition dynamics of hyperons.

This paper is organized as follows.
In Secs.~\ref{sec:mass_spectrum} and \ref{sec:semileptonic_framework}, we introduce the semirelativistic potential model and the relativistic three-quark model.
In Sec.~\ref{sec:results}, we present the numerical results for the form factors, branching fractions, angular observables, and compare them with the experimental data and other theoretical predictions. 
Finally, a summary is given in Sec.~\ref{sec:summary}.

\section{Baryon wave functions and mass spectrum}
\label{sec:mass_spectrum}

\subsection{Baryon wave functions}

In the constituent quark model, a baryon is described as a bound state of three constituent quarks. 
The total wave function takes the form
\begin{equation}
	\left|B,J^P\right\rangle
	=
	\left|\mathbf{1}_c\right\rangle
	\otimes
	\phi_B
	\otimes
	\left[\Psi\otimes\chi_S\right]_{J}.
\end{equation}
Here $\left|\mathbf{1}_c\right\rangle$ is the antisymmetric $\mathrm{SU}(3)_c$ color singlet wave function, and $\Psi$, $\phi_B$, and $\chi_S$ denote the spatial, flavor, and spin wave functions, respectively. 
The subscript $J$ denotes the coupling of the orbital angular momentum carried by $\Psi$ and the total quark spin $S$ to the baryon spin $J$. 
The parity $P$ is determined by the spatial wave function.

The common color factor is suppressed in the following.
Its antisymmetry requires the remaining spatial--spin--flavor wave function to be symmetric under permutations of identical quarks. 
For light baryons, this remaining part is classified by combining the flavor and spin wave functions into $\mathrm{SU}(6)$ spin--flavor wave functions, while the spatial wave functions are classified under $\mathrm{O}(3)$~\cite{Isgur:1978xj,Capstick:1986ter,Zhong:2024mnt}. 
For heavy baryons, unequal constituent quark masses break the exact $S_3$ permutation symmetry, so the basis is organized according to the exchange parity of a selected quark pair.

For the spatial wave function $\Psi$, the center-of-mass motion is first separated, and the two internal modes are described by the mass-independent Jacobi coordinates
\begin{align}
	\bm{\rho}  &= \frac{1}{\sqrt{2}}(\bm{r}_1 - \bm{r}_2),\nonumber \\
	\bm{\lambda} &= \frac{1}{\sqrt{6}}
	(\bm{r}_1+\bm{r}_2-2\bm{r}_3),\nonumber \\
	\bm{R}_{\mathrm{cm}} &= \frac{1}{3}(\bm{r}_1+\bm{r}_2+\bm{r}_3),
	\label{eq:jacobi_r}
\end{align}
where $\bm r_i$ is the position of quark $i$. 
The coordinate $\bm\rho$ describes the relative motion of quarks 1 and 2, $\bm\lambda$ describes the motion of quark 3 relative to the $(12)$ pair, and $\bm R_{\mathrm{cm}}$ describes the center-of-mass motion. 
The corresponding Jacobi momenta are
\begin{align}
	\bm{k}_\rho &= \frac{1}{\sqrt{2}}(\bm{k}_1-\bm{k}_2),\nonumber\\
	\bm{k}_\lambda &= \frac{1}{\sqrt{6}}
	(\bm{k}_1+\bm{k}_2-2\bm{k}_3),\nonumber\\
	\bm{P}_B &= \bm{k}_1+\bm{k}_2+\bm{k}_3,
	\label{eq:pi_jacobi}
\end{align}
where $\bm k_i$ is the momentum of quark $i$ and $\bm P_B$ is the baryon momentum. 
In the baryon rest frame, $\bm P_B=\bm0$, and the inverse transformation becomes
\begin{align*}
	\bm{k}_1 &= \frac{1}{\sqrt{2}}\,\bm{k}_\rho
	+ \frac{1}{\sqrt{6}}\,\bm{k}_\lambda,\\
	\bm{k}_2 &= -\frac{1}{\sqrt{2}}\,\bm{k}_\rho
	+ \frac{1}{\sqrt{6}}\,\bm{k}_\lambda,\\
	\bm{k}_3 &= -\sqrt{\frac{2}{3}}\,\bm{k}_\lambda.
\end{align*}
The rest-frame spatial wave function therefore depends only on the internal coordinates $(\bm\rho,\bm\lambda)$, or equivalently on the momenta $(\bm k_\rho,\bm k_\lambda)$.

In this work, we expand the spatial wave function in the simple harmonic oscillator (SHO) basis~\cite{Isgur:1978xj,Capstick:1986ter}. 
In coordinate space, the basis function for each Jacobi mode takes the form
\begin{equation}
	\psi_{n_\xi l_\xi m_\xi}(\bm{\xi})
	= \mathcal R_{n_\xi l_\xi}(\xi;\alpha_\xi)\,
	Y_{l_\xi m_\xi}(\hat{\bm{\xi}}),
	\label{eq:ho_mode}
\end{equation}
where $\bm\xi$ denotes either $\bm\rho$ or $\bm\lambda$, $\xi=|\bm\xi|$, and $\alpha_\xi$ is the corresponding oscillator parameter. 
The quantum numbers $n_\xi$, $l_\xi$, and $m_\xi$ denote the radial excitation, orbital angular momentum, and its projection, respectively. 
The normalized radial wave function is
\begin{align}
	\mathcal R_{n_\xi l_\xi}(\xi;\alpha_\xi)
	&= \alpha_\xi^{3/2}
	\left[\frac{2^{l_\xi+2-n_\xi}\,(2l_\xi+2n_\xi+1)!!}
	     {\sqrt{\pi}\,n_\xi!\,[(2l_\xi+1)!!]^2}\right]^{1/2}
	\left(\alpha_\xi\xi\right)^{l_\xi} \nonumber \\
	&\quad \times \exp\!\left[-\tfrac{1}{2}
	\left(\alpha_\xi\xi\right)^{2}\right]
	{}_1F_1\!\left(-n_\xi;\,l_\xi+\tfrac{3}{2};\,
	\left(\alpha_\xi\xi\right)^{2}\right).
	\label{eq:ho_explicit}
\end{align}
The function ${}_1F_1(a;b;z)$ is the confluent hypergeometric function.

The orbital angular momenta of the $\rho$ and $\lambda$ modes are coupled to a total orbital angular momentum $L$. 
The corresponding spatial basis function is defined by
\begin{align}
	\Psi_{N_{\mathrm{osc}}LM_L}^{(n_\rho l_\rho,n_\lambda l_\lambda)}
	(\bm\rho,\bm\lambda)
	&= \sum_{m_\rho,m_\lambda}
	\langle l_\rho m_\rho\,l_\lambda m_\lambda|LM_L\rangle
	\psi_{n_\rho l_\rho m_\rho}(\bm{\rho})
	\psi_{n_\lambda l_\lambda m_\lambda}(\bm{\lambda}).
	\label{eq:coupled_spatial_basis}
\end{align}
Here the oscillator shell number is $N_{\mathrm{osc}}=2n_\rho+l_\rho+2n_\lambda+l_\lambda$.
It counts the total number of oscillator quanta carried by the two internal modes. 
The parity of the baryon is determined by the internal orbital angular momenta, $P=(-1)^{l_\rho+l_\lambda}$.

For the ground state baryon spectrum, the spatial basis is restricted to $L=0$.
To include configuration mixing, the expansion is truncated at the first positive parity excited shell, $N_{\mathrm{osc}}\leq2$. 
This truncated space contains four primitive channels $(n_\rho l_\rho, n_\lambda l_\lambda)$: the $N_{\mathrm{osc}}=0$ ground state $(00,00)$, two $N_{\mathrm{osc}}=2$ radial excitations $(10,00)$ and $(00,10)$, and one $N_{\mathrm{osc}}=2$ internal orbital excitation $(01,01)$ where two $P$ waves ($l_\rho=l_\lambda=1$) couple to $L=0$.

These primitive functions are combined into spatial states with definite permutation symmetry:
\begin{align}
	\Psi_{000}^{s}
	&=\Psi_{000}^{(00,00)},\nonumber\\
	\Psi_{200}^{s}
	&=\frac{1}{\sqrt{2}}
	\left(\Psi_{200}^{(10,00)}+\Psi_{200}^{(00,10)}\right),\nonumber\\
	\Psi_{200}^{\rho}
	&=-\Psi_{200}^{(01,01)},\nonumber\\
	\Psi_{200}^{\lambda}
	&=\frac{1}{\sqrt{2}}
	\left(-\Psi_{200}^{(10,00)}+\Psi_{200}^{(00,10)}\right).
\end{align}
The superscripts $s$, $\rho$, and $\lambda$ specify the exchange symmetry of the first two quarks: $s$ denotes a totally symmetric state, while $\rho$ and $\lambda$ denote odd (antisymmetric) and even (symmetric) exchange parity, respectively. 
The functions $\Psi_{000}^{s}$, $\Psi_{200}^{s}$, and $\Psi_{200}^{\lambda}$ consist purely of $S$-wave modes. 
The function $\Psi_{200}^{\rho}$ carries the angular correlation absent from these pure $S$-wave products.

For the light octet, the lowest spin-$1/2$ state consistent with the Pauli principle is
\begin{align}
	\left|B_1^{\mathbf{8}},1/2^+\right\rangle
	&=\Phi_{\mathbf{56},\mathbf 8}^{s} \Psi_{000}^{s} \nonumber\\
	&=\frac{1}{\sqrt{2}}
	\left(
	\phi_{\mathbf 8}^{\rho}\chi_{1/2}^{\rho}
	+\phi_{\mathbf 8}^{\lambda}\chi_{1/2}^{\lambda}
	\right)\Psi_{000}^{s}.
	\label{eq:wf_octet}
\end{align}
Here $\Phi_{\mathbf{56},\mathbf 8}^{s}$ denotes the complete spin--flavor wave function of the flavor octet in the $\mathbf{56}$ representation of $\mathrm{SU}(6)$, while $\phi_{\mathbf 8}$ and $\chi_{1/2}$ denote the flavor-octet and spin-$1/2$ wave functions, respectively. 
Their explicit expressions for all light octet baryons are collected in Appendix~\ref{app:wavefunctions}.

At $N_{\mathrm{osc}}=2$, the mixed-symmetry spin--flavor functions in the $\mathbf{70}$ representation are
\begin{align}
	\Phi_{\mathbf{70}}^{\rho}
	&=\frac{1}{\sqrt{2}}
	\left(\phi_{\mathbf 8}^{\rho}\chi_{1/2}^{\lambda}
	+\phi_{\mathbf 8}^{\lambda}\chi_{1/2}^{\rho}\right),\nonumber\\
	\Phi_{\mathbf{70}}^{\lambda}
	&=\frac{1}{\sqrt{2}}
	\left(\phi_{\mathbf 8}^{\rho}\chi_{1/2}^{\rho}
	-\phi_{\mathbf 8}^{\lambda}\chi_{1/2}^{\lambda}\right).
	\label{eq:phi_70}
\end{align}
Combining the spatial and spin--flavor functions with the same permutation symmetry gives the remaining light octet basis states,
\begin{align}
	\left|B_2^{\mathbf{8}},1/2^+\right\rangle
	&=\Psi_{200}^{s}\Phi_{\mathbf{56},\mathbf 8}^{s},\nonumber\\
	\left|B_3^{\mathbf{8}},1/2^+\right\rangle
	&=\frac{1}{\sqrt{2}}
	\left(
	\Psi_{200}^{\rho}\Phi_{\mathbf{70}}^{\rho}
	+\Psi_{200}^{\lambda}\Phi_{\mathbf{70}}^{\lambda}
	\right).
\end{align}
The states $\left|B_1^{\mathbf{8}},1/2^+\right\rangle$ and $\left|B_2^{\mathbf{8}},1/2^+\right\rangle$ are direct products of totally symmetric spatial and spin--flavor wave functions. 
In $\left|B_3^{\mathbf{8}},1/2^+\right\rangle$, the $\rho$ and $\lambda$ spatial functions are paired with spin--flavor functions of the same exchange symmetry, giving a totally symmetric sum. 
The four spatial products reduce to three independent light octet configurations allowed by the Pauli principle.
A physical light octet baryon is expanded in these three configurations as
\begin{equation}
	\left|B^{\mathbf{8}},1/2^+\right\rangle
	=\sum_{\kappa=1}^{3}c_\kappa^{{\mathbf{8}},1/2}
	\left|B_\kappa^{\mathbf{8}},1/2^+\right\rangle .
	\label{eq:baryon_state_expansion}
\end{equation}
Here $\kappa=1,2,3$ labels $\left|B_1^{\mathbf{8}},1/2^+\right\rangle$, $\left|B_2^{\mathbf{8}},1/2^+\right\rangle$, and $\left|B_3^{\mathbf{8}},1/2^+\right\rangle$, and $c_\kappa^{{\mathbf{8}},1/2}$ is the corresponding mixing coefficient, satisfy $\sum_{\kappa=1}^{3}\left|c_\kappa^{{\mathbf{8}},1/2}\right|^2=1$.

For the light flavor decuplet, the flavor wave function $\phi_{\mathbf{10}}^{s}$ and the spin-$3/2$ wave function $\chi_{3/2}^{s}$ are both completely symmetric.
Their product gives the totally symmetric spin--flavor wave function in the $\mathbf{56}$ representation, $\Phi_{\mathbf{56},\mathbf{10}}^{s}=\phi_{\mathbf{10}}^{s}\chi_{3/2}^{s}$.
Because $\Phi_{\mathbf{56},\mathbf{10}}^{s}$ is totally symmetric, the Pauli principle restricts the spatial wave function to the symmetric modes $\Psi_{000}^{s}$ and $\Psi_{200}^{s}$. 
With $S=3/2$ and $L=0$, the total angular momentum is $J=S=3/2$, and the retained $N_{\mathrm{osc}}\leq2$ basis for a $J^P=3/2^+$ decuplet baryon is
\begin{align}
	\left|B_1^{\mathbf{10}},3/2^+\right\rangle
	&=\Psi_{000}^{s}\Phi_{\mathbf{56},\mathbf{10}}^{s}
	=\Psi_{000}^{s}\phi_{\mathbf{10}}^{s}\chi_{3/2}^{s},\nonumber\\
	\left|B_2^{\mathbf{10}},3/2^+\right\rangle
	&=\Psi_{200}^{s}\Phi_{\mathbf{56},\mathbf{10}}^{s}
	=\Psi_{200}^{s}\phi_{\mathbf{10}}^{s}\chi_{3/2}^{s}.
	\label{eq:light_decuplet_basis}
\end{align}
The decuplet component in the mixed-symmetry $\mathbf{70}$ representation has $S=1/2$; for $L=0$, it couples only to $J^P=1/2^+$, whereas a $J^P=3/2^+$ state requires $L=2$ and is thus excluded from the present truncation.
The physical decuplet state is therefore expanded in these two configurations as
\begin{equation}
	\left|B^{\mathbf{10}},3/2^+\right\rangle
	=\sum_{\kappa=1}^{2}c_\kappa^{\mathbf{10},3/2}
	\left|B_\kappa^{\mathbf{10}},3/2^+\right\rangle,
	\label{eq:decuplet_state_expansion}
\end{equation}
with the normalization $\sum_{\kappa=1}^{2}|c_\kappa^{\mathbf{10},3/2}|^2=1$.

For heavy baryons, the basis is organized by the exchange symmetry of quarks 1 and 2. 
Under the $1\leftrightarrow2$ permutation, the Jacobi coordinates transform as $\bm\rho \to -\bm\rho$ and $\bm\lambda \to \bm\lambda$, yielding the spatial exchange parity
\begin{equation}
	\widehat P_{12}
	\Psi_{N_{\mathrm{osc}}LM_L}^{(n_\rho l_\rho,n_\lambda l_\lambda)}
	=(-1)^{l_\rho}
	\Psi_{N_{\mathrm{osc}}LM_L}^{(n_\rho l_\rho,n_\lambda l_\lambda)}.
	\label{eq:spatial_exchange_parity}
\end{equation}
Thus, the spatial functions $\Psi_{000}^{(00,00)}$, $\Psi_{200}^{(10,00)}$, and $\Psi_{200}^{(00,10)}$ are symmetric, whereas $\Psi_{200}^{(01,01)}$ is antisymmetric. 
To satisfy the Pauli principle, each spatial function is paired with a spin--flavor state of the same exchange parity. Their product is symmetric under $1\leftrightarrow2$. Together with the antisymmetric color wave function, it gives an antisymmetric three-quark state.
For a singly heavy baryon, quarks 1 and 2 form the light pair.
The light pair flavor wave function is antisymmetric ($\phi_A$) for the $\overline{\mathbf 3}_F$ multiplet $(\Lambda_Q,\Xi_Q)$ and symmetric ($\phi_S$) for the $\mathbf 6_F$ multiplet $(\Sigma_Q,\Xi'_Q,\Omega_Q)$. 
For the $nsQ$ system, the constituent mass difference $m_s \neq m_n$ breaks the light flavor $\mathrm{SU}(3)_F$ symmetry, allowing the color-magnetic spin--spin interaction to mix the $\overline{\mathbf 3}_F$ and $\mathbf 6_F$ configurations. 
The physical $\Xi_Q$ and $\Xi'_Q$ states with $J^P=1/2^+$ are therefore obtained by diagonalizing the Hamiltonian in the combined eight-dimensional basis. 
For $J^P=3/2^+$, the three symmetric spatial functions are combined with $\phi_S\chi_{3/2}^{s}$. For $\Xi_Q^*$, the antisymmetric product $\Psi_{200}^{(01,01)}\phi_A\chi_{3/2}^{s}$ provides the fourth basis configuration.
For the doubly charmed baryons considered here, quarks 1 and 2 are identical charm quarks, so $\phi_A$ vanishes and only $\phi_S$ remains. 
The explicit basis states and the expansions for all heavy baryons are given in Appendix~\ref{app:heavy_baryon_bases}.

\subsection{Mass spectrum}

The physical baryon mass spectrum and configuration-mixed wave functions are determined within a semirelativistic quark potential model~\cite{Capstick:1986ter,Zhong:2024mnt} using the $N_{\mathrm{osc}}\leq2$ basis constructed above. 
The Hamiltonian takes the form
\begin{equation}
	\mathcal H = \sum_{i=1}^{3} \sqrt{\bm{k}_i^{\,2} + m_i^{2}}
	+ \sum_{i<j} V(r_{ij}) + C_0,
	\label{eq:H_sr}
\end{equation}
where $m_i$ is the constituent mass of quark $i$, $r_{ij}=|\bm r_i-\bm r_j|$, $V(r_{ij})$ is the quark--quark interaction, and $C_0$ is a common zero-point energy. 
The interaction $V(r_{ij})$ contains a Cornell potential and a contact hyperfine interaction~\cite{Eichten:1978tg,Capstick:1986ter},
\begin{equation}
	V(r_{ij})=V_{\mathrm{corn}}(r_{ij})+V_{\mathrm{hyp}}(r_{ij}).
	\label{eq:V_decomp}
\end{equation}
The Cornell potential is
\begin{equation}
	V_{\mathrm{corn}}(r_{ij})
	=-\frac{2}{3}\frac{\alpha_s^{ij}}{r_{ij}}+\frac{b}{2}r_{ij},
	\label{eq:V_Corn}
\end{equation}
where $b$ is the string tension of the linear confinement. The flavor-dependent Coulomb coupling is $\alpha_s^{ij}=g_i^{\mathrm{OGE}}g_j^{\mathrm{OGE}}$~\cite{Huang:2015nja}, and $g_i^{\mathrm{OGE}}$ is the one-gluon-exchange (OGE) coupling for quark flavor $i$.
The spin-dependent term $V_{\mathrm{hyp}}(r_{ij})$ is the OGE contact interaction smeared by a Gaussian function,
\begin{equation}
	V_{\mathrm{hyp}}(r_{ij})
	=\frac{16\pi\alpha_s^{ij}}{9m_im_j}
	\frac{\exp[-r_{ij}^{2}/r_{0}^{2}(ij)]}
	{\pi^{3/2}r_{0}^{3}(ij)}
	\bigl(\bm S_i\cdot\bm S_j\bigr),
	\label{eq:V_SS}
\end{equation}
where $\bm S_i$ is the spin operator of quark $i$. The reduced mass of the interacting pair is $\mu_{ij}=\frac{m_im_j}{m_i+m_j}$, and the smearing radius is parametrized as $r_0(ij)=A_{\mathrm{sm}}\bigl(2\mu_{ij}\bigr)^{-B_{\mathrm{sm}}}$~\cite{Silvestre-Brac:1996myf}. 
It should be emphasized that spin--orbit and tensor interactions are omitted in the present Hamiltonian, so the spatial basis is restricted to the $L=0$ subspace.

For a baryon with spin $J$, the mass eigenvalue $M_B$ and the mixing coefficients $c_\kappa^{B,J}$ satisfy the eigenvalue equation
\begin{equation}
	\sum_{\kappa'=1}^{n_B}
	\left(\mathcal H_{\kappa\kappa'}^{B,J}
	-M_B\delta_{\kappa\kappa'}\right)c_{\kappa'}^{B,J}
	=0,
	\label{eq:mass_eigenproblem}
\end{equation}
where $n_B$ is the basis dimension and the Hamiltonian matrix element is $\mathcal H_{\kappa\kappa'}^{B,J}
= \left\langle B_\kappa,J^P\right| \mathcal H \left|B_{\kappa'},J^P\right\rangle$.
To calculate the Hamiltonian matrix elements, the harmonic oscillator size parameters $\alpha_\rho$ and $\alpha_\lambda$ in Eq.~\eqref{eq:ho_mode} are parameterized in terms of the harmonic oscillator stiffness factor $\mathcal K$,
\begin{align}
	\alpha_\rho=(3\mathcal K\mu_\rho)^{1/4}, \quad 
	\alpha_\lambda=(3\mathcal K\mu_\lambda)^{1/4},
	\label{eq:sho_widths}
\end{align}
where $\mu_\rho=2m_1m_2/(m_1+m_2)$ and $\mu_\lambda=3(m_1+m_2)m_3/[2(m_1+m_2+m_3)]$ are the reduced masses for the $\rho$ and $\lambda$ Jacobi coordinates, respectively. 
Here $\mathcal K$ serves as a variational basis parameter rather than an adjustable parameter in the Hamiltonian. 
For a given set of potential parameters, $\mathcal K$ is determined by minimizing the proton ground state energy of the spin-independent Hamiltonian in the $N_{\mathrm{osc}}\leq2$ space. 
With $\mathcal K$ fixed by the proton, the size parameters $\alpha_\rho$ and $\alpha_\lambda$ for all other baryons are directly determined by their corresponding reduced masses. 
Diagonalizing the Hamiltonian matrix in this optimized basis yields the physical baryon masses and configuration mixing coefficients.

The Hamiltonian parameters are determined by minimizing $\chi^2$ against the reference masses of light, singly heavy, and doubly charmed baryons. 
The fitted potential parameters are listed in Table~\ref{tab:potential_params}, and the calculated mass spectrum is compared with experimental and lattice data in Table~\ref{tab:baryon_spectrum_fit}. 
In the truncated harmonic oscillator basis ($N_{\mathrm{osc}}\leq2$, $L=0$), we obtain the physical baryon states as configuration mixtures of the basis states. 
The fitted constituent quark masses and these configuration mixed baryon wave functions are directly adopted in our subsequent calculations of the hyperon semileptonic decay form factors and decay observables. 
The explicit configuration mixing coefficients for all light, singly heavy, and doubly heavy baryons included in the spectrum fit are collected in Table~\ref{tab:baryon_configuration_mixing} of Appendix~\ref{app:baryon_mixing_coefficients}.

\begin{table}[t!]
	\centering
	\renewcommand{\arraystretch}{1.12}
	\setlength{\tabcolsep}{8pt}
	\caption{Parameters of the semirelativistic potential model. 
	}
	\label{tab:potential_params}
	\begin{tabular}{cccc}
			\hline\hline
			$m_{u/d}$ (MeV) & $m_s$ (MeV) & $m_c$ (MeV) & $m_b$ (MeV) \\
			368.5 & 460.7 & 1550.8 & 5005.7 \\
			\hline
			$g_{u/d}^{\mathrm{OGE}}$ &
			$g_s^{\mathrm{OGE}}$ &
			$g_c^{\mathrm{OGE}}$ &
			$g_b^{\mathrm{OGE}}$ \\
			0.9445 & 0.7592 & 0.3888 & 0.4245 \\
			\hline
			$A_{\mathrm{sm}}$ (GeV$^{B_{\mathrm{sm}}-1}$) &
			$B_{\mathrm{sm}}$ & $b$ (GeV$^{2}$) & $C_0$ (MeV) \\
			0.9425 & 0.8869 & 0.1472 & $-654.7$ \\
			\hline\hline
	\end{tabular}
\end{table}

\begin{table*}[t!]
	\centering
	\setlength{\tabcolsep}{15pt}
	\renewcommand{\arraystretch}{1.15}
	\caption{Ground state baryon masses obtained with the semirelativistic potential model, compared with the reference masses used in the fit. 
		All masses are in MeV. 
		The 12 model parameters are determined by minimizing $\chi^2=\sum_h (M_{h,\mathrm{thy.}}-M_{h,\mathrm{ref.}})^2/\sigma_h^2$ for the 26 reference masses listed in the table. 
		A common fitting uncertainty $\sigma_h=5$ MeV is used for all states. 
		The experimental masses of the 23 observed light and singly heavy baryons are taken from the RPP~\cite{ParticleDataGroup:2024cfk}. 
		The Lattice QCD result $M_{\Omega_b^{*}}\simeq 6085$ MeV~\cite{Brown:2014ena} is used for the unobserved $\Omega_b^{*}$. 
		For the doubly charmed baryons, the reference mass of the isospin symmetric $\Xi_{cc}$ state is set to the measured $\Xi_{cc}^{++}$ mass, $M_{\Xi_{cc}^{++}}\simeq 3622$ MeV~\cite{LHCb:2020XiCCMass}. 
		The preliminary result of the LHCb Collaboration, $M_{\Omega_{cc}^{+}}\simeq 3726$ MeV~\cite{Xu:2026OmegaCC}, is used for $\Omega_{cc}$. 
		The fit gives $\chi^2/\mathrm{d.o.f.}=148.76/14=10.63$.}
	\label{tab:baryon_spectrum_fit}
	\begin{tabular}{ccccccccccc}
		\hline\hline
		\multicolumn{3}{c}{Light baryons} &&
		\multicolumn{3}{c}{Charmed baryons} &&
		\multicolumn{3}{c}{Bottom baryons} \\
		\cmidrule(lr){1-3}\cmidrule(lr){5-7}\cmidrule(lr){9-11}
		States & $M_{\mathrm{thy.}}$ & $M_{\mathrm{ref.}}$ &&
		States & $M_{\mathrm{thy.}}$ & $M_{\mathrm{ref.}}$ &&
		States & $M_{\mathrm{thy.}}$ & $M_{\mathrm{ref.}}$ \\
		\hline
		$p$          & 947  & 938  && $\Lambda_c$     & 2281 & 2286 && $\Lambda_b$     & 5611 & 5620 \\
		$\Delta$     & 1251 & 1235 && $\Sigma_c$       & 2448 & 2453 && $\Sigma_b$       & 5803 & 5811 \\
		$\Lambda$    & 1123 & 1116 && $\Sigma_c^{*}$   & 2498 & 2519 && $\Sigma_b^{*}$   & 5827 & 5830 \\
		$\Sigma$     & 1180 & 1193 && $\Xi_c$          & 2465 & 2469 && $\Xi_b$          & 5799 & 5794 \\
		$\Sigma^{*}$ & 1397 & 1385 && $\Xi_c^{\prime}$ & 2580 & 2579 && $\Xi_b^{\prime}$ & 5936 & 5935 \\
		$\Xi$        & 1319 & 1318 && $\Xi_c^{*}$      & 2624 & 2646 && $\Xi_b^{*}$      & 5956 & 5952 \\
		$\Xi^{*}$    & 1535 & 1533 && $\Omega_c$       & 2706 & 2695 && $\Omega_b$       & 6062 & 6046 \\
		$\Omega$     & 1666 & 1672 && $\Omega_c^{*}$   & 2743 & 2766 && $\Omega_b^{*}$   & 6079 & 6085 \\
		             &      &      && $\Xi_{cc}$       & 3633 & 3622 &&                  &      &      \\
		             &      &      && $\Omega_{cc}$    & 3753 & 3726 &&                  &      &      \\
		\hline\hline
	\end{tabular}
\end{table*}

\section{Semileptonic decay}
\label{sec:semileptonic_framework}

In the R3QM framework, we calculate all kinematically allowed octet-to-octet hyperon semileptonic decays.
The constituent quark masses and baryon rest-frame wave functions are taken from the spectrum calculation in Sec.~\ref{sec:mass_spectrum}, leaving no additional free parameters in the weak matrix elements.
The transition matrix elements are calculated in the impulse approximation, with final-state recoil and Wigner spin rotations implemented through the Bakamjian--Thomas boost~\cite{Bakamjian:1953kh,Faustov:1972rp}. 
From these helicity amplitudes, we obtain the decay widths by phase space integration and extract the corresponding octet-to-octet form factors.

\subsection{Weak transition amplitudes}
\label{sec:quarkmodel_calc}

At tree level, the leptonic and hadronic weak currents are connected by the virtual $W$ boson propagator.
The transition amplitude is then given by~\cite{ParticleDataGroup:2024cfk},
\begin{equation}
	\mathcal{M}(s_\ell,s_{\bar{\nu}};\lambda_f,\lambda_i)
	=
	\frac{G_F}{\sqrt{2}}\,V_{\text{CKM}}\,
	L^\mu_-(s_\ell,s_{\bar{\nu}})\,
	g_{\mu\nu}\,
	H^\nu(\lambda_f,\lambda_i;P_i,P_f),
	\label{eq:Amplitude}
\end{equation}
where $G_F$ is the Fermi constant and $V_{\text{CKM}}$ is the Cabibbo--Kobayashi--Maskawa (CKM) matrix element for the selected flavor transition. 
Here $L^\mu_-$ is the charge-lowering leptonic current, while $H^\nu$ is the hadronic weak current matrix element.
The labels $\lambda_i$ and $\lambda_f$ are the initial and final baryon helicities, while $s_\ell$ and $s_{\bar{\nu}}$ are the charged lepton and antineutrino spin labels.

The leptonic current is
\begin{equation}
	L_-^\mu(s_\ell,s_{\bar{\nu}})
	=
	\bar{u}_\ell(p_\ell,s_\ell)\,
	\gamma^\mu(1-\gamma^5)\,
	v_{\bar{\nu}}(p_\nu,s_{\bar{\nu}}).
\end{equation}
Here $p_\ell=(E_\ell,\bm{p}_\ell)$ and $p_\nu=(E_\nu,\bm{p}_\nu)$ are the charged lepton and antineutrino four momenta, $m_\ell$ is the charged lepton mass, and $E_\nu=|\bm{p}_\nu|$ is used for the massless antineutrino. 
The charge raising case follows by interchanging the charged lepton and neutrino spinors together with the corresponding momentum relabeling.

In the impulse approximation, the hadronic weak current is taken as a sum of one-body quark currents, and the baryon transition matrix element is defined by
\begin{equation}
	H^\nu(\lambda_f,\lambda_i;P_i,P_f)
	\equiv
	\sum_{j=1}^{3} \langle B_f(P_f,\lambda_f)\,| J^\nu_j(0)\,|\,B_i(P_i,\lambda_i)\rangle.
	\label{eq:H_def}
\end{equation}
Here $P_i$ and $P_f$ denote the initial and final baryon four momenta. For each term in the sum, $J^\nu_j$ acts on quark line $j$, while the remaining two quark lines are spectators.
The spatial overlap is calculated in the initial baryon rest frame, $P_i=(M_i,\mathbf{0})$, with the initial spatial wave function expressed in the Jacobi momenta $(\bm{k}_\rho,\bm{k}_\lambda)$.

In the Bakamjian--Thomas (BT) formalism~\cite{Bakamjian:1953kh,Faustov:1972rp}, the final baryon spatial wave function in the moving frame is related to the rest-frame wave function by
~\cite{Ping:2002uj,Ping:2004wz,Ping:2004sh}
\begin{equation}
	\psi_f^{(\bm{P}_f)}(\{\bm{p}'_a\})
	=
	\sqrt{\mathcal{J}_f(\{\bm{p}'_a\},\bm{P}_f)}\;
	\psi_f^{(\mathbf{0})}(\{\bm{k}'_a\}).
	\label{eq:boosted_wf}
\end{equation}
Here $p'_a=(e'_a,\bm{p}'_a)$ and $k'_a=(\epsilon'_a,\bm{k}'_a)$ denote the constituent quark four momenta in the final baryon moving frame and rest frame, respectively, with $k_a^{\prime\mu}=[\Lambda^{-1}(P_f)]^\mu{}_{\nu}p_a^{\prime\nu}$. The Jacobi variables on the two sides are defined with the same convention. The corresponding boost Jacobian is
\begin{equation}
	\mathcal{J}_f(\{\bm{p}'_a\},\bm{P}_f)
	=
	\frac{E_f}{M_f}\prod_{a=1}^{3}\frac{\epsilon'_a}{e'_a},
	\label{eq:jacobian}
\end{equation}
where $a=1,2,3$ runs over the three constituent quark lines in the final baryon, $e'_a=\sqrt{(m'_a)^2+(\bm{p}'_a)^2}$, $\epsilon'_a=\sqrt{(m'_a)^2+(\bm{k}'_a)^2}$, and $E_f=\sqrt{M_f^2+\bm{P}_f^2}$.

The boost in Eq.~\eqref{eq:boosted_wf} carries the final baryon spatial wave function to the recoil frame.
It also acts on the spin part of the baryon wave function.
Since the canonical boost of the baryon is not, in general, collinear with the boosts of the individual quarks~\cite{Wigner:1939cj}, the spin wave function acquires Wigner rotations~\cite{Faustov:1970af,Andreadis:1974ey}.
For a constituent quark of mass $m$ and a baryon of mass $M$, the corresponding spinor boost matrices are
\begin{align}
	S_m(\bm{p})
	=
	\frac{
		(e+m)\mathbb{I}_2
		+\bm{\sigma}\!\cdot\!\bm{p}
	}{
		\sqrt{2m(e+m)}
	},
	~~
	S_M(\bm{P})
	=
	\frac{
		(E+M)\mathbb{I}_2
		+\bm{\sigma}\!\cdot\!\bm{P}
	}{
		\sqrt{2M(E+M)}
	}.
	\label{eq:boost_op}
\end{align}
where $\mathbb{I}_2$ denotes the $2\times2$ identity matrix.
For a constituent quark with rest frame momentum $\bm{k}$ in a baryon of momentum $\bm{P}$, the corresponding moving frame momentum is
\begin{equation}
	\bm{p}
	=
	\bm{k}
	+
	\frac{\bm{P}}{M}
	\left(
	\epsilon+\frac{\bm{P}\!\cdot\!\bm{k}}{E+M}
	\right).
	\label{eq:mom_constraint}
\end{equation}
The Wigner rotation is then the mismatch between the direct constituent quark boost and the boost induced through the baryon~\cite{Ebert:2004ck,Ebert:2006rp},
\begin{equation}
	D(\bm{k},\bm{P};m,M)
	\equiv
	S_m^{-1}(\bm{p})\,S_M(\bm{P})\,S_m(\bm{k}).
\end{equation}
Multiplying the three $2\times2$ boost matrices gives the Pauli form
\begin{align}
	D(\bm{k},\bm{P};m,M)
	&=
	\frac{\sqrt{E+M}}
	{(e+\epsilon)\sqrt{2M(e+m)(\epsilon+m)}} \nonumber \\
	&\quad\times \left\{
	\left[
	(e+m)(\epsilon+m)
	+\bm{p}\!\cdot\!\bm{k}
	\right]\mathbb{I}_2
	+i\bm{\sigma}\!\cdot\!
	(\bm{p}\times\bm{k})
	\right\}.
	\label{eq:D_explicit}
\end{align}

For the active transition $q_i\to q_f$, the weak vertex is taken in the $V-A$ form
\begin{equation}
	[\mathcal{O}^\nu(\bm{p}'_j,\bm{p}_j)]^{s'_j s_j}
	=
	\bar{u}_{q_f}(p'_j,s'_j)\,
	\gamma^\nu(1-\gamma^5)\,
	u_{q_i}(p_j,s_j),
	\label{eq:quark_va_current}
\end{equation}
where $p_j=(e_j,\bm{p}_j)$ and $p'_j=(e'_j,\bm{p}'_j)$ are the initial and final four momenta of quark line $j$.
The Wigner rotations are then attached to the initial and final spin lines of quark $j$,
\begin{align}
	&[\mathcal{O}^\nu_{W,j}]^{s'_j s_j} \nonumber\\
	&=
	\left[
	D^\dagger(\bm{k}'_j,\bm{P}_f;m'_j,M_f)\,
	\mathcal{O}^\nu(\bm{p}'_j,\bm{p}_j)
	D(\bm{k}_j,\bm{P}_i;m_j,M_i)
	\right]^{s'_j s_j}.
	\label{eq:Gamma_eff_full}
\end{align}
For a spectator quark $r\neq j$ with constituent quark mass $m_r$, the three momentum is unchanged in the calculation frame, $\bm{p}'_r=\bm{p}_r$, whereas $\bm{k}_r$ and $\bm{k}'_r$ are defined in the initial and final baryon rest frames.
The spectator spin line is represented by the normalized scalar overlap
\begin{equation}
	\mathcal{S}_r
	=
	\frac{1}{2}\,
	\mathrm{Tr}
	\left[
	D^\dagger(\bm{k}'_r,\bm{P}_f;m_r,M_f)\,
	D(\bm{k}_r,\bm{P}_i;m_r,M_i)
	\right],
	\label{eq:spectator_overlap}
\end{equation}
where the factor $\tfrac{1}{2}$ normalizes the trace over the two-dimensional spectator spin space.
%

Substituting the configuration expansion [Eq.~\eqref{eq:baryon_state_expansion}] into Eq.~\eqref{eq:H_def}, the hadronic transition matrix element is expressed as
\begin{equation}
	H^\nu(\lambda_f,\lambda_i;P_i,P_f)
	=
	\sum_{\kappa_f=1}^{3}\sum_{\kappa_i=1}^{3}
	(c_{\kappa_f}^{B_f,1/2})^*c_{\kappa_i}^{B_i,1/2}\,
	H_{\kappa_f\kappa_i}^\nu(\lambda_f,\lambda_i;P_i,P_f),
	\label{eq:configuration_mixed_hadronic_sum}
\end{equation}
where $H_{\kappa_f\kappa_i}^\nu$ is the transition matrix element between the unmixed basis configurations $|B_{i,\kappa_i}\rangle$ and $|B_{f,\kappa_f}\rangle$,
\begin{equation}
	H_{\kappa_f\kappa_i}^\nu(\lambda_f,\lambda_i;P_i,P_f)
	\equiv
	\sum_{j=1}^{3}
	\left\langle
	B_{f,\kappa_f}(P_f,\lambda_f)
	\left|
	J_j^\nu(0)
	\right|
	B_{i,\kappa_i}(P_i,\lambda_i)
	\right\rangle.
	\label{eq:unmixed_configuration_matrix_element}
\end{equation}

In the impulse approximation, the single-quark transition matrix element for a given spatial--spin--flavor component product takes the form
\begin{widetext}
\begin{align}
	\left\langle
	\Psi_f\Phi_f(\lambda_f)
	\left|
	J_j^\nu(0)
	\right|
	\Psi_i\Phi_i(\lambda_i)
	\right\rangle
	={}&
	\sqrt{2E_i\,2E_f}
	\int \frac{d^3\bm{k}_\rho}{(2\pi)^3}
	\frac{d^3\bm{k}_\lambda}{(2\pi)^3}
	\frac{\sqrt{\mathcal{J}_f(\{\bm{p}'_1,\bm{p}'_2,\bm{p}'_3\},\bm{P}_f)}}{\sqrt{4e_je'_j}}
	\left(\prod_{r\neq j}\mathcal{S}_r\right)
	\nonumber\\
	&\times
	\widetilde\Psi_f^{*}
	(\bm{k}'_\rho,\bm{k}'_\lambda)\,
	\widetilde\Psi_i
	(\bm{k}_\rho,\bm{k}_\lambda)
	\sum_{\alpha,\beta}
	\left\langle
	\Phi_f(\lambda_f)
	\left|
	\hat F_j^{q_fq_i}\hat S_j^{\alpha\beta}
	\right|
	\Phi_i(\lambda_i)
	\right\rangle
	[\mathcal{O}^\nu_{W,j}]^{\alpha\beta}.
	\label{eq:master_integral}
\end{align}
\end{widetext}
Here, $\widetilde\Psi_i
(\bm{k}_\rho,\bm{k}_\lambda)$ and $\widetilde\Psi_f^{*}
(\bm{k}'_\rho,\bm{k}'_\lambda)$ are the momentum space spatial wave functions obtained by Fourier transforming the coordinate space wave functions determined from the baryon mass spectrum in Sec.~\ref{sec:mass_spectrum}.
The factor $\sqrt{2E_i\,2E_f}$ arises from the covariant normalization of the external baryon states, while the boost Jacobian $\mathcal J_f$ and spectator spin overlap $\mathcal S_r$ ($r\neq j$) are given in Eqs.~\eqref{eq:jacobian} and \eqref{eq:spectator_overlap}, respectively. 
The flavor operator $\hat{F}^{q_fq_i}_{j}=\hat{b}_{q_f}^\dagger(j)\hat{b}_{q_i}(j)$ and spin projector $\hat{S}^{\alpha\beta}_{j}=|\alpha\rangle_j\,{}_j\langle\beta|$ ($\alpha,\beta\in\{\uparrow,\downarrow\}$) act on the active quark line $j$, and $[\mathcal O_{W,j}^\nu]^{\alpha\beta}$ is the effective weak vertex in Eq.~\eqref{eq:Gamma_eff_full}.

The flavor overlap and spin transition matrix elements in Eq.~\eqref{eq:master_integral} are defined as
\begin{align}
	I_{\zeta_f\zeta_i}^{fi,j}
	&\equiv
	\left\langle\phi_{\mathbf 8,B_f}^{\zeta_f}\right|
	\hat F_j^{q_fq_i}
	\left|\phi_{\mathbf 8,B_i}^{\zeta_i}\right\rangle,
	\label{eq:flavor_overlap_definition}
	\\
	(\mathcal{C}^{\sigma_f\sigma_i,j}_{\lambda_f\lambda_i})_{\alpha\beta}
	&\equiv
	\left\langle\chi_{1/2,\lambda_f}^{\sigma_f}\right|
	\hat S_j^{\alpha\beta}
	\left|\chi_{1/2,\lambda_i}^{\sigma_i}\right\rangle,
	\label{eq:spin_recoupling_definition}
\end{align}
where $\zeta_f,\zeta_i\in\{\rho,\lambda\}$ and $\sigma_f,\sigma_i\in\{\rho,\lambda\}$ denote the flavor and spin components, respectively. 
Using the explicit flavor wave functions in Appendix~\ref{app:wavefunctions}, the flavor overlap factors $I_{\zeta_f\zeta_i}^{fi,j}$ for various decay channels are calculated and listed in Table~\ref{tab:flavor_light_optimized}. 
More details on the spin transition matrix elements, the $\mathrm{SU}(6)$ spin--flavor recoupling, and the explicit expressions for the blocks $\mathbf C_+$, $\mathbf C_\times$, and $\mathbf C_-$ are presented in Appendix~\ref{app:spin_matrices}.

Since the total wave functions are symmetric under the permutation of identical quarks, the sum over the three quark lines in Eq.~\eqref{eq:unmixed_configuration_matrix_element} reduces to three times the representative contribution from quark line $j=3$. 
Combining the nine configuration matrix elements $H_{\kappa_f\kappa_i}^\nu$ through Eq.~\eqref{eq:configuration_mixed_hadronic_sum}, one obtains the total hadronic matrix element $H^\nu(\lambda_f,\lambda_i;P_i,P_f)$, which determines the decay amplitude $\mathcal M$ in Eq.~\eqref{eq:Amplitude}.

\subsection{Octet to octet form factors in the helicity basis}
\label{sec:ff_representation}

To compare the R3QM matrix elements with other form factor calculations, we project them onto a covariant basis.
For spin $\frac12\to\frac12$ octet transitions, we separate the hadronic weak current into its vector and axial-vector parts,
\begin{equation}
	J^\nu_{\mathrm{had}}=J_V^\nu-J_A^\nu,
\end{equation}
and define the corresponding current matrix elements by
\begin{equation}
	H^\nu_{V/A}(\lambda_f,\lambda_i;P_i,P_f)
	\equiv
	\langle B_f(P_f,\lambda_f)|J^\nu_{V/A}(0)|B_i(P_i,\lambda_i)\rangle .
\end{equation}
The full current matrix element is then $H^\nu=H_V^\nu-H_A^\nu$.
With $q^\nu=P_i^\nu-P_f^\nu$ and $M=M_i$, the on shell decompositions are given by~\cite{Cabibbo:2003cu,Schlumpf:1994fb},
\begin{widetext}
\begin{align}
	H_V^\nu(\lambda_f,\lambda_i;P_i,P_f)
	&=
	\bar u_f(P_f,\lambda_f)
	\left[
	f_1(q^2)\gamma^\nu
	-i\frac{f_2(q^2)}{M}\sigma^{\nu\alpha}q_\alpha
	+\frac{f_3(q^2)}{M}q^\nu
	\right]
	u_i(P_i,\lambda_i),
	\label{eq:ff_vector_current}
	\\
	H_A^\nu(\lambda_f,\lambda_i;P_i,P_f)
	&=
	\bar u_f(P_f,\lambda_f)
	\left[
	g_1(q^2)\gamma^\nu\gamma^5
	-i\frac{g_2(q^2)}{M}\sigma^{\nu\alpha}q_\alpha\gamma^5
	+\frac{g_3(q^2)}{M}q^\nu\gamma^5
	\right]
	u_i(P_i,\lambda_i),
	\label{eq:ff_axial_current}
\end{align}	
\end{widetext}
where $\sigma^{\nu\alpha}\equiv i[\gamma^\nu,\gamma^\alpha]/2$.
%
%

At each $q^2$, the R3QM overlap is calculated for fixed initial and final baryon helicities and for one Cartesian component of the weak current in the initial baryon rest frame.
The boost of the final baryon, the Wigner rotations of the constituent quark spins, and the spatial overlap of the baryon wave functions are all contained in this matrix element.
The direct output of the calculation is the current matrix element $H^\nu_{V/A}(\lambda_f,\lambda_i)$.
The form factors are obtained after these current matrix elements are matched to the covariant decompositions in Eqs.~\eqref{eq:ff_vector_current} and \eqref{eq:ff_axial_current}.
The three vector components determine $f_1$, $f_2$, and $f_3$, while the corresponding axial-vector components determine $g_1$, $g_2$, and $g_3$, giving two $3\times3$ linear systems.

The same current matrix element is used differently in related approaches.
In light front quark model calculations, one usually chooses a light front frame and the good current $J^+$, whose spin conserving and spin flip matrix elements give compact projections for $f_1$, $f_2$, $g_1$, and $g_2$~\cite{Schlumpf:1994fb,Geng:2020gjh}.
In helicity amplitude descriptions of decay rates and angular distributions, the same current matrix element is used at a later step.
After the weak current is expressed in terms of form factors, its Lorentz index is contracted with the virtual $W$ polarization vectors, $H_{\lambda_f,\lambda_W}=\epsilon_\nu^*(\lambda_W)H^\nu(\lambda_f,\lambda_i)$, to form the $W$ helicity amplitudes~\cite{Singleton:1990ye,Korner:1991ph,Kadeer:2005aq,Korner:2014bca}.
This projection reorganizes the decay amplitude, but it is not the projection adopted here for the form factor extraction.
In the present instant form calculation, the independent inputs are Cartesian components of $H^\nu$ in the initial baryon rest frame, and these components are matched directly to the vector and axial-vector form factor decompositions.

For the projection below, we introduce the following recoil variables:
\begin{align}
	E_f&=\frac{M_i^2+M_f^2-q^2}{2M_i},~
	|\bm{P}_f|=\frac{\lambda^{1/2}(M_i^2,M_f^2,q^2)}{2M_i}, \nonumber \\
	R&=\frac{|\bm{P}_f|}{E_f+M_f},~
	\mathcal N=\sqrt{2M_i(E_f+M_f)},
\end{align}
where $\lambda(a,b,c)=a^2+b^2+c^2-2ab-2ac-2bc$.
The coefficients in the projection equations are fixed by the corresponding Dirac bilinears of these spinors.
For example, the time component of the vector current without helicity flip gives
\begin{align}
	\bar u_f\gamma^0u_i=\mathcal N,
	\quad
	\bar u_fu_i&=\mathcal N,
	\quad
	\bar u_f\left(-i\sigma^{0\alpha}q_\alpha\right)u_i=\mathcal N(M_f-E_f),
\end{align}
where the tensor bilinear follows from the Gordon identity for $-i\sigma^{\nu\alpha}q_\alpha$, with $P^\nu=P_i^\nu+P_f^\nu$.
With $q^0=M_i-E_f$ in the decay convention, the $f_1$, $f_2$, and $f_3$ terms in $H_V^0(\tfrac12,\tfrac12)$ contribute $\mathcal N f_1(q^2)$, $-\mathcal N(E_f-M_f)f_2(q^2)/M$, and $\mathcal N(M_i-E_f)f_3(q^2)/M$, respectively.
The longitudinal component without helicity flip and the transverse component with helicity flip are matched by the same spinor algebra.
The six Cartesian components of the vector and axial-vector currents for the projection are
\begin{align}
	H_V^0\!\left(\tfrac12,\tfrac12\right)
	&=
	\mathcal N
	\left[
	f_1(q^2)-\frac{E_f-M_f}{M}f_2(q^2)
	+\frac{M_i-E_f}{M}f_3(q^2)
	\right],
\end{align}
\begin{align}
	H_V^3\!\left(\tfrac12,\tfrac12\right)
	&=
	\mathcal N
	\left[
	Rf_1(q^2)+R\frac{M_i-E_f}{M}f_2(q^2)
	-\frac{|\bm{P}_f|}{M}f_3(q^2)
	\right],
\end{align}
\begin{align}
	H_V^{1}\!\left(-\tfrac12,\tfrac12\right)
	&=
	-\mathcal N R
	\left[
	f_1(q^2)+\frac{M_i+M_f}{M}f_2(q^2)
	\right],
\end{align}
\begin{align}
	H_A^0\!\left(\tfrac12,\tfrac12\right)
	&=
	\mathcal N
	\left[
	Rg_1(q^2)+\frac{|\bm{P}_f|}{M}g_2(q^2)
	-R\frac{M_i-E_f}{M}g_3(q^2)
	\right],
\end{align}
\begin{align}
	H_A^3\!\left(\tfrac12,\tfrac12\right)
	&=
	\mathcal N
	\left[
	g_1(q^2)-\frac{M_i-E_f}{M}g_2(q^2)
	+R\frac{|\bm{P}_f|}{M}g_3(q^2)
	\right],
\end{align}
\begin{align}
	H_A^{1}\!\left(-\tfrac12,\tfrac12\right)
	&=
	\mathcal N
	\left[
	g_1(q^2)-\frac{M_i-M_f}{M}g_2(q^2)
	\right].
\end{align}

Inverting the six vector and axial-vector currents relations above gives
\begin{widetext}
\begin{align}
	f_2(q^2)
	&=
	-\frac{M}{2M_i\mathcal N}
	\left[
	H_V^0\!\left(\tfrac12,\tfrac12\right)
	+\frac{M_i - E_f}{|\bm{P}_f|}H_V^3\!\left(\tfrac12,\tfrac12\right) 
	+\left(\frac{1}{R}+\frac{M_i - E_f}{|\bm{P}_f|}\right)
	H_V^{1}\!\left(-\tfrac12,\tfrac12\right)
	\right],
	\label{eq:ff_extract_f2}
	\\
	f_1(q^2)
	&=
	-\frac{M_i+M_f}{M}f_2(q^2)
	-\frac{H_V^{1}\!\left(-\tfrac12,\tfrac12\right)}{\mathcal N R},
	\label{eq:ff_extract_f1}
	\\
	f_3(q^2)
	&=
	\frac{M}{M_i - E_f}
	\left[
	\frac{H_V^0\!\left(\tfrac12,\tfrac12\right)}{\mathcal N}
	-f_1(q^2)
	+\frac{E_f-M_f}{M}f_2(q^2)
	\right].
	\label{eq:ff_extract_f3}\\
	g_2(q^2)
	&=
	\frac{M}{2M_i\mathcal N}
	\left[
	\frac{1}{R}H_A^0\!\left(\tfrac12,\tfrac12\right)
	+\frac{M_i - E_f}{R|\bm{P}_f|}H_A^3\!\left(\tfrac12,\tfrac12\right) 
	-\left(1+\frac{M_i - E_f}{R|\bm{P}_f|}\right)
	H_A^{1}\!\left(-\tfrac12,\tfrac12\right)
	\right],
	\label{eq:ff_extract_g2}
	\\
	g_1(q^2)
	&=
	\frac{H_A^{1}\!\left(-\tfrac12,\tfrac12\right)}{\mathcal N}
	+\frac{M_i-M_f}{M}g_2(q^2),
	\label{eq:ff_extract_g1}
	\\
	g_3^{\mathrm{np}}(q^2)
	&=
	\frac{M}{R|\bm{P}_f|}
	\left[
	\frac{H_A^3\!\left(\tfrac12,\tfrac12\right)}{\mathcal N}
	-g_1(q^2)
	+\frac{M_i-E_f}{M}g_2(q^2)
	\right].
	\label{eq:ff_extract_g3}
\end{align}
\end{widetext}
The above equations are the algebraic inverse of the selected current component system. 
This inverse cannot be evaluated literally at the zero recoil endpoint.
At $q^2=(M_i-M_f)^2$, one has $|\bm{P}_f|=0$ and $R|\bm{P}_f|=0$, so the selected components are no longer independent.
We therefore extract the form factors at nonzero recoil and use their smooth limit for the endpoint values.

The inversion above gives the regular overlap part of the octet form factors.
We use it for $f_1$, $f_2$, $f_3$, $g_1$, and $g_2$.
The induced pseudoscalar form factor needs a separate treatment, because the
axial-vector divergence contains the corresponding Goldstone boson pole term.
Thus, we determine the physical $g_3$ from the divergence of the
axial-vector current.
Contracting Eq.~\eqref{eq:ff_axial_current} with $q_\nu$ and using the on-shell Dirac equations gives
\begin{align}
	&q_\nu H_A^\nu(\lambda_f,\lambda_i) \nonumber\\
	&=
	\left[
	-(M_i+M_f)g_1(q^2)
	+\frac{q^2}{M}g_3(q^2)
	\right]
	\bar u_f(P_f,\lambda_f)\gamma^5 u_i(P_i,\lambda_i) .
	\label{eq:kinematic_div}
\end{align}
Keeping the pseudoscalar pole contribution, the axial-vector divergence is approximated by~\cite{Goity:1999by}
\begin{equation}
	q_\nu H_A^\nu(\lambda_f,\lambda_i)
	\simeq
	\frac{f_{\mathcal P} m_{\mathcal P}^2 g_{\mathcal P B_iB_f}}
	{m_{\mathcal P}^2-q^2}\,
	\bar u_f(P_f,\lambda_f)\gamma^5 u_i(P_i,\lambda_i),
	\label{eq:pcac_div}
\end{equation}
where $\mathcal P=\pi$ for $|\Delta S|=0$ and $\mathcal P=K$ for
$|\Delta S|=1$.
Using the pseudoscalar pole prescription and the generalized Goldberger--Treiman relation~\cite{Goldberger:1958vp}, the pole part of the induced pseudoscalar form factor is fixed by the axial form factor $g_1$ and takes the generalized pseudoscalar pole form~\cite{Guadagnoli:2006gj,Sasaki:2008ha}:
\begin{equation}
	g_3(q^2)
	\simeq
	-\frac{M(M_i+M_f)}{m_{\mathcal P}^2-q^2}\,g_1(q^2).
	\label{eq:g3_reconstructed}
\end{equation}
The $g_3$ in the decay amplitudes below is therefore reconstructed by the pole prescription in Eq.~\eqref{eq:g3_reconstructed}, whereas $g_3^{\mathrm{np}}$ denotes only the smooth overlap contribution without the pole.

\subsection{Static spin--flavor coefficients and Cabibbo mapping}
\label{sec:cabibbo_connection}

In the Cabibbo theory, the hyperon semileptonic decay matrix elements at zero momentum transfer are parametrized by the symmetric and antisymmetric $\mathrm{SU}(3)$ couplings $D$ and $F$~\cite{Cabibbo:2003cu,Gaillard:1984ny,Garcia:1985xz}. 
To establish the connection between our quark model calculation and the Cabibbo framework, we calculate the leading form factors $f_1$ and $g_1$ in the static $\mathrm{SU}(6)$ limit. 
In this limit, each baryon is approximated by its unmixed ground state configuration $\left|B_1^{\mathbf 8}, 1/2^+\right\rangle$ in Eq.~\eqref{eq:wf_octet}, the spatial wave function overlap is set to unity, and relativistic corrections (such as the boost Jacobian and Wigner rotations) are neglected.

Substituting the ground state wave function $\left|B_1^{\mathbf 8}, 1/2^+\right\rangle = \frac{1}{\sqrt{2}}\left(\phi_{\mathbf 8}^{\rho}\chi_{1/2}^{\rho} + \phi_{\mathbf 8}^{\lambda}\chi_{1/2}^{\lambda}\right)\Psi_{000}^{s}$ into the hadronic matrix element, the single-quark weak current operator acting on quark line $j=3$ gives
\begin{equation}
	\left\langle B_{f,1}\left|\hat O_3\right|B_{i,1}\right\rangle
	=
	\frac{1}{2}
	\left(
	I_{\rho\rho}^{fi}\langle \chi_{1/2}^{\rho}|\hat S_3|\chi_{1/2}^{\rho}\rangle
	+
	I_{\lambda\lambda}^{fi}\langle \chi_{1/2}^{\lambda}|\hat S_3|\chi_{1/2}^{\lambda}\rangle
	\right),
\end{equation}
where the cross terms between the $\rho$ and $\lambda$ components vanish because the single quark operator preserves the exchange symmetry of the $(12)$ spectator pair. 
Here, $I_{\rho\rho}^{fi}\equiv I_{\rho\rho}^{fi,3}$ and $I_{\lambda\lambda}^{fi}\equiv I_{\lambda\lambda}^{fi,3}$ denote the flavor overlap integrals for the third quark line ($j=3$) defined in Eq.~\eqref{eq:flavor_overlap_definition}. 
Taking into account the sum over the three identical constituent quarks, the total static matrix element acquires an overall prefactor of $3\times(1/2) = 3/2$.

For the vector form factor $f_1$, the active quark spin operator is the unit operator, yielding $\langle \chi_{1/2}^{\rho}|\hat{\mathbb I}|\chi_{1/2}^{\rho}\rangle = \langle \chi_{1/2}^{\lambda}|\hat{\mathbb I}|\chi_{1/2}^{\lambda}\rangle = 1$. 
For the axial-vector form factor $g_1$, the spin operator $\sigma_z$ gives $+1$ for the spin-singlet ($s_{12}=0$) spectator pair in $\chi_{1/2}^{\rho}$ and $-1/3$ for the spin-triplet ($s_{12}=1$) spectator pair in $\chi_{1/2}^{\lambda}$. 
The static form factors are therefore given by
\begin{equation}
	f_1^{\mathrm{stat.}}=\frac{3}{2}\left(I_{\rho\rho}^{fi}+I_{\lambda\lambda}^{fi}\right),\qquad
	g_1^{\mathrm{stat.}}=\frac{3}{2}\left(I_{\rho\rho}^{fi}-\frac{1}{3}I_{\lambda\lambda}^{fi}\right).
	\label{eq:static_mapping}
\end{equation}
The flavor overlap factors $I_{\rho\rho}^{fi}$ and $I_{\lambda\lambda}^{fi}$ for all hyperon decay channels are listed in Table~\ref{tab:flavor_light_optimized}.
In the static $\mathrm{SU}(6)$ limit, Eq.~\eqref{eq:static_mapping} reproduces the standard $\mathrm{SU}(6)$ relations $F=2/3$ and $D=1$~\cite{Gursey:1964keh,LeYaouanc:1988fx}. 
For instance, for $\Xi^0 \to \Sigma^+ \ell^- \bar{\nu}_\ell$, one has $I_{\rho\rho}^{fi}=1$ and $I_{\lambda\lambda}^{fi}=-1/3$, which gives $f_1^{\mathrm{stat.}}=1$ and $g_1^{\mathrm{stat.}}=5/3=F+D$. 
The corresponding static coefficients expressed in terms of $F$ and $D$ are listed in the last two columns of Table~\ref{tab:flavor_light_optimized} for comparison. 
In our full R3QM calculation, the physical form factors depart from these static values due to the configuration mixing in the baryon wave functions, the momentum dependent spatial wave function overlaps, and the relativistic recoil and Wigner rotation effects.

%
%
%
%
%

For $|\Delta S|=1$ transitions, the Ademollo--Gatto theorem protects the vector form factor $f_1(0)$ against first order SU$(3)$ breaking corrections~\cite{Ademollo:1964sr}. 
By contrast, the axial-vector form factor $g_1(0)$ receives no such protection. 
Its deviation from the static SU$(6)$ reference value is generated by the lower components of the Dirac spinors, the mass-dependent Wigner rotations, and the spatial mismatch between the initial and boosted final baryon wave functions~\cite{Schlumpf:1994fb}.

Some coefficients in Table~\ref{tab:flavor_light_optimized} differ by a common sign between the static columns and the Cabibbo comparison columns, since we follow the Chen convention~\cite{Chen:1979qz} which is different from the Cabibbo--Rabl convention~\cite{Cabibbo:2003cu,Rabl:1975zy}.
In the Chen convention, it carries an extra minus sign for the $\Sigma^+$, $\Lambda$, and $\Xi^0$ fields~\cite{Lu:2024ajt}. 
Specifically, the $\Lambda\to p$, $\Sigma^-\to\Lambda$, $\Sigma^0\to\Sigma^+$, $\Xi^-\to\Lambda$, and $\Xi^-\to\Xi^0$ channels carry a common minus sign, while the other listed octet channels show no sign difference. 
This common sign flips $f_1$ and $g_1$ simultaneously and has no effect on decay widths or ratios such as $g_1/f_1$.
For the $\Sigma^0\to\Sigma^+$ channel, we take $f_1^{\text{Cab.}}=-\sqrt{2}$ in the Cabibbo comparison column. 
This sign preserves the isospin relation with the $\Sigma^-\to\Sigma^0$ transition, and the $+\sqrt{2}$ sign given in Ref.~\cite{Cabibbo:2003cu} is not adopted in the present comparison.

\subsection{The experimental observation variables}

For an unpolarized initial hyperon, averaging $|\mathcal{M}|^2$ over the initial spin and summing over the final state spins gives the differential decay width~\cite{ParticleDataGroup:2024cfk}
\begin{align}
	&d\Gamma
	=
	\frac{(2\pi)^4}{2M_i}\,
	\frac{1}{2J_i+1}
	\sum_{\lambda_i,\lambda_f,s_\ell,s_{\bar{\nu}}}
	|\mathcal{M}|^2\, \delta^{(4)}(P_i-P_f-p_\ell-p_\nu)
	\nonumber\\
	&\quad\times 
	\frac{d^3\bm{P}_f}{(2\pi)^3\,2E_f}
	\frac{d^3\bm{p}_\ell}{(2\pi)^3\,2E_\ell}
	\frac{d^3\bm{p}_\nu}{(2\pi)^3\,2E_\nu}.
	\label{eq:differential_width}
\end{align}
Integrating Eq.~\eqref{eq:differential_width} over the three-body phase space yields the partial decay width $\Gamma$.
The branching fraction $\mathcal{B} = \Gamma \tau$ is then obtained by combining this partial width with the experimental hyperon lifetime $\tau$ from the RPP~\cite{ParticleDataGroup:2024cfk}.
It should be emphasized that we do not theoretically calculate the total decay width or lifetime of the initial hyperon; rather, the experimental lifetime is adopted as an external input to convert the predicted semileptonic partial width into the corresponding branching fraction.

Besides the decay width and branching fraction, the angular distributions of the decay products in hyperon semileptonic decays provide important observables to probe the structure of the weak current~\cite{Cabibbo:2003cu,Kadeer:2005aq}.
In the initial hyperon rest frame, these angular observables include the lepton--antineutrino correlation $\alpha_{\ell\nu}$ for an unpolarized hyperon, and three spin asymmetry parameters $\alpha_\ell$, $\alpha_\nu$, and $\alpha_B$ when the initial hyperon is polarized~\cite{Garcia:1971wc,Garcia:1985xz,Cabibbo:2003cu}.
The opening angle between the charged lepton and the antineutrino is defined as
\begin{equation}
	\cos\theta_{\ell\nu} = \hat{\bm p}_\ell \cdot \hat{\bm p}_\nu,
	\label{eq:cos_lnu}
\end{equation}
where $\hat{\bm p}_\ell$ and $\hat{\bm p}_\nu$ are the unit three momenta of the charged lepton and antineutrino, respectively.
The differential decay rate is parametrized as~\cite{Garcia:1971wc}
\begin{equation}
	\frac{d\Gamma}{d\cos\theta_{\ell\nu}} = \frac{\Gamma}{2} \left(1 + \alpha_{\ell\nu} \cos\theta_{\ell\nu}\right).
	\label{eq:dist_lnu}
\end{equation}
Integrating the differential width over the three-body phase space, the correlation parameter $\alpha_{\ell\nu}$ is extracted from the forward--backward asymmetry~\cite{Garcia:1971wc,Garcia:1985xz}:
\begin{equation}
	\alpha_{\ell\nu} = 2\, \frac{\Gamma_{\ell\nu}^{F} - \Gamma_{\ell\nu}^{B}}{\Gamma_{\ell\nu}^{F} + \Gamma_{\ell\nu}^{B}},
	\label{eq:alpha_lnu_def}
\end{equation}
where $\Gamma_{\ell\nu}^{F}$ and $\Gamma_{\ell\nu}^{B}$ denote the partial decay widths integrated over the forward ($\cos\theta_{\ell\nu}>0$) and backward ($\cos\theta_{\ell\nu}<0$) hemispheres, respectively.

For a polarized initial hyperon, the three spin asymmetry parameters $\alpha_\ell$, $\alpha_\nu$, and $\alpha_B$ characterize the angular correlations between the hyperon spin polarization and the momentum directions of the final-state particles.
Taking the polarization vector along the $\hat{\bm Z}$ axis, $\bm{\mathcal P}_i = \mathcal P_i \hat{\bm Z}$, the directional cosines of the decay products are defined by
\begin{equation}
	\cos\theta_\ell = \hat{\bm Z} \cdot \hat{\bm p}_\ell, \quad
	\cos\theta_\nu = \hat{\bm Z} \cdot \hat{\bm p}_\nu, \quad
	\cos\theta_B = \hat{\bm Z} \cdot \hat{\bm p}_f.
	\label{eq:cos_polarized}
\end{equation}
With the initial spin density matrix $\rho_i = \frac{1}{2}(\mathbb I_2 + \bm{\mathcal P}_i \cdot \bm\sigma)$~\cite{Kadeer:2005aq}, the differential decay rate for each emission angle takes the form~\cite{Garcia:1971wc,Kadeer:2005aq}
\begin{equation}
	\frac{d\Gamma}{d\cos\theta_X} = \frac{\Gamma}{2} \left(1 + \mathcal P_i\,\alpha_X \cos\theta_X\right),
	\label{eq:dist_X}
\end{equation}
where $X \in \{\ell, \nu, B\}$.
The asymmetry parameter $\alpha_X$ is then determined by the forward--backward asymmetry~\cite{Garcia:1971wc,Flores-Mendieta:2001fcy}
\begin{equation}
	\alpha_X = \frac{2}{\mathcal P_i}\, \frac{\Gamma_X^{F} - \Gamma_X^{B}}{\Gamma_X^{F} + \Gamma_X^{B}},
	\label{eq:alpha_X_def}
\end{equation}
where $\Gamma_X^{F}$ and $\Gamma_X^{B}$ are the partial decay widths integrated over the forward ($\cos\theta_X>0$) and backward ($\cos\theta_X<0$) hemispheres, respectively.
%

The lepton asymmetry parameters $\alpha_\ell$ and $\alpha_\nu$ are directly calculated from Eq.~\eqref{eq:alpha_X_def} by integrating the polarized decay distribution over the forward and backward hemispheres. 
For the final baryon asymmetry $\alpha_B$, it can be conveniently expressed in terms of the helicity amplitudes $H_{\lambda_f,\lambda_W}$~\cite{Korner:1991ph,Kadeer:2005aq}. 
In the initial hyperon rest frame, by choosing the $z$ axis along the final baryon momentum $\hat{\bm p}_f$, the reduced decay rates for the initial spin states $\lambda_i = \pm 1/2$ at a given $q^2$ are expressed as
\begin{equation}
	R_\pm = \left(1+\frac{m_\ell^2}{2q^2}\right) \left( \left|H_{\pm\frac12,0}\right|^2 + \left|H_{\mp\frac12,\mp1}\right|^2 \right) + \frac{3m_\ell^2}{2q^2} \left|H_{\pm\frac12,t}\right|^2.
	\label{eq:Rpm_def}
\end{equation}
The differential asymmetry $\alpha_B(q^2)$ is then given by
\begin{equation}
	\alpha_B(q^2) = \frac{R_+(q^2) - R_-(q^2)}{R_+(q^2) + R_-(q^2)}.
	\label{eq:alphaB_q2}
\end{equation}
Integrating $\alpha_B(q^2)$ weighted with the differential decay width $d\Gamma/dq^2$ over the physical $q^2$ region, we obtain the total integrated asymmetry:
\begin{equation}
	\alpha_B = \frac{1}{\Gamma} \int_{m_\ell^2}^{(M_i-M_f)^2} dq^2\, \frac{d\Gamma}{dq^2}\, \alpha_B(q^2).
	\label{eq:alphaB_integrated}
\end{equation}

To illustrate the connection between these angular observables and the leading form factors $f_1$ and $g_1$, we consider the leading-order allowed approximation~\cite{Garcia:1971wc,Garcia:1985xz}. 
Neglecting recoil and lepton mass effects, for the $B_i\to B_f\ell^-\bar{\nu}_\ell$ mode these observables reduce to the simple analytical relations:
\begin{align}
	\alpha_{\ell\nu} &\simeq \frac{f_1^2 - g_1^2}{f_1^2 + 3g_1^2}, \label{eq:alpha_lnu_approx} \\
	\alpha_\ell &\simeq \frac{2g_1(f_1 - g_1)}{f_1^2 + 3g_1^2}, \label{eq:alpha_l_approx} \\
	\alpha_\nu &\simeq \frac{2g_1(f_1 + g_1)}{f_1^2 + 3g_1^2}, \label{eq:alpha_nu_approx} \\
	\alpha_B &\simeq -\frac{5 f_1 g_1}{2\left(f_1^2 + 3g_1^2\right)}. \label{eq:alpha_B_approx}
\end{align}
For the corresponding positron mode $B_i\to B_f\ell^+\nu_\ell$, the $V-A$ chiral projector remains unchanged, whereas replacing the external lepton spinors by their charge-conjugate counterparts reverses the sign of the antisymmetric Levi-Civita term in the leptonic tensor~\cite{Kadeer:2005aq}.
In the allowed approximation, this sign reversal directly interchanges the expressions for $\alpha_\ell$ and $\alpha_\nu$:
\begin{align*}
	\alpha_\ell^{(+)}
	&\simeq
	\frac{2g_1(f_1+g_1)}{f_1^2+3g_1^2},
	&
	\alpha_\nu^{(+)}
	&\simeq
	\frac{2g_1(f_1-g_1)}{f_1^2+3g_1^2},
\end{align*}
while $\alpha_{\ell\nu}$ and $\alpha_B$ remain unchanged.
These relations clearly show that:
\begin{enumerate}
	\item[(i)] The lepton--neutrino correlation $\alpha_{\ell\nu}$ depends only on the ratio $|g_1/f_1|$, approaching $+1$ for a pure vector current ($g_1=0$) and $-1/3$ for a pure axial-vector current ($f_1=0$).
	\item[(ii)] The lepton asymmetries $\alpha_\ell$ and $\alpha_\nu$ directly probe the vector and axial-vector interference. For $B_i\to B_f\ell^-\bar{\nu}_\ell$, $\alpha_\ell$ vanishes when $f_1=g_1$, while $\alpha_\nu$ vanishes when $f_1=-g_1$. For $B_i\to B_f\ell^+\nu_\ell$, these two conditions are interchanged.
	\item[(iii)] The baryon asymmetry $\alpha_B$ arises from the parity-violating vector--axial interference. It vanishes if either $f_1=0$ or $g_1=0$, and its sign is determined by the relative sign of $g_1/f_1$.
\end{enumerate}
Since the common CKM factor cancels out in these angular observables, they are free from CKM uncertainties and provide clean probes of the form factor ratio $g_1/f_1$.

It should be emphasized that the analytical relations in Eqs.~\eqref{eq:alpha_lnu_approx}--\eqref{eq:alpha_B_approx} hold only in the leading-order allowed approximation. 
In our practical calculations, all angular observables are obtained by integrating the complete transition amplitudes over the full three-body phase space. 
Therefore, relativistic effects, including finite recoil, boost Jacobians, Wigner spin rotations, and nonzero lepton masses, are fully incorporated into our numerical results.

\begin{table*}[t]
	\centering
	\renewcommand\arraystretch{1.40} 
	\setlength{\tabcolsep}{10pt}
	\caption{
		Static SU$(6)$ spin--flavor coefficients for semileptonic decays of the hyperon octet.
		The reduced overlaps $I_{\rho\rho}^{fi}$ and $I_{\lambda\lambda}^{fi}$ are obtained from the $\rho\rho$ and $\lambda\lambda$ components of the octet wave function in Eq.~\eqref{eq:wf_octet}. 
		Their spin parts correspond to spectator pair spins $s_{12}=0$ and $s_{12}=1$, respectively.
		%
		Here $f_1^{\mathrm{stat.}}$ is the corresponding SU$(3)$ Clebsch--Gordan coefficient, and $g_1^{\mathrm{stat.}}$ is the Cabibbo $F$-$D$ combination~\cite{Cabibbo:2003cu,Gaillard:1984ny,Garcia:1985xz} evaluated at the static SU$(6)$ values $F=2/3$ and $D=1$~\cite{Gursey:1964keh,LeYaouanc:1988fx}.
		The static columns follow the Chen phase convention adopted in the present work~\cite{Chen:1979qz}. 
		The Cabibbo columns follow the Cabibbo--Rabl convention~\cite{Cabibbo:2003cu,Rabl:1975zy}.
	}
	\label{tab:flavor_light_optimized}
	\begin{tabular}{cccccccccc}
		\hline\hline
	     Decay Process & Type & Scale & $I_{\rho\rho}^{fi}$ & $I_{\lambda\lambda}^{fi}$ & $f_1^{\mathrm{stat.}}$ & $g_1^{\mathrm{stat.}}$ & $f_1^{\text{Cab.}}$ & $g_1^{\text{Cab.}}$ \\
		\hline
		
		 $\Lambda \to p $ & $\Delta S=1$ & $V_{us}$ & $\sqrt{\frac{2}{3}}$ & $0$
		& $\sqrt{\frac{3}{2}}$ & $\sqrt{\frac{3}{2}}$
		& $-\sqrt{\frac{3}{2}}$ & $-\sqrt{\frac{3}{2}}\big(F+\frac{D}{3}\big)$ \\
		\hline
		 $\Sigma^- \to n $ & $\Delta S=1$ & $V_{us}$ & $0$ & $-\frac{2}{3}$
		& $-1$ & $\frac{1}{3}$
		& $-1$ & $D-F$ \\
		 $\Sigma^- \to \Lambda$ & $\Delta S=0$ & $V_{ud}$ & $-\frac{1}{\sqrt{6}}$ & $\frac{1}{\sqrt{6}}$
		& $0$ & $-\sqrt{\frac{2}{3}}$
		& $0$ & $\sqrt{\frac{2}{3}}D$ \\
		 $\Sigma^- \to \Sigma^0 $ & $\Delta S=0$ & $V_{ud}$ & $\frac{1}{\sqrt{2}}$ & $\frac{1}{3\sqrt{2}}$
		& $\sqrt{2}$ & $\frac{2\sqrt{2}}{3}$
		& $\sqrt{2}$ & $\sqrt{2}F$ \\
		 $\Sigma^0 \to \Sigma^+ $ & $\Delta S=0$ & $V_{ud}$ & $\frac{1}{\sqrt{2}}$ & $\frac{1}{3\sqrt{2}}$
		& $\sqrt{2}$ & $\frac{2\sqrt{2}}{3}$
		& $-\sqrt{2}$ & $-\sqrt{2}F$ \\
		 $\Sigma^0 \to p \, $ & $\Delta S=1$ & $V_{us}$ & $0$ & $-\frac{\sqrt{2}}{3}$
		& $-\frac{1}{\sqrt{2}}$ & $\frac{1}{3\sqrt{2}}$
		& $-\frac{1}{\sqrt{2}}$ & $\frac{1}{\sqrt{2}}(D-F)$ \\
		 $\Sigma^+ \to \Lambda $ & $\Delta S=0$ & $V_{ud}$ & $\frac{1}{\sqrt{6}}$ & $-\frac{1}{\sqrt{6}}$
		& $0$ & $\sqrt{\frac{2}{3}}$
		& $0$ & $\sqrt{\frac{2}{3}}D$ \\
		\hline
		 $\Xi^- \to \Lambda $ & $\Delta S=1$ & $V_{us}$ & $-\frac{1}{\sqrt{6}}$ & $-\frac{1}{\sqrt{6}}$
		& $-\sqrt{\frac{3}{2}}$ & $-\frac{1}{\sqrt{6}}$
		& $\sqrt{\frac{3}{2}}$ & $-\frac{1}{\sqrt{6}}(D-3F)$ \\
		 $\Xi^- \to \Xi^0 $ & $\Delta S=0$ & $V_{ud}$ & $0$ & $\frac{2}{3}$
		& $1$ & $-\frac{1}{3}$
		& $-1$ & $D-F$ \\
		 $\Xi^- \to \Sigma^0 $ & $\Delta S=1$ & $V_{us}$ & $\frac{1}{\sqrt{2}}$ & $-\frac{1}{3\sqrt{2}}$
		& $\frac{1}{\sqrt{2}}$ & $\frac{5}{3\sqrt{2}}$
		& $\frac{1}{\sqrt{2}}$ & $\frac{1}{\sqrt{2}}(D+F)$ \\
		 $\Xi^0 \to \Sigma^+ $ & $\Delta S=1$ & $V_{us}$ & $1$ & $-\frac{1}{3}$
		& $1$ & $\frac{5}{3}$
		& $1$ & $D+F$ \\
		\hline\hline
	\end{tabular}
\end{table*}

\subsection{Parameters and form factor conventions}
\label{sec:convention_alignment}

In the numerical calculation of the weak transition amplitudes, the wave functions and constituent quark masses are fixed by the spectrum calculation, with no additional adjustable parameters in the weak matrix element. 
The constituent quark masses $m_q$ are taken from Table~\ref{tab:potential_params}, and the spatial wave functions are obtained variationally from the same potential model. 
The physical baryon masses and lifetimes, the charged lepton and pseudoscalar meson masses, the Fermi constant $G_F=1.16636\times10^{-5}\,\text{GeV}^{-2}$, and the CKM elements are taken from the RPP~\cite{ParticleDataGroup:2024cfk} for the decay widths and branching fractions.

We compare the calculated form factors $f_1$, $g_1$, $f_2$, $g_2$, $f_3$, and $g_3$ at $q^2=0$ with the available lattice QCD results and other theoretical predictions. 
For a uniform comparison, all form factors reported in the literature are converted to the present sign and mass denominator conventions. 
The conversion is made as follows.
\begin{enumerate}
\item[(i)] For channels in which the baryon fields differ by a common minus sign between the Chen and Cabibbo--Rabl phase conventions, all form factors for that channel are multiplied by the corresponding sign factor.
\item[(ii)] For works that use $(M_i+M_f)$ in the denominator of the subleading form factors, the listed $f_2$, $f_3$, $g_2$, and $g_3$ are rescaled to the $M_i$ convention before they are included in the comparison tables.
\item[(iii)] For works that define the momentum transfer as $q^\nu=P_f^\nu-P_i^\nu$ with the same Lorentz decomposition, the signs of $f_3$ and $g_3$ are reversed before comparison.
\end{enumerate}
Furthermore, in our R3QM calculation, the induced pseudoscalar form factor $g_3$ is obtained from the pseudoscalar pole form in Eq.~\eqref{eq:g3_reconstructed}. For comparisons with other calculations, the literature values of $g_3$ are converted only to the present baryon phase, momentum transfer, and mass denominator conventions. No extra pole contribution is added.

\section{Results and discussion}
\label{sec:results}

\begin{figure*}[t]
	\centering
	\includegraphics[width=17.5cm]{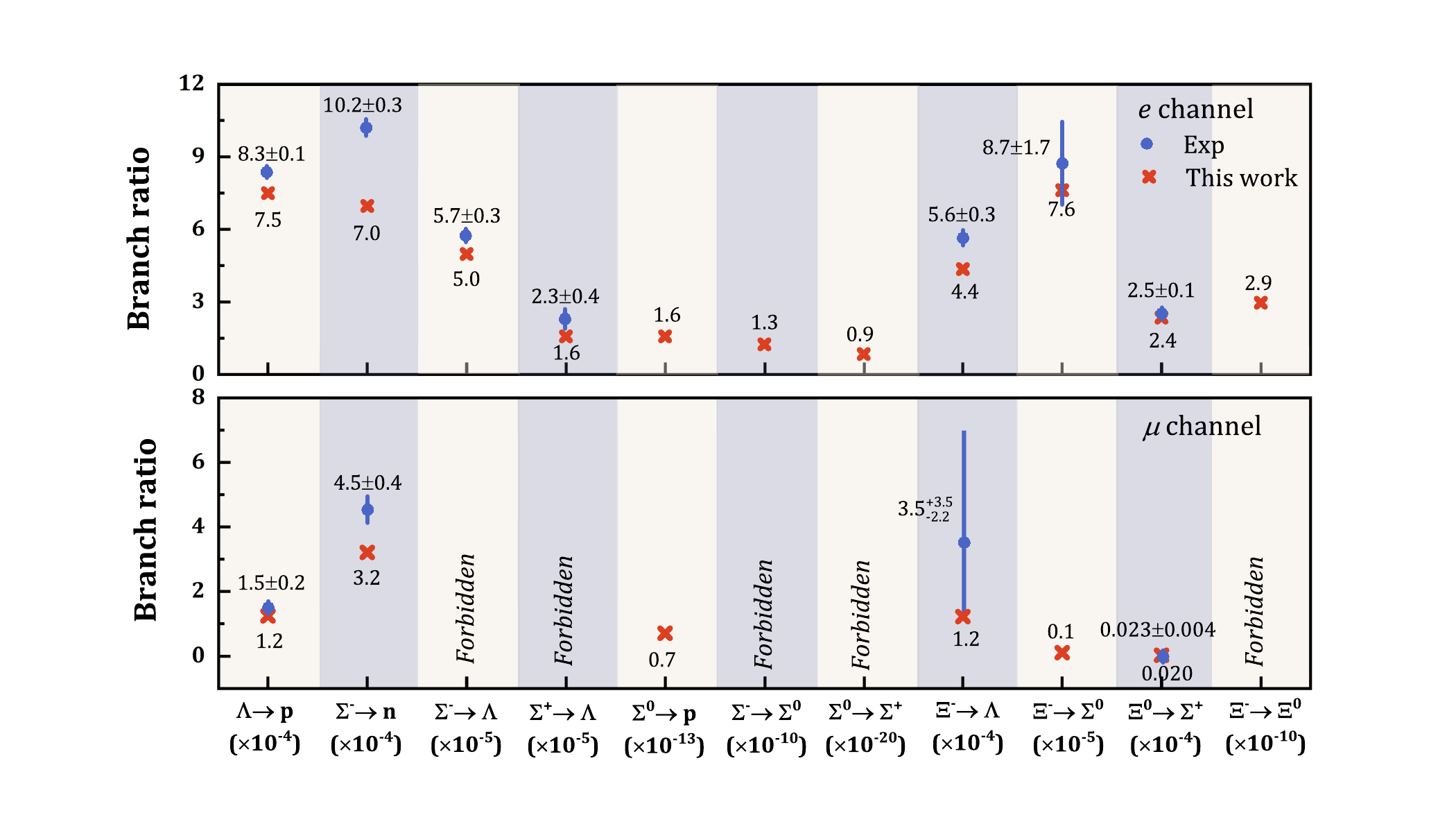}
	\vspace{-0.8 cm} \caption{
		Branching fractions for the hyperon semileptonic decays. The upper panel shows the electron channels, and the lower panel presents the muon channels. The blue circles with error bars denote the experimental averages from RPP~\cite{ParticleDataGroup:2024cfk}. The red crosses represent the theoretical predictions of this work. 
	}\label{fig:branching_ratios}
\end{figure*}

\begin{figure*}[!t]
	\centering
	\includegraphics[width=\textwidth]{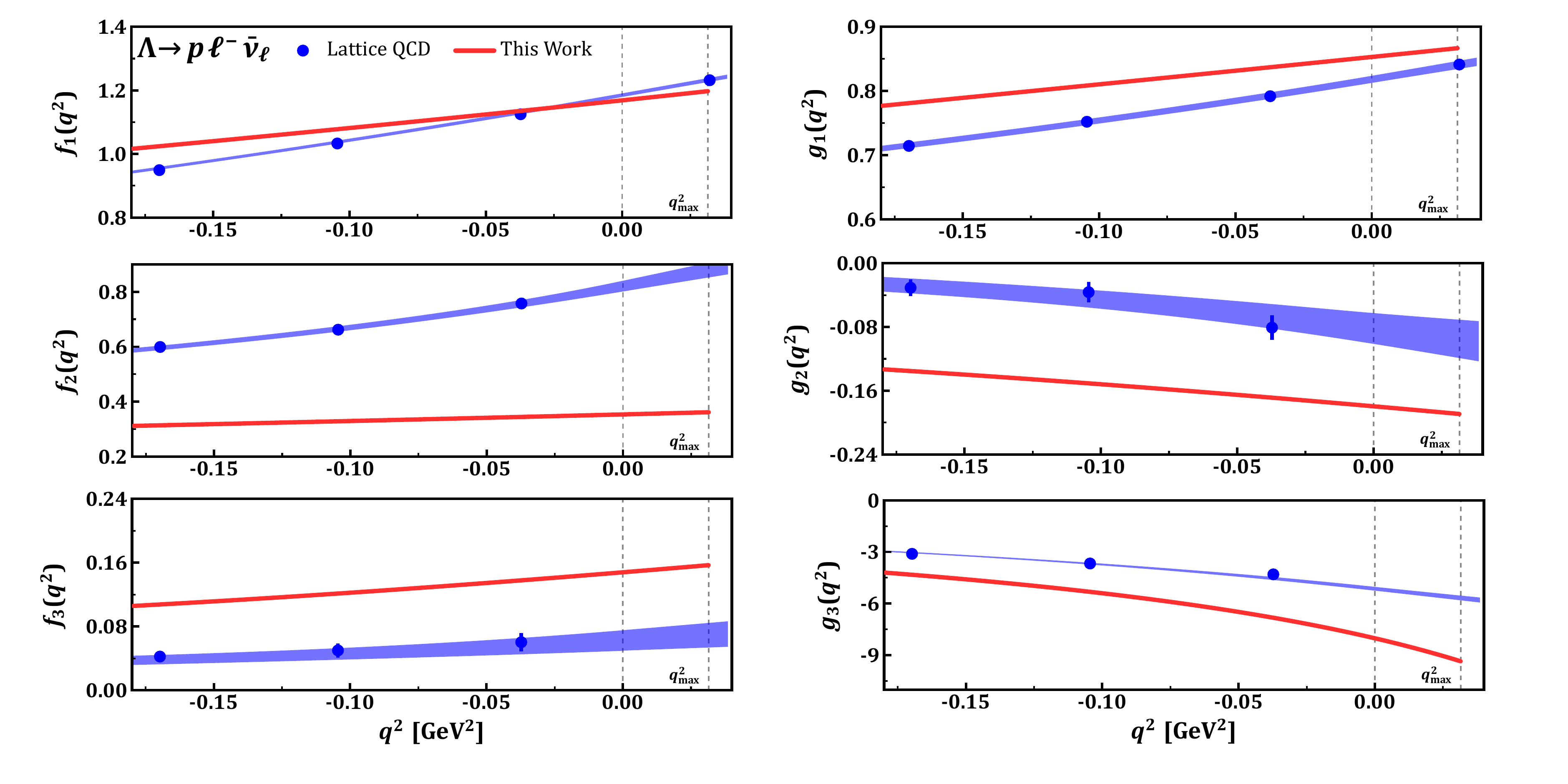}%
	\caption{Form factors $f_i(q^2)$ and $g_i(q^2)$ in the standard Lorentz decomposition for the $\Lambda\to p\ell^-\bar\nu_\ell$ decays. The solid lines denote our R3QM results. The data points with error bands denote the lattice QCD results~\cite{Bacchio:2025auj} for comparison.}
	\label{fig:Lambda_p_form_factors_vs_LQCD}
\end{figure*}

In this section, we present the calculated semileptonic decay widths $\Gamma$, branching fractions, LFU ratios, and six form factors for the $\Lambda$, $\Sigma$, and $\Xi$ semileptonic decays, and discuss them in comparison with available experimental data and theoretical predictions.
The predicted decay widths and branching fractions for all channels are systematically compiled in Tables~\ref{tab:Hyperon_Decays_Observables}, \ref{tab:Hyperon_Decays_Sigma}, and \ref{tab:Hyperon_Decays_Xi} alongside other theoretical calculations and experimental data.
The branching fractions are shown in Fig.~\ref{fig:branching_ratios}, where they are compared with the average values listed in RPP~\cite{ParticleDataGroup:2024cfk}.
Except for the $\Sigma \to n e^- \bar{\nu}_e$ channel, our results are consistent with the experimental data.
In addition, Fig.~\ref{fig:Lambda_p_form_factors_vs_LQCD} shows that the $\Lambda\to p$ form factors are consistent with the recent lattice QCD results~\cite{Bacchio:2025auj}. 

The empirical Cabibbo parametrization in the last column of Table~\ref{tab:flavor_light_optimized} provides a compact way to reorganize the leading axial-vector form factors.
In this comparison, the vector form factor does not introduce an additional $F$--$D$ structure.
Its static reference value is fixed by the SU$(3)$ Clebsch--Gordan coefficient listed in Table~\ref{tab:flavor_light_optimized}.
The Cabibbo parametrization is applied only to the leading axial-vector form factor $g_1(0)$.
The two calculated axial charges in $\Lambda\to p$ and $\Sigma^-\to n$ form an independent basis for the two reduced couplings.
In this basis, $g_1(\Lambda\to p)=0.853$ corresponds to $\sqrt{3/2}(F_{\rm eff}+D_{\rm eff}/3)$, while $g_1(\Sigma^-\to n)=0.218$ corresponds to $D_{\rm eff}-F_{\rm eff}$.
Solving these two equations gives $F_{\rm eff}=0.468$ and $D_{\rm eff}=0.686$.
Using these two effective couplings $F_{\rm eff}$ and $D_{\rm eff}$ and the relations given in the last column of Table~\ref{tab:flavor_light_optimized}, we derive the $g_1(0)$ values for all transition channels. 
The largest discrepancy between these $\mathrm{SU}(3)$-derived values and the direct R3QM results is merely 6~\%, confirming that the R3QM calculation preserves excellent consistency with the leading-order flavor symmetry contributions.

With these experimental and lattice QCD comparisons included, our model is quite successful in describing the weak current structure of the hyperon octet, with the remaining differences mainly appearing in selected weak magnetism amplitudes.
The channel by channel comparisons below further compare the decay widths, branching fractions, LFU ratios, vector charges, and subleading form factors with the available experimental data and lattice QCD results.

\begin{table*}[t]
	\centering
	\setlength{\tabcolsep}{6pt}
	\renewcommand{\arraystretch}{1.20}
	\caption{
			Partial decay widths $\Gamma$, branching fractions $\mathcal{B}$, and the lepton flavor universality (LFU) ratio $R_{\mu e}$ for $\Lambda\to p\ell^-\bar{\nu}_\ell$ ($\ell=e,\mu$), compared with other theoretical calculations and available experimental data.
			For theoretical calculations that report only decay widths (or branching fractions), the corresponding branching fractions (or widths) are obtained via $\Gamma=\mathcal{B}/\tau_\Lambda$ using the $\Lambda$ lifetime $\tau_{\Lambda}=2.617\times10^{-10}\,\mathrm{s}$ from RPP~\cite{ParticleDataGroup:2024cfk}.
	}
	\label{tab:Hyperon_Decays_Observables}
	\begin{tabular*}{\textwidth}{@{\extracolsep{\fill}}lccccc@{}}
		\hline\hline
		\multirow{2}{*}{Method}
		& \multicolumn{2}{c}{$\Gamma\; (10^{6}\,\mathrm{s}^{-1})$}
		& \multicolumn{2}{c}{$\mathcal{B}\; (10^{-4})$}
		& \multirow{2}{*}{$R_{\mu e}$} \\
		\cmidrule(lr){2-3}\cmidrule(lr){4-5}
		& $\ell=e$ & $\ell=\mu$
		& $\ell=e$ & $\ell=\mu$
		& \\
		\hline
		SU$(3)$ IRA~\cite{Wang:2019alu}        & $3.18(5)$ & $0.53(1)$ & $8.32(14)$    & $1.40(2)$     & $0.168(4)$ \\
		LFQM~\cite{Schlumpf:1994fb}            & $3.51$    & $0.58$    & $9.19$        & $1.52$        & $0.165$ \\
		CCQM~\cite{Faessler:2008ix}            & $3.28$    & $0.57$    & $8.58$        & $1.49$        & $0.174$ \\
		NR3QM~\cite{Wu:2013kla}                & $4.09$    & $0.70$    & $10.70$       & $1.82$        & $0.170$ \\
		                                       & $4.51$    & $0.75$    & $11.8$        & $1.95$        & $0.165$ \\
		QCDSR~\cite{Zhang:2024ick}             & $2.95(24)$& $0.52(4)$ & $7.72(64)$    & $1.35(11)$    & $0.175(31)$ \\
		                                       & $2.72(27)$& $0.44(4)$ & $7.12(70)$    & $1.15(11)$    & $0.162(35)$ \\
		LFD~\cite{Lih:2026dcj}                 & $3.18({}^{+12}_{-21})$ & $0.53({}^{+2}_{-3})$ & $8.32({}^{+32}_{-54})$ & $1.38({}^{+5}_{-9})$ & $0.166({}^{+9}_{-15})$ \\
		                                       & $3.18({}^{+12}_{-21})$ & $0.50({}^{+2}_{-3})$ & $8.32({}^{+32}_{-54})$ & $1.31({}^{+4}_{-8})$ & $0.158({}^{+7}_{-14})$ \\
		LQCD~\cite{Bacchio:2025auj}            & $2.93(18)$& $0.51(6)$ & $7.68(48)$    & $1.33(16)$    & $0.1735(98)$ \\
		BESIII~\cite{BESIII:2025hgj}           & --- & --- & $8.16(22)(15)$& ---           & --- \\
		BESIII~\cite{BESIII:2021ynj}           & --- & --- & ---           & $1.48(21)(8)$ & $0.178(28)$ \\
		LHCb~\cite{LHCb:2025wld}               & --- & --- & ---           & $1.462(16)(100)(11)$ & $0.175(12)$ \\
		RPP~\cite{ParticleDataGroup:2024cfk}   & --- & --- & $8.34(14)$    & $1.51(19)$    & $0.181(23)$ \\
		\textbf{This work}
		& $\mathbf{2.85}$ & $\mathbf{0.47}$
		& $\mathbf{7.46}$ & $\mathbf{1.24}$
		& $\mathbf{0.166}$ \\
		\hline\hline
	\end{tabular*}
\end{table*}

\begin{table*}[t]
	\centering
	\setlength{\tabcolsep}{5pt}
	\renewcommand{\arraystretch}{1.25}
	\caption{
		Our predicted transition form factors at $q^2=0$ for $\Lambda \to p\ell^-\bar{\nu}_\ell$, compared with available experimental information and other theoretical results.
		All values are expressed in the unified convention of Sec.~\ref{sec:convention_alignment}.
		The last two columns give the ratios to the static SU$(6)$ reference values, where $|f_1^{\mathrm{stat.}}|=|g_1^{\mathrm{stat.}}|=\sqrt{3/2}$.
		The labels $\chi$QM, $\chi$QSM, and $\chi$CQM denote the chiral quark, chiral quark-soliton, and chiral constituent quark models, respectively.
	}
	\label{tab:Hyperon_Decays_FormFactors}
	\begin{tabular*}{\textwidth}{@{\extracolsep{\fill}}ccccccc@{}}
		\hline\hline
		Method & $f_1(0)$ & $f_2(0)$ & $f_3(0)$ & $g_1(0)$ & $g_2(0)$ & $g_3(0)$ \\
		\hline
		LFQM~\cite{Schlumpf:1994fb}                      & $1.190$      & $0.950$      & ---          & $0.990$      & $0.028$      & --- \\
		$\chi$QM~\cite{Ohlsson:1998bk}                   & $1.220$      & $0.913$      & $-0.337$     & $0.820$      & $0.054$      & $-11.407$ \\
		CCQM~\cite{Faessler:2008ix}                      & $1.226$      & $1.226$      & $-0.067$     & $0.888$      & $0.072$      & $-6.760$ \\
		$\chi$QSM~\cite{Ledwig:2008ku}                   & $1.225$      & $0.870$      & ---          & $0.830$      & ---          & --- \\
		$\chi$CQM~\cite{Sharma:2009hg}                   & $1.225$      & $0.563$      & $-0.225$     & $0.909$      & $0.092$      & $-11.244$ \\
		QCDSR~\cite{Zhang:2024ick}                       & $1.179(75)$  & $0.888(74)$  & ---          & $0.843(10)$  & ---          & --- \\
		LFD~\cite{Lih:2026dcj}                           & $1.305({}^{+15}_{-38})$ & $1.139({}^{+32}_{-13})$ & $-0.917({}^{+144}_{-182})$ & $0.924({}^{+12}_{-24})$ & $0.269({}^{+101}_{-115})$ & $-7.487({}^{+1377}_{-1169})$ \\
		LQCD~\cite{Bacchio:2025auj}                      & $1.185(6)$   & $0.821(19)$  & $0.062(13)$  & $0.818(6)$   & $-0.082(19)$ & $-5.14(11)$ \\
		\textbf{This work}                               & $\mathbf{1.169}$ & $\mathbf{0.354}$ & $\mathbf{0.148}$ & $\mathbf{0.853}$ & $\mathbf{-0.180}$ & $\mathbf{-8.017}$ \\
		\hline\hline
	\end{tabular*}\par\vspace{-0.2em}
	
	\begin{tabular*}{\textwidth}{@{\extracolsep{\fill}}cccccc@{}}
		Method & $f_2(0)/f_1(0)$ & $g_1(0)/f_1(0)$ & $g_2(0)/f_1(0)$ & $|f_1(0)|/|f_{1}^{\mathrm{stat.}}|$ & $|g_1(0)|/|g_{1}^{\mathrm{stat.}}|$ \\
		\hline
		LFQM~\cite{Schlumpf:1994fb}                    & $0.798$     & $0.826$     & $0.023$      & $0.972$     & $0.808$ \\
		$\chi$QM~\cite{Ohlsson:1998bk}                 & $0.748$     & $0.672$     & $0.044$      & $0.996$     & $0.670$ \\
		CCQM~\cite{Faessler:2008ix}                    & $1.000$     & $0.724$     & $0.059$      & $1.001$     & $0.725$ \\
		$\chi$QSM~\cite{Ledwig:2008ku}                 & $0.710$     & $0.678$     & ---          & $1.000$     & $0.678$ \\
		$\chi$CQM~\cite{Sharma:2009hg}                 & $0.460$     & $0.742$     & $0.075$      & $1.000$     & $0.742$ \\
		QCDSR~\cite{Zhang:2024ick}                     & $0.752(74)$ & $0.708(47)$ & ---          & $0.963(61)$ & $0.688(8)$ \\
		LFD~\cite{Lih:2026dcj}                         & $0.872({}^{+35}_{-14})$ & $0.708({}^{+12}_{-27})$ & $0.206({}^{+77}_{-89})$ & $1.065({}^{+12}_{-27})$ & $0.755({}^{+10}_{-20})$ \\
		LQCD~\cite{Bacchio:2025auj}                    & $0.693(17)$ & $0.690(4)$  & $-0.069(16)$ & $0.967(5)$  & $0.668(4)$ \\
		BESIII~\cite{BESIII:2025hgj}                   & $0.77({}^{+53}_{-49})$ & $0.706({}^{+89}_{-86})$ & $-0.19({}^{+65}_{-63})$ & --- & --- \\
		RPP~\cite{ParticleDataGroup:2024cfk}           & ---          & $0.718(15)$ & ---          & ---          & --- \\
		\textbf{This work}                             & $\mathbf{0.303}$ & $\mathbf{0.730}$ & $\mathbf{-0.154}$ & $\mathbf{0.954}$ & $\mathbf{0.696}$ \\
		\hline\hline
	\end{tabular*}
\end{table*}

\subsection{$\Lambda \to p \ell^- \bar{\nu}_\ell$}
\label{sec:lambda_decay_discussion}

The $\Lambda \to p \ell^- \bar{\nu}_\ell$ transitions are the most precisely measured hyperon semileptonic channels. 
In Table~\ref{tab:Hyperon_Decays_Observables} and Fig.~\ref{fig:branching_ratios}, we present the predicted branching fractions.
These values are consistent with the experimental averages compiled in RPP~\cite{ParticleDataGroup:2024cfk} and with the recent measurements by the BESIII Collaboration~\cite{BESIII:2025hgj,BESIII:2021ynj} and the LHCb Collaboration~\cite{LHCb:2025wld}. 
They also agree with the lattice QCD~\cite{Bacchio:2025auj} and QCD sum rule~\cite{Zhang:2024ick} results listed in Table~\ref{tab:Hyperon_Decays_Observables}. 
The $g_1/f_1$ ratio is also consistent with the RPP average value and the recent BESIII determination in the electron channel~\cite{ParticleDataGroup:2024cfk,BESIII:2025hgj}.

Indeed, $f_1$ and $g_1$ give the dominant contribution to the branching fractions and the LFU ratio $R_{\mu e}$. 
The recoil and charged lepton mass factors suppress the subleading terms in the total decay rates. 
The weak magnetism and induced tensor terms, $f_2$ and $g_2$, appear in the transition current through the recoil order structure $\sigma^{\nu\alpha}q_\alpha/M_i$. 
Although these two terms carry the transverse magnetic response and the second-class axial-vector current, respectively, the recoil factor keeps their contributions small in the decay rates. 
The induced scalar and induced pseudoscalar structures, $f_3 q^\nu/M_i$ and $g_3 q^\nu\gamma^5/M_i$, are further suppressed by the charged lepton mass. 
Since these terms are proportional to $q^\nu$, their contraction with the lepton current gives a factor $q_\mu L^\mu=m_\ell\bar u_\ell(1-\gamma^5)v_{\bar\nu}$ for a massless antineutrino. 
Their effects are negligible in the electron channels and remain subleading in the muon channels. 
For the decay width, the $g_3$ term is still suppressed, even though the PCAC relation and the pseudoscalar meson pole increase the magnitude of $g_3$. 
Thus, the absolute rates mainly follow the leading vector and axial-vector part of the helicity amplitudes together with the available phase space.
In Table~\ref{tab:Hyperon_Decays_FormFactors}, we list the values of six form factors at $q^2=0$, compared with various model and experimental data.

For the vector form factor $f_1(0)$, flavor SU$(3)$ breaking is protected by the Ademollo--Gatto theorem~\cite{Ademollo:1964sr}. 
The first-order correction vanishes, so the vector charge can deviate from its SU$(3)$ value only at second order. 
In the present R3QM calculation, we obtain $|f_1(0)|/|f_1^{\mathrm{stat.}}|\simeq 0.954$, where $f_1^{\mathrm{stat.}}$ is the SU$(3)$ Clebsch--Gordan vector charge listed in Table~\ref{tab:flavor_light_optimized}.
This $4.6\%$ reduction is mainly caused by the constituent quark mass difference, $m_s/m_u\approx 1.3$, included in the baryon wave functions.
The size of this correction is close to the octet baryon chiral perturbation theory~\cite{Lacour:2007wm} value ($-5.7 \pm 2.1\%$) and is also compatible with the recent lattice QCD~\cite{Bacchio:2025auj} result $|f_1(0)/f_1^{\mathrm{stat.}}| = 0.9674\pm0.0047$.

The axial-vector form factor $g_1(0)$ shows a larger deviation from its static spin--flavor value than $f_1(0)$. 
The axial-vector weak current contains spin operators, and its matrix element therefore depends directly on the internal spin polarization of baryons and on the spatial wave function mismatch induced by $m_s > m_u$.
The lower components of the Dirac spinors and mass-dependent Wigner rotations further suppress the spin--flavor overlap.
All these effects are automatically incorporated in our calculation through the initial and final baryon wave functions.
The ratio of axial-vector to vector form factors at zero momentum transfer is $g_1(0)/f_1(0) \simeq 0.730$, in good agreement with the PDG average $0.718 \pm 0.015$~\cite{ParticleDataGroup:2024cfk}, the measurement by Bourquin \emph{et al.}~\cite{Bristol-Geneva-Heidelberg-Orsay-Rutherford-Strasbourg:1983rna} of $0.70 \pm 0.03$, and the recent BESIII value $0.706^{+0.089}_{-0.086}$ obtained with the weak electricity term included in the fit~\cite{BESIII:2025hgj}.
The strength of the flavor symmetry breaking effect can be quantified by the ratio of $g_1(0)$ to the static $\mathrm{SU}(6)$ reference value $g_1^{\text{stat.}}$ listed in Table~\ref{tab:flavor_light_optimized}. Our calculation gives a ratio of $0.696$, which falls within the $0.668$--$0.808$ range of theoretical predictions presented in Table~\ref{tab:Hyperon_Decays_FormFactors}.

For the subleading form factors, the values of $f_3(0)$ and $g_3(0)$ in our calculation are still in the range of various theoretical models, while $f_2(0)$ and $g_2(0)$ are both much smaller than those of others.
Furthermore, in Fig.~\ref{fig:Lambda_p_form_factors_vs_LQCD}, we compare the calculated $q^2$ evolution of the six form factors for the $\Lambda\to p\ell^-\bar{\nu}_\ell$ decays with the recent lattice QCD results~\cite{Bacchio:2025auj}. 
The leading form factors $f_1$ and $g_1$ show good agreement with the lattice QCD results in both magnitude and $q^2$ slope. 
For the subleading form factors, $f_3$, $g_2$, and $g_3$ also have the same signs and comparable magnitudes, although their values differ slightly from the lattice QCD results. 
The largest separation appears in the weak magnetism form factor $f_2$. 
Our result is much smaller than that of the lattice calculation. 
The recent BESIII Collaboration~\cite{BESIII:2025hgj} measurement gives $\langle f_2/f_1\rangle_{exp.}=0.77^{+0.53}_{-0.49}$. 
Its central value is also larger than our prediction, but the present experimental uncertainty is still too large to distinguish the two results.

The suppressed value of $f_2$ reflects how the transverse magnetic current is generated in the present calculation. 
The constituent quarks are treated as pointlike Dirac particles, and neither anomalous quark magnetic moments nor explicit exchange currents are introduced. 
With these assumptions, the weak magnetism form factor is generated by recoil kinematics and Wigner spin rotations in the valence three-quark wave functions, as in instant-form calculations with structureless quarks~\cite{Julia-Diaz:2004yqv}. 
A related issue appears in baryon magnetic moments and electromagnetic magnetic form factors, where pion cloud effects or $qqqq\bar q$ components can add magnetic strength beyond the valence quark contribution~\cite{Lu:1997sd, Lu:2001yc, Ramalho:2011pp, an:2006zf}. 
The comparison with other calculations in Table~\ref{tab:Hyperon_Decays_FormFactors} makes the same point for weak magnetism.
The axial ratio $g_1/f_1$ remains in the common range of existing calculations.
For the weak magnetism ratio, the present value $f_2/f_1$ is much smaller than most model results, mainly because the weak magnetism strength is treated differently.
In the other models, larger $f_2$ values usually receive contributions from chiral dynamics, collective motion, configuration mixing, or phenomenological magnetic inputs, which add to or effectively absorb magnetic strength beyond the valence three-quark core~\cite{Schlumpf:1994fb,Ohlsson:1998bk,Faessler:2008ix,Ledwig:2008ku,Sharma:2009hg}.
The small $f_2/f_1$ in the $\Lambda\to p$ channel then points to transverse current strength beyond the three-quark core, which may be supplied by meson exchange currents, meson cloud dressing, or five-quark Fock components.

The scalar form factor $f_3$ and the induced tensor form factor $g_2$ test the flavor breaking part of the overlap. 
Since they vanish in the exact SU$(3)$ limit, their nonzero values in the present calculation arise from the spatial wave function mismatch induced by flavor symmetry breaking. 

The induced pseudoscalar form factor $g_3$ has a separate origin, being governed mainly by the PCAC relation and the pseudoscalar meson pole. 
The direct overlap extraction gives only a smooth contribution, $g_3^{\mathrm{np}}(0)\approx -1.1$, and does not contain the corresponding kaon pole in the $t$ channel. 
We reconstruct this pole contribution using the generalized Goldberger--Treiman relation in Eq.~\eqref{eq:g3_reconstructed}, obtaining $g_3(0)\simeq -8.017$.
This value has the same sign and order of magnitude as the lattice QCD result, $g_3^{\mathrm{LQCD}}(0)\simeq-5.14\pm0.11$~\cite{Bacchio:2025auj}, and sets the correct scale of $g_3$ in the physical decay region, as shown in Fig.~\ref{fig:Lambda_p_form_factors_vs_LQCD}.

As a whole, the $\Lambda\to p$ transition serves as the reference channel for the present R3QM calculation. 
Our calculated branching fractions, LFU ratio, and leading form factors $f_1$ and $g_1$ are compatible with the available experimental data and lattice QCD results. 
For the second-class form factors, the present three-quark calculation relates $f_3$ and $g_2$ to the spatial overlap generated by SU$(3)$ breaking. 
The induced pseudoscalar form factor $g_3$ is obtained from the PCAC pole relation. 
For weak magnetism, the calculated $f_2$ is smaller than the lattice QCD value, suggesting that the transverse weak current may not be fully saturated by the valence three-quark core in this channel. 
Meson cloud effects or five-quark components may therefore be relevant for a more complete description of $f_2$.

\begin{table*}[t!]
	\centering
	\setlength{\tabcolsep}{2.5pt}
	\renewcommand{\arraystretch}{1.15}
	\caption{
		Partial decay widths $\Gamma$, branching fractions $\mathcal{B}$, and LFU ratios $R_{\mu e}$ for semileptonic $\Sigma$ hyperon decays, compared with available experimental data and other theoretical results.
		For theoretical sources that report only branching fractions, the displayed widths are obtained via $\Gamma=\mathcal{B}/\tau$ using the hyperon lifetimes $\tau_{\Sigma^-}=1.479\times10^{-10}\,\mathrm{s}$, $\tau_{\Sigma^0}=7.4\times10^{-20}\,\mathrm{s}$, and $\tau_{\Sigma^+}=0.8018\times10^{-10}\,\mathrm{s}$ from RPP~\cite{ParticleDataGroup:2024cfk}.
		The experimental values are taken from RPP~\cite{ParticleDataGroup:2024cfk}; dashes denote unavailable data, while ``Forbidden'' denotes muon modes that are kinematically inaccessible.
	}
	\label{tab:Hyperon_Decays_Sigma}
	\begin{tabular*}{\textwidth}{@{\extracolsep{\fill}}ccccccc@{}}
		\hline\hline
		\multirow{2}{*}{Method}
		& \multicolumn{2}{c}{$\Gamma\; (10^{6}\,\mathrm{s}^{-1})$}
		&
		& \multicolumn{2}{c}{$\mathcal{B}$}
		& \multirow{2}{*}{$R_{\mu e}$} \\
		\cline{2-3}\cline{5-6}
		& $\ell=e$ & $\ell=\mu$ & & $\ell=e$ & $\ell=\mu$ & \\
		\hline
		\multicolumn{7}{@{}l}{\textit{$\Sigma^- \to n \ell^- \bar{\nu}_\ell$}\quad [$\mathcal{B}$ in units of $10^{-4}$]} \\
		SU$(3)$ IRA~\cite{Wang:2019alu}       & $6.88(23)$ & $3.06(10)$ & & $10.17(34)$ & $4.53(15)$ & $0.445$ \\
		LFQM~\cite{Schlumpf:1994fb}           & $5.74$    & $2.54$    & & $8.49$      & $3.76$     & $0.443$ \\
		CCQM~\cite{Faessler:2008ix}           & $6.50$    & $3.15$    & & $9.61$      & $4.66$     & $0.485$ \\
		NR3QM~\cite{Wu:2013kla}               & $5.52$    & $2.67$    & & $8.16$      & $3.95$     & $0.484$ \\
		                                      & $7.00$    & $3.28$    & & $10.35$     & $4.85$     & $0.469$ \\
		QCDSR~\cite{Zhang:2024ick}            & $6.76(122)$ & $3.20(59)$ & & $10.00(180)$ & $4.73(88)$ & $0.473$ \\
		                              & $5.88(95)$  & $2.53(42)$ & & $8.70(140)$  & $3.74(62)$ & $0.430$ \\
		RPP~\cite{ParticleDataGroup:2024cfk}  & --- & --- & & $10.17(34)$ & $4.5(4)$ & $0.442(56)$ \\
		\textbf{This work} & $\mathbf{4.73}$ & $\mathbf{2.17}$ & & $\mathbf{6.99}$ & $\mathbf{3.22}$ & $\mathbf{0.460}$ \\
		\hline
		\multicolumn{7}{@{}l}{\textit{$\Sigma^- \to \Lambda e^- \bar{\nu}_e$}\quad [$\mathcal{B}$ in units of $10^{-5}$]} \\
		SU$(3)$ IRA~\cite{Wang:2019alu}       & $0.39(2)$ & --- & & $5.73(27)$ & --- & --- \\
		LFQM~\cite{Schlumpf:1994fb}           & $0.39$    & --- & & $5.75$     & --- & --- \\
		CCQM~\cite{Faessler:2008ix}           & $0.43$    & --- & & $6.36$     & --- & --- \\
		RPP~\cite{ParticleDataGroup:2024cfk}  & --- & --- & & $5.73(27)$ & --- & --- \\
		\textbf{This work} & $\mathbf{0.34}$ & \textbf{Forbidden} & & $\mathbf{4.98}$ & \textbf{Forbidden} & --- \\[0.2em]
		\multicolumn{7}{@{}l}{\textit{$\Sigma^+ \to \Lambda e^+ \nu_e$}\quad [$\mathcal{B}$ in units of $10^{-5}$]} \\
		SU$(3)$ IRA~\cite{Wang:2019alu}       & $0.23(1)$ & --- & & $1.88(11)$ & --- & --- \\
		LFQM~\cite{Schlumpf:1994fb}           & $0.24$    & --- & & $1.92$     & --- & --- \\
		CCQM~\cite{Faessler:2008ix}           & $0.26$    & --- & & $2.08$     & --- & --- \\
		RPP~\cite{ParticleDataGroup:2024cfk}  & --- & --- & & $2.3(4)$   & --- & --- \\
		\textbf{This work} & $\mathbf{0.20}$ & \textbf{Forbidden} & & $\mathbf{1.62}$ & \textbf{Forbidden} & --- \\
		\hline
		\multicolumn{7}{@{}l}{\textit{$\Sigma^0 \to p \ell^- \bar{\nu}_\ell$}\quad [$\mathcal{B}$ in units of $10^{-13}$]} \\
		SU$(3)$ IRA~\cite{Wang:2019alu}       & $3.26(43)$ & $1.42(19)$ & & $2.41(32)$ & $1.05(14)$ & $0.436$ \\
		LFQM~\cite{Schlumpf:1994fb}           & $2.72$     & $1.18$     & & $2.01$     & $0.87$     & $0.434$ \\
		NR3QM~\cite{Wu:2013kla}          & $2.54$     & $1.19$     & & $1.88$     & $0.88$     & $0.469$ \\
		                                  & $3.22$     & $1.46$     & & $2.38$     & $1.08$     & $0.453$ \\
		QCDSR~\cite{Zhang:2024ick}            & $3.38(95)$ & $1.59(46)$ & & $2.50(70)$ & $1.18(34)$ & $0.472$ \\
		                              & $2.95(78)$ & $1.27(34)$ & & $2.18(58)$ & $0.94(25)$ & $0.431$ \\
		\textbf{This work} & $\mathbf{2.22}$ & $\mathbf{1.00}$ & & $\mathbf{1.64}$ & $\mathbf{0.74}$ & $\mathbf{0.449}$ \\[0.2em]
		\multicolumn{7}{@{}l}{\textit{$\Sigma^- \to \Sigma^0 e^- \bar{\nu}_e$}\quad [$\mathcal{B}$ in units of $10^{-10}$]} \\
		SU$(3)$ IRA~\cite{Wang:2019alu}       & $2.95(271)\times10^{-6}$ & --- & & $4.36(401)$ & --- & --- \\
		LFQM~\cite{Schlumpf:1994fb}           & $1.47\times10^{-6}$ & --- & & $2.17$ & --- & --- \\
		\textbf{This work} & $\mathbf{8.49\times10^{-7}}$ & \textbf{Forbidden} & & $\mathbf{1.26}$ & \textbf{Forbidden} & --- \\[0.3em]
		\multicolumn{7}{@{}l}{\textit{$\Sigma^0 \to \Sigma^+ e^- \bar{\nu}_e$}\quad [$\mathcal{B}$ in units of $10^{-20}$]} \\
		SU$(3)$ IRA~\cite{Wang:2019alu}       & $4.61(432)\times10^{-7}$ & --- & & $3.41(320)$ & --- & --- \\
		LFQM~\cite{Schlumpf:1994fb}           & $3.65\times10^{-8}$ & --- & & $0.27$ & --- & --- \\
		\textbf{This work} & $\mathbf{1.17\times10^{-7}}$ & \textbf{Forbidden} & & $\mathbf{0.86}$ & \textbf{Forbidden} & --- \\
		\hline\hline
	\end{tabular*}
\end{table*}

\begin{table*}[t]
	\centering
	\setlength{\tabcolsep}{4pt}
	\renewcommand{\arraystretch}{1.20}
	\caption{
		Our predicted transition form factors at $q^2=0$ for $\Sigma^- \to n\ell^-\bar{\nu}_\ell$ and $\Sigma^0 \to p\ell^-\bar{\nu}_\ell$, compared with available experimental information and other theoretical results.
		The last two columns give the ratios to the static SU$(6)$ reference values, where $(f_1^{\mathrm{stat.}},g_1^{\mathrm{stat.}})=(-1,1/3)$ for $\Sigma^-\to n$ and $(-1/\sqrt{2},1/(3\sqrt{2}))$ for $\Sigma^0\to p$.
	}
	\label{tab:Sigma_np_FormFactors}
	\begin{tabular*}{\textwidth}{@{\extracolsep{\fill}}ccccccc@{}}
		\hline\hline
		Method & $f_1(0)$ & $f_2(0)$ & $f_3(0)$ & $g_1(0)$ & $g_2(0)$ & $g_3(0)$ \\
		\hline
		\multicolumn{7}{@{}l}{\textit{$\Sigma^- \to n\ell^-\bar{\nu}_\ell$}} \\
		LFQM~\cite{Schlumpf:1994fb}                      & $-0.970$     & $0.742$      & ---          & $0.270$      & $0.007$      & --- \\
		$\chi$QM~\cite{Ohlsson:1998bk}                    & $-1.000$     & $1.020$      & $0.471$      & $0.220$      & $-0.017$     & $-4.034$ \\
		CCQM~\cite{Faessler:2008ix}                       & $-1.009$     & $0.971$      & $0.055$      & $0.262$      & $-0.078$     & $-2.199$ \\
		$\chi$QSM~\cite{Ledwig:2008ku}                    & $-1.000$     & $0.960$      & ---          & $0.270$      & ---          & --- \\
		$\chi$CQM~\cite{Sharma:2009hg}                    & $-1.000$     & $1.016$      & $0.345$      & $0.314$      & $0.010$      & $-5.155$ \\
		QCDSR~\cite{Zhang:2024ick}                        & $-0.993$     & $1.025$      & ---          & $0.324$      & ---          & --- \\
		LQCD~\cite{Sasaki:2017jue}                        & $-0.957(10)$ & ---          & ---          & ---          & ---          & --- \\
		LQCD~\cite{Guadagnoli:2006gj}                     & $-0.988(49)$ & $0.841(450)$ & $0.232(122)$ & $0.284(53)$  & $-0.349(145)$ & $-3.376(1834)$ \\
		\textbf{This work}                                & $\mathbf{-0.929}$ & $\mathbf{0.780}$ & $\mathbf{0.011}$ & $\mathbf{0.218}$ & $\mathbf{-0.050}$ & $\mathbf{-2.292}$ \\[0.3em]
		\multicolumn{7}{@{}l}{\textit{$\Sigma^0 \to p\ell^-\bar{\nu}_\ell$}} \\
		LFQM~\cite{Schlumpf:1994fb}       & $-0.690$ & $0.524$ & --- & $0.190$ & $0.005$ & --- \\
		QCDSR~\cite{Zhang:2024ick}        & $-0.702$ & $0.725$ & --- & $0.229$ & ---     & --- \\
		\textbf{This work}                & $\mathbf{-0.658}$ & $\mathbf{0.552}$ & $\mathbf{0.007}$ & $\mathbf{0.155}$ & $\mathbf{-0.035}$ & $\mathbf{-1.612}$ \\
		\hline\hline
	\end{tabular*}\par\vspace{-0.25em}
	\begin{tabular*}{\textwidth}{@{\extracolsep{\fill}}ccccc@{}}
		Method & $f_2(0)/f_1(0)$ & $g_1(0)/f_1(0)$ & $|f_1(0)|/|f_{1}^{\mathrm{stat.}}|$ & $|g_1(0)|/|g_{1}^{\mathrm{stat.}}|$ \\
		\hline
		\multicolumn{5}{@{}l}{\textit{$\Sigma^- \to n\ell^-\bar{\nu}_\ell$}} \\
		LFQM~\cite{Schlumpf:1994fb}                         & $-0.765$     & $-0.275$     & $0.970$     & $0.811$ \\
		$\chi$QM~\cite{Ohlsson:1998bk}                      & $-1.020$     & $-0.220$     & $1.000$     & $0.661$ \\
		CCQM~\cite{Faessler:2008ix}                         & $-0.962$     & $-0.260$     & $1.009$     & $0.787$ \\
		$\chi$QSM~\cite{Ledwig:2008ku}                      & $-0.960$     & $-0.270$     & $1.000$     & $0.811$ \\
		$\chi$CQM~\cite{Sharma:2009hg}                      & $-1.016$     & $-0.314$     & $1.000$     & $0.943$ \\
		QCDSR~\cite{Zhang:2024ick}                          & $-1.042(90)$ & $-0.327(46)$ & $0.993$     & $0.973$ \\
		LQCD~\cite{Sasaki:2017jue}                          & ---          & ---          & $0.957(10)$ & --- \\
		LQCD~\cite{Guadagnoli:2006gj}                       & $-0.852(454)$ & $-0.287(52)$ & $0.988(49)$ & $0.851(160)$ \\
		RPP~\cite{ParticleDataGroup:2024cfk}                & $-0.97(14)$ & $-0.340(17)$ & ---          & --- \\
		\textbf{This work}                                  & $\mathbf{-0.841}$ & $\mathbf{-0.235}$ & $\mathbf{0.929}$ & $\mathbf{0.655}$ \\[0.3em]
		\multicolumn{5}{@{}l}{\textit{$\Sigma^0 \to p\ell^-\bar{\nu}_\ell$}} \\
		LFQM~\cite{Schlumpf:1994fb}         & $-0.759$     & $-0.275$     & $0.976$     & $0.805$ \\
		QCDSR~\cite{Zhang:2024ick}          & $-1.042(90)$ & $-0.327(46)$ & $0.993$     & $0.970$ \\
		\textbf{This work}                  & $\mathbf{-0.839}$ & $\mathbf{-0.235}$ & $\mathbf{0.930}$ & $\mathbf{0.656}$ \\
		\hline\hline
	\end{tabular*}
\end{table*}

\subsection{$\Sigma$ hyperon}
\label{sec:Sigma_decay_discussion}

The $\Sigma$ semileptonic decays separate three symmetry effects. 
The first is the $|\Delta S|=1$ structure in $\Sigma^- \to n \ell^- \bar{\nu}_\ell$ and $\Sigma^0\to p\ell^-\bar{\nu}_\ell$, where the vector charge is fixed by the SU$(3)$ Clebsch--Gordan coefficient and the axial-vector strength is given by the small Cabibbo combination $D-F$.
The neutral mode $\Sigma^0\to p$ shares the same weak current structure with $\Sigma^-\to n$, but its branching fraction is strongly suppressed by the short $\Sigma^0$ lifetime.
A different limit is reached in the $\Delta S=0$ transitions $\Sigma \to \Lambda \ell\nu$, which isolate a structure dominated by the axial-vector current because the leading vector form factor $f_1$ vanishes in the exact SU$(3)$ limit. 
The remaining $\Sigma\to\Sigma$ transitions stay inside the isotriplet.
Their very small phase space makes the isospin breaking effects visible.

The experimental information is also uneven among these three groups of $\Sigma$ transitions. 
For the measured $\Sigma^- \to n \ell^- \bar{\nu}_\ell$, $\Sigma^- \to \Lambda e^- \bar{\nu}_e$, and $\Sigma^+ \to \Lambda e^+ \nu_e$ channels, Table~\ref{tab:Hyperon_Decays_Sigma} shows that our branching fractions are slightly lower than the RPP averages, whereas $R_{\mu e}$ for $\Sigma^- \to n \ell^- \bar{\nu}_\ell$ stays close to the listed value. 
It should be emphasized that the present rate comparison is still limited by the experimental data, since the RPP averages for these channels are largely inherited from legacy fixed target measurements around 40 years ago~\cite{ParticleDataGroup:2024cfk,Bristol-Geneva-Heidelberg-Orsay-Rutherford-Strasbourg:1981uuv,Bristol-Geneva-Heidelberg-Orsay-Rutherford-Strasbourg:1983jpz}. 
Future measurements of the branching fractions and decay distributions would test whether the lower $\Sigma$ rates and the axial-vector matrix elements discussed below are supported by data.

\subsubsection{$\Sigma^- \to n \ell^- \bar{\nu}_\ell$ and $\Sigma^0 \to p \ell^- \bar{\nu}_\ell$}

The $\Sigma^-\to n$ and $\Sigma^0\to p$ channels are two isospin projections of the same $s\to u$ weak matrix element between a $\Sigma$ hyperon and a nucleon.
The charged channel, $\Sigma^-\to n$, carries the available comparison with experiment and lattice QCD.
The neutral channel, $\Sigma^0\to p$, gives the corresponding neutral isospin projection of this matrix element.
In the exact isospin limit, its form factors are related to those of $\Sigma^-\to n$ by the Clebsch--Gordan factor $1/\sqrt{2}$.
%

This isospin relation appears directly in the leading form factors.
For $\Sigma^-\to n$, the static coefficients are $f_1^{\mathrm{stat.}}=-1$ and $g_1^{\mathrm{stat.}}=D-F=1/3$.
For $\Sigma^0\to p$, they are $f_1^{\mathrm{stat.}}=-1/\sqrt{2}$ and $g_1^{\mathrm{stat.}}=1/(3\sqrt{2})$.
At $q^2=0$, the present calculation gives $f_1(0)=-0.929$ and $g_1(0)=0.218$ for $\Sigma^-\to n$, and $f_1(0)=-0.658$ and $g_1(0)=0.155$ for $\Sigma^0\to p$.
The corresponding ratios to the static SU$(6)$ reference values are $|f_1(0)|/|f_1^{\mathrm{stat.}}|\simeq0.929$ and $|g_1(0)|/|g_1^{\mathrm{stat.}}|\simeq0.655$ for the charged channel, and $0.930$ and $0.656$ for the neutral channel.
The near equality of these ratios shows that the neutral channel mainly changes the flavor Clebsch--Gordan projection.

For the $\Sigma\to N$ pair, the charged channel $\Sigma^-\to n$ gives the direct comparison with lattice QCD and with the measured $g_1/f_1$ ratio.
The vector charge is close to its static SU$(6)$ reference value, with $f_1(0)=-0.929$ and $|f_1(0)|/|f_1^{\mathrm{stat.}}|\simeq0.929$.
This ratio is slightly below the lattice QCD central values $0.957(10)$~\cite{Sasaki:2017jue} and $0.988(49)$~\cite{Guadagnoli:2006gj}, with the size of a moderate Ademollo--Gatto protected correction~\cite{Ademollo:1964sr}.
The clearer difference appears in the axial ratio.
The calculated $|g_1(0)/f_1(0)|\simeq0.235$ lies below the RPP average $0.340\pm0.017$~\cite{ParticleDataGroup:2024cfk} and the older Tanenbaum result $0.435\pm0.035$~\cite{Tanenbaum:1975bp}.
This ratio is governed by the small $D-F$ matrix element in the axial current.
The value $g_1(0)=0.218$ is also slightly below the lattice QCD central value $0.284(53)$, while $g_1(0)/f_1(0)\simeq -0.235$ remains compatible with the lattice QCD result $-0.287(52)$ within the present uncertainty~\cite{Guadagnoli:2006gj}.
Future measurements would give a stricter test of the present model.
More precise branching fractions would fix the rate scale, and decay distributions would determine whether the lower rate is accompanied by a smaller $D-F$ axial contribution through $g_1/f_1$.

Among the subleading form factors, the weak magnetism term $f_2$ is the one with direct lattice QCD and RPP comparison in $\Sigma^-\to n$.
It also gives the clearest comparison with the $\Lambda\to p$ channel.
The present calculation gives $f_2(0)=0.780$, which falls within the theoretical range $0.742$--$1.025$ listed in Table~\ref{tab:Sigma_np_FormFactors}.
After division by $f_1$, this gives $f_2(0)/f_1(0)\simeq -0.841$, also within the corresponding theoretical range ($-1.042$ to $-0.765$) and close to the lattice QCD central value, $-0.852$~\cite{Guadagnoli:2006gj}, as listed in Table~\ref{tab:Sigma_np_FormFactors}.
With the same sign convention, it is also compatible with the RPP average $(f_2/f_1)_{\mathrm{exp.}}=-0.97\pm0.14$~\cite{ParticleDataGroup:2024cfk}.
Unlike in $\Lambda\to p$, where the present value is much smaller than the lattice QCD result~\cite{Bacchio:2025auj}, the weak magnetism ratio in $\Sigma^-\to n$ remains close to both comparison values.

The spectator spin decomposition in Table~\ref{tab:flavor_light_optimized} may explain why $f_2/f_1$ is less suppressed in $\Sigma^-\to n$ than in $\Lambda\to p$.
The $\Lambda\to p$ flavor overlap comes mainly from the $\rho\rho$ component, where the spectator pair has $s_{12}=0$.
The $\Sigma^-\to n$ flavor overlap instead comes mainly from the $\lambda\lambda$ component with $s_{12}=1$.
A spin-1 spectator may retain more of the spin recoupling induced by recoil, so the transverse magnetic response in the present valence three-quark calculation with pointlike quark weak currents need not be as suppressed as in scalar spectator channels such as $\Lambda\to p$ and $\Xi^0\to\Sigma^+$ (discussed below).
The $\Sigma^-\to n$ channel then shows how an axial-vector spectator component may keep a sizable weak magnetism contribution without introducing additional nonvalence transverse currents.

\begin{table*}[t]
	\centering
	\setlength{\tabcolsep}{3pt}
	\renewcommand{\arraystretch}{1.20}
	\caption{
		Our predicted transition form factors at $q^2=0$ for $\Sigma\to\Lambda$ semileptonic hyperon decays, compared with available experimental information and other theoretical results.
		The last two columns give $f_2(0)/g_1(0)$ and $g_1(0)/g_1^{\mathrm{stat.}}$, where $g_1^{\mathrm{stat.}}=-\sqrt{2/3}$ for $\Sigma^-\to\Lambda$ and $g_1^{\mathrm{stat.}}=\sqrt{2/3}$ for $\Sigma^+\to\Lambda$.
	}
	\label{tab:Sigma_Lambda_FormFactors}
	\begin{tabular*}{\textwidth}{@{\extracolsep{\fill}}ccccccccc@{}}
		\hline\hline
		Method & $f_1(0)$ & $f_2(0)$ & $f_3(0)$ & $g_1(0)$ & $g_2(0)$ & $g_3(0)$ & $f_2(0)/g_1(0)$ & $g_1(0)/g_1^{\mathrm{stat.}}$ \\
		\hline
		\multicolumn{9}{@{}l}{\textit{$\Sigma^- \to \Lambda e^- \bar{\nu}_e$}} \\
		LFQM~\cite{Schlumpf:1994fb}    & $0$          & $-1.245$     & ---          & $-0.600$     & $0$          & ---       & $2.08$ & $0.735$ \\
		$\chi$QM~\cite{Ohlsson:1998bk} & $0$          & $-1.413$     & $-0.052$     & $-0.550$     & $0.129$      & $124.200$ & $2.57$ & $0.674$ \\
		CCQM~\cite{Faessler:2008ix}    & $0.002$      & $-1.206$     & $0.016$      & $-0.633$     & $-0.013$     & $84.446$  & $1.91$ & $0.775$ \\
		$\chi$QSM~\cite{Ledwig:2008ku} & $0$          & $-1.240$     & ---          & $-0.600$     & ---          & ---       & $2.07$ & $0.735$ \\
		$\chi$CQM~\cite{Sharma:2009hg} & $0$          & $-1.173$     & $-0.041$     & $-0.646$     & $0.079$      & $140.500$ & $1.82$ & $0.791$ \\
		RPP~\cite{ParticleDataGroup:2024cfk} & --- & --- & --- & --- & --- & --- & $2.4(17)$ & --- \\
		\textbf{This work}             & $\mathbf{0.001}$ & $\mathbf{-0.722}$ & $\mathbf{-0.030}$ & $\mathbf{-0.562}$ & $\mathbf{0.121}$ & $\mathbf{79.981}$ & $\mathbf{1.28}$ & $\mathbf{0.689}$ \\[0.3em]
		\multicolumn{9}{@{}l}{\textit{$\Sigma^+ \to \Lambda e^+ \nu_e$}} \\
		LFQM~\cite{Schlumpf:1994fb}    & $0$          & $1.237$      & ---          & $0.600$      & $0$          & ---        & $2.06$ & $0.735$ \\
		$\chi$QM~\cite{Ohlsson:1998bk} & $0$          & $1.404$      & $0.046$      & $0.550$      & $-0.114$     & $-108.400$ & $2.55$ & $0.674$ \\
		$\chi$CQM~\cite{Sharma:2009hg} & $0$          & $1.165$      & $0.037$      & $0.646$      & $-0.070$     & $-126.900$ & $1.80$ & $0.791$ \\
		\textbf{This work}             & $\mathbf{-0.001}$ & $\mathbf{0.720}$ & $\mathbf{0.028}$ & $\mathbf{0.564}$ & $\mathbf{-0.119}$ & $\mathbf{-79.342}$ & $\mathbf{1.28}$ & $\mathbf{0.690}$ \\
		\hline\hline
	\end{tabular*}
\end{table*}

\subsubsection{$\Sigma^- \to \Lambda e^- \bar{\nu}_e$ and $\Sigma^+ \to \Lambda e^+ \nu_e$}

The $\Delta S=0$ transitions $\Sigma^- \to \Lambda e^- \bar{\nu}_e$ and $\Sigma^+ \to \Lambda e^+ \nu_e$ are related by isospin reflection, and the leading vector charge vanishes in this pair.
In the exact SU$(3)$ flavor limit, the static coefficients are $f_1^{\mathrm{stat.}}=0$ and $g_1^{\mathrm{stat.}}=\mp\sqrt{2/3}$, so the leading transition strength is axial-vector dominated.
The two charge channels share the same weak transition dynamics, and the different rates listed in Table~\ref{tab:Hyperon_Decays_Sigma} mainly reflect the longer lifetime and the larger available phase space of the $\Sigma^-$ channel.

The leading form factors display the same isospin reflection.
With the present phase convention, the calculation gives $f_1(0)=+0.001$ and $g_1(0)=-0.562$ for $\Sigma^-\to\Lambda$, and $f_1(0)=-0.001$ and $g_1(0)=+0.564$ for $\Sigma^+\to\Lambda$; the two vector charges are numerically consistent with the symmetry limit value zero.
The corresponding axial-vector ratios are $|g_1(0)|/|g_1^{\mathrm{stat.}}|\simeq0.689$ and $0.690$, respectively.
The nearly equal ratios show that the two charge channels have the same axial-vector scale, with the relative sign fixed by the charge convention.
For the measured $\Sigma^-\to\Lambda$ channel, the RPP average $(f_1/g_1)_{\mathrm{exp.}}=0.01\pm0.10$~\cite{ParticleDataGroup:2024cfk} and the WA2 value $+0.034\pm0.080$~\cite{Bristol-Geneva-Heidelberg-Orsay-Rutherford-Strasbourg:1981uuv} both support a small vector charge contribution.
The $\Sigma^+\to\Lambda$ channel has no corresponding form factor measurement, but the same suppression follows from the matrix element related by isospin reflection.

The weak magnetism term $f_2$ is the main vector response left after the vector charge vanishes.
We obtain $f_2(0)=-0.722$ and $f_2(0)/g_1(0)\simeq1.28$ for $\Sigma^-\to\Lambda$, while $f_2(0)=+0.720$ and $f_2(0)/g_1(0)\simeq1.28$ for $\Sigma^+\to\Lambda$.
Our charged channel ratio is compatible with the measured value $(f_2/g_1)_{\mathrm{exp.}}=2.4\pm1.7$~\cite{Bristol-Geneva-Heidelberg-Orsay-Rutherford-Strasbourg:1981uuv,ParticleDataGroup:2024cfk}, but the present experimental uncertainty is still too large to determine the transverse magnetic strength.
Compared with CVC and other model estimates, the main difference is in the size of $f_2$ itself.
Our value $|f_2(0)|\simeq0.72$ is below the CVC estimate based on octet magnetic moments and also below the model values around $1.2$--$1.5$~\cite{Cabibbo:2003cu,Schlumpf:1994fb,Ohlsson:1998bk,Faessler:2008ix,Ledwig:2008ku}.
This smaller weak magnetism strength suggests that the present valence three-quark calculation with pointlike quark weak currents may not fully saturate the transverse magnetic response, leaving room for meson cloud or exchange current contributions.
A more precise measurement of $f_2/g_1$ would determine whether the transverse magnetic response in $\Sigma\to\Lambda$ is below the CVC estimate.

The remaining terms have two separate roles.
The induced scalar and induced tensor pieces are small, with $f_3(0)=-0.030$ and $g_2(0)=+0.121$ for $\Sigma^-\to\Lambda$, and $f_3(0)=+0.028$ and $g_2(0)=-0.119$ for $\Sigma^+\to\Lambda$.
They are SU$(3)$ breaking second class components, and their contributions to the electron widths are suppressed by recoil kinematics and the $q^\nu$ contraction.
The induced pseudoscalar form factor is different.
In the $\Delta S=0$ $\Sigma\to\Lambda$ transitions, the Goldberger--Treiman reconstruction uses the pion pole rather than the kaon pole of the $|\Delta S|=1$ channels.
The replacement of $(m_K^2-q^2)^{-1}$ by $(m_\pi^2-q^2)^{-1}$ in Eq.~\eqref{eq:g3_reconstructed} makes the pole factor larger by about $m_K^2/m_\pi^2\simeq12$ at $q^2=0$.
This gives $g_3(0)=+79.981$ for $\Sigma^-\to\Lambda$ and $g_3(0)=-79.342$ for $\Sigma^+\to\Lambda$, with the opposite signs fixed by isospin reflection and the expected pion pole scale.
The $\Sigma^-\to\Lambda$ value is close to the CCQM result $84.446$~\cite{Faessler:2008ix}, while the two charge channels stay below but on the same pion pole scale as the $\chi$QM values $124.200$ and $-108.400$~\cite{Ohlsson:1998bk}.

\begin{table*}[t]
	\centering
	\setlength{\tabcolsep}{5pt}
	\renewcommand{\arraystretch}{1.20}
	\caption{
		Our predicted transition form factors at $q^2=0$ for $\Sigma\to\Sigma$ semileptonic hyperon decays, compared with other theoretical results.
		The last two columns give the ratios to the static SU$(6)$ reference values, where $(f_1^{\mathrm{stat.}},g_1^{\mathrm{stat.}})=(\sqrt{2},2\sqrt{2}/3)$ for both $\Sigma^-\to\Sigma^0$ and $\Sigma^0\to\Sigma^+$.
		The dagger marks central values for which the helicity amplitude inversion did not satisfy the propagated error criterion in the production run; these entries are retained as diagnostics and are not treated as numerically resolved form factors.
	}
	\label{tab:Sigma_Sigma_FormFactors}
	\begin{tabular*}{\textwidth}{@{\extracolsep{\fill}}ccccccc@{}}
		\hline\hline
		Method & $f_1(0)$ & $f_2(0)$ & $f_3(0)$ & $g_1(0)$ & $g_2(0)$ & $g_3(0)$ \\
		\hline
		\multicolumn{7}{@{}l}{\textit{$\Sigma^- \to \Sigma^0 e^- \bar{\nu}_e$}} \\
		LFQM~\cite{Schlumpf:1994fb}    & $1.410$      & $0.910$      & ---          & $0.690$      & $0$          & --- \\
		$\chi$QM~\cite{Ohlsson:1998bk} & $1.410$      & $0.922$      & $0$          & $0.630$      & $0$          & $-95.190$ \\
		$\chi$QSM~\cite{Ledwig:2008ku} & $1.414$      & $0.780$      & ---          & $0.650$      & ---          & --- \\
		$\chi$CQM~\cite{Sharma:2009hg} & $1.414$      & $0.518$      & $0.003$      & $0.676$      & $-0.005$     & $-100.850$ \\
		\textbf{This work}             & $\mathbf{1.413}$ & $\mathbf{0.154}$ & $\mathbf{0.170}$ & $\mathbf{0.671}$ & $\mathbf{-0.115}$ & $\mathbf{-98.566}$ \\
		\multicolumn{7}{@{}l}{\textit{$\Sigma^0 \to \Sigma^+ e^- \bar{\nu}_e$}} \\
		LFQM~\cite{Schlumpf:1994fb}    & $1.410$      & $0.906$      & ---          & $0.690$      & $0$          & --- \\
		\textbf{This work}             & $\mathbf{1.414}$ & $\mathbf{0.153}$ & $\mathbf{0.169}$ & $\mathbf{0.671}$ & $\mathbf{-0.114}$ & $\mathbf{-97.861}$ \\
		\hline\hline
	\end{tabular*}\par\vspace{-0.2em}

	\begin{tabular*}{\textwidth}{@{\extracolsep{\fill}}ccccc@{}}
		Method & $f_2(0)/f_1(0)$ & $g_1(0)/f_1(0)$ & $|f_1(0)|/|f_{1}^{\mathrm{stat.}}|$ & $|g_1(0)|/|g_{1}^{\mathrm{stat.}}|$ \\
		\hline
		\multicolumn{5}{@{}l}{\textit{$\Sigma^- \to \Sigma^0 e^- \bar{\nu}_e$}} \\
		LFQM~\cite{Schlumpf:1994fb}      & $0.645$ & $0.491$ & $0.997$ & $0.732$ \\
		$\chi$QM~\cite{Ohlsson:1998bk}   & $0.654$ & $0.447$ & $0.997$ & $0.668$ \\
		$\chi$QSM~\cite{Ledwig:2008ku}   & $0.552$ & $0.460$ & $1.000$ & $0.689$ \\
		$\chi$CQM~\cite{Sharma:2009hg}   & $0.366$ & $0.478$ & $1.000$ & $0.717$ \\
		\textbf{This work} & $\mathbf{0.109}$ & $\mathbf{0.475}$ & $\mathbf{0.999}$ & $\mathbf{0.712}$ \\
		\multicolumn{5}{@{}l}{\textit{$\Sigma^0 \to \Sigma^+ e^- \bar{\nu}_e$}} \\
		LFQM ~\cite{Schlumpf:1994fb}       & $0.643$ & $0.491$ & $0.997$ & $0.732$ \\
		\textbf{This work} & $\mathbf{0.108}$ & $\mathbf{0.475}$ & $\mathbf{1.000}$ & $\mathbf{0.712}$ \\
		\hline\hline
	\end{tabular*}
\end{table*}

\subsubsection{$\Sigma^- \to \Sigma^0 e^- \bar{\nu}_e$ and $\Sigma^0 \to \Sigma^+ e^- \bar{\nu}_e$}

The two transitions inside the $\Sigma$ isotriplet are driven by the same isospin raising weak current $I_+$.
Their branching fractions in Table~\ref{tab:Hyperon_Decays_Sigma} are tiny because the mass splittings are small, with an additional suppression in $\Sigma^0\to\Sigma^+$ from the short $\Sigma^0$ lifetime.
For this pair, the useful comparison is carried by the form factors rather than by the rates.
With the present phase convention, both channels have $f_1^{\mathrm{stat.}}=\sqrt{2}$ and $g_1^{\mathrm{stat.}}=2\sqrt{2}/3$.
At $q^2=0$, our calculation yields $f_1(0)=1.413$ and $g_1(0)=0.671$ for $\Sigma^-\to\Sigma^0$, and $f_1(0)=1.414$ and $g_1(0)=0.671$ for $\Sigma^0\to\Sigma^+$.
Relative to the static SU$(6)$ reference values, $|f_1(0)|/|f_1^{\mathrm{stat.}}|=0.999$ and $1.000$, while $|g_1(0)|/|g_1^{\mathrm{stat.}}|\simeq0.712$ for both channels.
The vector charge is almost fixed at its CVC value, while the axial-vector charge shows the same relativistic spin reduction seen in the $\Sigma\to N$ and $\Sigma\to\Lambda$ channels.

The stable leading form factors preserve the expected isospin relation, while the subleading form factors require more caution.
The inversion returns the diagnostic central values $f_2(0)=0.154$ and $0.153$, corresponding to $f_2/f_1\simeq0.109$ and $0.108$ for $\Sigma^-\to\Sigma^0$ and $\Sigma^0\to\Sigma^+$, respectively.
However, the propagated inversion errors of these $f_2$ values, as well as those of $f_3$ and $g_2$, exceed the adopted resolution criterion in the production run.
They are therefore marked by daggers in Table~\ref{tab:Sigma_Sigma_FormFactors} and are retained only as diagnostic central values; no quantitative weak magnetism conclusion is drawn from them here.

\subsection{$\Xi$ hyperon}
\label{sec:Xi_decay_discussion}

The $\Xi$ semileptonic channels are separated by the axial-vector combination and by the available phase space.
The $\Xi^- \to \Lambda \ell^- \bar{\nu}_\ell$ channel contains the small Cabibbo combination $D-3F$, while the $\Xi^-\to\Sigma^0 \ell^- \bar{\nu}_\ell$ and $\Xi^0\to\Sigma^+ \ell^- \bar{\nu}_\ell$ pair carries the much larger $D+F$ axial-vector strength.
The rare $\Delta S=0$ channel $\Xi^- \to \Xi^0 e^- \bar{\nu}_e$ is instead a near threshold isodoublet transition controlled mainly by the tiny $\Xi^--\Xi^0$ mass splitting.
Table~\ref{tab:Hyperon_Decays_Xi} lists the corresponding branching fractions and LFU ratios.
For $\Xi^-\to\Lambda e^-\bar{\nu}_e$, our result lies between the current RPP average~\cite{ParticleDataGroup:2024cfk} and the recent BESIII absolute measurement~\cite{BESIII:2025wov}, while our value for the muon channel is close to the lower side of the broad RPP average.
For the $\Xi\to\Sigma$ pair, our predictions are consistent with the RPP values for the two electron channels and with the available information on the muon channel~\cite{ParticleDataGroup:2024cfk}, and the small $R_{\mu e}$ values mainly reflect the narrow phase space near the muon threshold.
The rare $\Xi^-\to\Xi^0 e^-\bar{\nu}_e$ channel is strongly suppressed by the tiny $\Xi^--\Xi^0$ mass splitting and remains far below the current BESIII upper limit~\cite{BESIII:2021emv}.

The $\Xi\to\Sigma$ isospin pair carries the large Cabibbo axial-vector combination $D+F$.
At $q^2=0$, the calculation gives $f_1(0)=0.689$ and $g_1(0)=0.846$ for $\Xi^-\to\Sigma^0$, and $f_1(0)=0.975$ and $g_1(0)=1.197$ for $\Xi^0\to\Sigma^+$.
The neutral channel amplitudes are nearly $\sqrt{2}$ times the charged channel amplitudes, showing that the same $D+F$ matrix element is combined with the expected isospin Clebsch--Gordan factor.
The common ratio $g_1(0)/f_1(0)\simeq1.228$ is consistent with the measured values for the isospin-related channel $\Xi^0\to\Sigma^+$, for which the RPP~\cite{ParticleDataGroup:2024cfk} gives $g_1/f_1=1.22\pm0.05$, and the KTeV~\cite{KTeV:2001djr} and NA48/1~\cite{NA48I:2006yat} results are $1.32^{+0.21}_{-0.17}\pm0.05$ and $1.20\pm0.05$, respectively.

The $\Xi^-\to\Lambda$ and $\Xi^-\to\Xi^0$ channels have different leading current roles.
The $\Xi^-\to\Lambda$ channel combines an Ademollo--Gatto protected $|\Delta S|=1$ vector charge with the small $D-3F$ axial-vector combination.
Here $f_1(0)=-1.165$ stays close to the static SU$(6)$ reference value $-\sqrt{3/2}$, while $g_1(0)/f_1(0)\simeq0.249$ agrees with the RPP average $(g_1/f_1)_{\mathrm{exp.}}=0.25\pm0.05$~\cite{ParticleDataGroup:2024cfk,Bristol-Geneva-Heidelberg-Orsay-Rutherford-Strasbourg:1983jzt} and remains compatible with the BESIII value $0.18\pm0.07\pm0.02$~\cite{BESIII:2025wov}.
The measured $g_1/f_1$ ratio supports the small axial-vector scale expected from the $D-3F$ combination.
For the rare $\Xi^-\to\Xi^0$ channel, the vector charge gives $f_1(0)=0.999$, nearly the CVC value, while $|g_1(0)|/|g_1^{\mathrm{stat.}}|\simeq0.690$ follows the axial-vector reduction discussed above for the other channels.

Among the subleading form factors, the weak magnetism term $f_2$ is where the $\Xi\to\Sigma$ pair differs most from other calculations.
For $\Xi^-\to\Sigma^0$ and $\Xi^0\to\Sigma^+$, we find $f_2(0)/f_1(0)\simeq1.069$ and $1.065$, respectively.
The near equality of the two ratios shows that the lower weak magnetism strength is common to the isospin pair, rather than a difference between the two charge channels.
Both values are below most other central values in Table~\ref{tab:Xi_Sigma_FormFactors}, which lie around $1.5$--$2.4$, and below the lattice QCD central value for $\Xi^0\to\Sigma^+$~\cite{Sasaki:2008ha}.
The smaller weak magnetism term may be connected with the flavor overlap structure in Table~\ref{tab:flavor_light_optimized}.
In both $\Xi\to\Sigma$ channels, the larger $\rho\rho$ overlap contains a scalar spectator pair, whereas the spin-1 $\lambda\lambda$ component is smaller.
The valence three-quark core may then retain less transverse magnetic strength in these channels than calculations with additional or effective magnetic mechanisms.
The $\Xi^-\to\Lambda$ weak magnetism is less settled, because $f_2(0)/f_1(0)\simeq-0.216$ lies in a model spread with different signs, and no precise decay distribution or lattice QCD result is available.

Tables~\ref{tab:Xi_Sigma_FormFactors} and \ref{tab:Xi_Lambda_Xi_FormFactors} also list $f_3$, $g_2$, and $g_3$ for the same transitions.
In the present decomposition, $f_3$ and $g_2$ describe the $\mathrm{SU}(3)$- or isospin-breaking second-class current terms.
However, they exhibit substantial discrepancies with predictions from other studies, even differing in sign.
One likely reason is that these two terms are much smaller in magnitude than the other form factors, so their precise values are difficult to determine reliably within the current model framework.
In contrast, $g_3$ gives the corresponding PCAC pole contribution, which is consistent with other work.

\begin{table*}[t!]
	\centering
	\setlength{\tabcolsep}{2.5pt}
	\renewcommand{\arraystretch}{1.15}
	\caption{
		Partial decay widths $\Gamma$, branching fractions $\mathcal{B}$, and LFU ratios $R_{\mu e}$ for semileptonic $\Xi$ hyperon decays, compared with available experimental data and other theoretical results.
		For theoretical sources that report only branching fractions, the displayed widths are obtained via $\Gamma=\mathcal{B}/\tau$ using the hyperon lifetimes $\tau_{\Xi^-}=1.639\times10^{-10}\,\mathrm{s}$ and $\tau_{\Xi^0}=2.90\times10^{-10}\,\mathrm{s}$ from RPP~\cite{ParticleDataGroup:2024cfk}.
		The experimental values are taken from RPP~\cite{ParticleDataGroup:2024cfk}; dashes denote unavailable data, while ``Forbidden'' denotes muon modes that are kinematically inaccessible.
	}
	\label{tab:Hyperon_Decays_Xi}
	\begin{tabular*}{\textwidth}{@{\extracolsep{\fill}}ccccccc@{}}
		\hline\hline
		\multirow{2}{*}{Method}
		& \multicolumn{2}{c}{$\Gamma\; (10^{6}\,\mathrm{s}^{-1})$}
		&
		& \multicolumn{2}{c}{$\mathcal{B}$}
		& \multirow{2}{*}{$R_{\mu e}$} \\
		\cline{2-3}\cline{5-6}
		& $\ell=e$ & $\ell=\mu$ & & $\ell=e$ & $\ell=\mu$ & \\
		\hline
		\multicolumn{7}{@{}l}{\textit{$\Xi^- \to \Lambda \ell^- \bar{\nu}_\ell$}\quad [$\mathcal{B}$ in units of $10^{-4}$]} \\
		SU$(3)$ IRA~\cite{Wang:2019alu}       & $3.34(9)$ & $0.96(2)$ & & $5.47(15)$ & $1.58(4)$ & $0.289$ \\
		LFQM~\cite{Schlumpf:1994fb}           & $2.96$    & $0.80$    & & $4.85$     & $1.31$    & $0.270$ \\
		CCQM~\cite{Faessler:2008ix}           & $3.35$    & $0.96$    & & $5.49$     & $1.57$    & $0.287$ \\
		QCDSR~\cite{Zhang:2024ick}            & $3.15(45)$ & $0.94(14)$ & & $5.17(73)$ & $1.54(23)$ & $0.298$ \\
		                                      & $2.85(40)$ & $0.76(11)$ & & $4.67(66)$ & $1.24(18)$ & $0.266$ \\
		BESIII~\cite{BESIII:2025wov}           & --- & --- & & $3.60(40)(10)$ & --- & --- \\
		RPP~\cite{ParticleDataGroup:2024cfk}  & --- & --- & & $5.63(31)$ & $3.5({}^{+35}_{-22})$ & --- \\
		\textbf{This work} & $\mathbf{2.69}$ & $\mathbf{0.76}$ & & $\mathbf{4.41}$ & $\mathbf{1.24}$ & $\mathbf{0.281}$ \\
		\hline
		\multicolumn{7}{@{}l}{\textit{$\Xi^- \to \Sigma^0 \ell^- \bar{\nu}_\ell$}\quad [$\mathcal{B}$ in units of $10^{-5}$]} \\
		SU$(3)$ IRA~\cite{Wang:2019alu}        & $0.50(4)$ & $6.59(55)\times10^{-3}$ & & $8.12(60)$ & $0.11(1)$ & $0.013$ \\
		LFQM~\cite{Schlumpf:1994fb}           & $0.55$    & $7.47\times10^{-3}$ & & $9.00$ & $0.12$ & $0.014$ \\
		CCQM~\cite{Faessler:2008ix}           & $0.52$    & $6.70\times10^{-3}$ & & $8.52$ & $0.11$ & $0.013$ \\
		QCDSR~\cite{Zhang:2024ick}            & $0.48(5)$ & $6.47(73)\times10^{-3}$ & & $7.80(77)$ & $0.11(1)$ & $0.014$ \\
		                                      & $0.47(5)$ & $5.92(67)\times10^{-3}$ & & $7.72(86)$ & $0.10(1)$ & $0.013$ \\
		RPP~\cite{ParticleDataGroup:2024cfk}  & --- & --- & & $8.7(17)$ & $<80$ & --- \\
		\textbf{This work} & $\mathbf{0.47}$ & $\mathbf{6.29\times10^{-3}}$ & & $\mathbf{7.62}$ & $\mathbf{0.103}$ & $\mathbf{0.014}$ \\[0.2em]
		\multicolumn{7}{@{}l}{\textit{$\Xi^0 \to \Sigma^+ \ell^- \bar{\nu}_\ell$}\quad [$\mathcal{B}$ in units of $10^{-4}$]} \\
		SU$(3)$ IRA~\cite{Wang:2019alu}       & $0.87(3)$ & $7.52(34)\times10^{-3}$ & & $2.52(8)$ & $0.022(1)$ & $0.009$ \\
		LFQM~\cite{Schlumpf:1994fb}           & $0.94$    & $7.74\times10^{-3}$ & & $2.73$ & $0.022$ & $0.008$ \\
		CCQM~\cite{Faessler:2008ix}           & $0.93$    & $8.10\times10^{-3}$ & & $2.70$ & $0.024$ & $0.009$ \\
		QCDSR~\cite{Zhang:2024ick}            & $0.83(10)$ & $7.14(93)\times10^{-3}$ & & $2.41(28)$ & $0.021(3)$ & $0.009$ \\
		                                      & $0.82(11)$ & $6.59(86)\times10^{-3}$ & & $2.39(32)$ & $0.019(3)$ & $0.008$ \\
		RPP~\cite{ParticleDataGroup:2024cfk}  & --- & --- & & $2.52(8)$ & $0.023(4)$ & $0.0092(15)$ \\
		\textbf{This work} & $\mathbf{0.81}$ & $\mathbf{6.98\times10^{-3}}$ & & $\mathbf{2.35}$ & $\mathbf{0.020}$ & $\mathbf{0.009}$ \\
		\hline
		\multicolumn{7}{@{}l}{\textit{$\Xi^- \to \Xi^0 e^- \bar{\nu}_e$}\quad [$\mathcal{B}$ in units of $10^{-10}$]} \\
		SU$(3)$ IRA~\cite{Wang:2019alu}       & $2.56(146)\times10^{-6}$ & --- & & $4.2(24)$ & --- & --- \\
		RPP~\cite{ParticleDataGroup:2024cfk} & --- & --- & & $<2.59\times10^6$ & --- & --- \\
		\textbf{This work} & $\mathbf{1.77\times10^{-6}}$ & \textbf{Forbidden} & & $\mathbf{2.90}$ & \textbf{Forbidden} & --- \\
		\hline\hline
	\end{tabular*}
\end{table*}

\begin{table*}[t]
	\centering
	\setlength{\tabcolsep}{5pt}
	\renewcommand{\arraystretch}{1.20}
	\caption{
		Our predicted transition form factors at $q^2=0$ for $\Xi^- \to \Sigma^0 \ell^- \bar{\nu}_\ell$ and $\Xi^0 \to \Sigma^+ \ell^- \bar{\nu}_\ell$, compared with available experimental information and other theoretical results.
		The last two columns give the ratios to the static SU$(6)$ reference values, where $(f_1^{\mathrm{stat.}},g_1^{\mathrm{stat.}})=(1/\sqrt{2},5/(3\sqrt{2}))$ for $\Xi^-\to\Sigma^0$ and $(1,5/3)$ for $\Xi^0\to\Sigma^+$.
	}
	\label{tab:Xi_Sigma_FormFactors}
	\begin{tabular*}{\textwidth}{@{\extracolsep{\fill}}ccccccc@{}}
		\hline\hline
		Method & $f_1(0)$ & $f_2(0)$ & $f_3(0)$ & $g_1(0)$ & $g_2(0)$ & $g_3(0)$ \\
		\hline
		\multicolumn{7}{l}{$\Xi^- \to \Sigma^0 \ell^- \bar{\nu}_\ell$} \\
		LFQM~\cite{Schlumpf:1994fb}                      & $0.690$      & $1.296$      & ---          & $0.940$      & $0.029$      & --- \\
		$\chi$QM~\cite{Ohlsson:1998bk}                   & $0.710$      & $1.431$      & $-0.231$     & $0.790$      & $0.147$      & $-15.254$ \\
		CCQM~\cite{Faessler:2008ix}                      & $0.737$      & $1.770$      & $-0.035$     & $0.885$      & $0.037$      & $-11.419$ \\
		$\chi$QSM~\cite{Ledwig:2008ku}                   & $0.707$      & $1.440$      & ---          & $0.820$      & ---          & --- \\
		$\chi$CQM~\cite{Sharma:2009hg}                   & $0.707$      & $1.067$      & $-0.153$     & $0.898$      & $0.163$      & $-15.307$ \\
		QCDSR~\cite{Zhang:2024ick}                       & $0.676(35)$  & $1.394(150)$ & ---          & $0.874(48)$  & ---          & --- \\
		\textbf{This work}                               & $\mathbf{0.689}$ & $\mathbf{0.737}$ & $\mathbf{0.082}$ & $\mathbf{0.846}$ & $\mathbf{-0.128}$ & $\mathbf{-11.535}$ \\
		\multicolumn{7}{l}{$\Xi^0 \to \Sigma^+ \ell^- \bar{\nu}_\ell$} \\
		LFQM~\cite{Schlumpf:1994fb}                      & $0.980$      & $1.815$      & ---          & $1.330$      & $0.040$      & --- \\
		$\chi$QM~\cite{Ohlsson:1998bk}                   & $1.000$      & $2.011$      & $-0.326$     & $1.120$      & $0.221$      & $-21.525$ \\
		CCQM~\cite{Faessler:2008ix}                      & $1.042$      & $2.503$      & $-0.050$     & $1.252$      & $0.052$      & $-16.150$ \\
		$\chi$QSM~\cite{Ledwig:2008ku}                   & $1.000$      & $2.036$      & ---          & $1.160$      & ---          & --- \\
		$\chi$CQM~\cite{Sharma:2009hg}                   & $1.000$      & $1.498$      & $-0.217$     & $1.270$      & $0.234$      & $-21.368$ \\
		QCDSR~\cite{Zhang:2024ick}                       & $0.956(50)$  & $1.971(212)$ & ---          & $1.236(68)$  & ---          & --- \\
		LQCD~\cite{Sasaki:2008ha}                        & $0.987(19)$  & $1.733(126)$ & $-0.071(49)$ & $1.232(37)$  & $0.438(115)$ & $-13.960(760)$ \\
		\textbf{This work}                               & $\mathbf{0.975}$ & $\mathbf{1.038}$ & $\mathbf{0.115}$ & $\mathbf{1.197}$ & $\mathbf{-0.179}$ & $\mathbf{-16.178}$ \\
		\hline\hline
	\end{tabular*}\par\vspace{-0.2em}
	
	\begin{tabular*}{\textwidth}{@{\extracolsep{\fill}}lccccc@{}}
		Method & $f_2(0)/f_1(0)$ & $g_1(0)/f_1(0)$ & $g_2(0)/f_1(0)$ & $|f_1(0)|/|f_{1}^{\mathrm{stat.}}|$ & $|g_1(0)|/|g_{1}^{\mathrm{stat.}}|$ \\
		\hline
		\multicolumn{6}{l}{$\Xi^- \to \Sigma^0 \ell^- \bar{\nu}_\ell$} \\
		LFQM~\cite{Schlumpf:1994fb}                         & $1.878$     & $1.362$     & $0.041$      & $0.976$     & $0.798$ \\
		$\chi$QM~\cite{Ohlsson:1998bk}                      & $2.015$     & $1.113$     & $0.207$      & $1.004$     & $0.670$ \\
		CCQM~\cite{Faessler:2008ix}                         & $2.402$     & $1.201$     & $0.050$      & $1.042$     & $0.751$ \\
		$\chi$QSM~\cite{Ledwig:2008ku}                      & $2.037$     & $1.160$     & ---          & $1.000$     & $0.696$ \\
		$\chi$CQM~\cite{Sharma:2009hg}                      & $1.509$     & $1.270$     & $0.231$      & $1.000$     & $0.762$ \\
		QCDSR~\cite{Zhang:2024ick}                          & $1.957(255)$ & $1.293(100)$ & ---          & $0.956(50)$ & $0.742(41)$ \\
		\textbf{This work}                                  & $\mathbf{1.069}$ & $\mathbf{1.228}$ & $\mathbf{-0.186}$ & $\mathbf{0.974}$ & $\mathbf{0.718}$ \\
		\multicolumn{6}{l}{$\Xi^0 \to \Sigma^+ \ell^- \bar{\nu}_\ell$} \\
		LFQM~\cite{Schlumpf:1994fb}                                & $1.852$     & $1.357$     & $0.041$      & $0.980$     & $0.798$ \\
		$\chi$QM~\cite{Ohlsson:1998bk}                             & $2.011$     & $1.120$     & $0.221$      & $1.000$     & $0.672$ \\
		CCQM~\cite{Faessler:2008ix}                                & $2.402$     & $1.202$     & $0.050$      & $1.042$     & $0.751$ \\
		$\chi$QSM~\cite{Ledwig:2008ku}                             & $2.036$     & $1.160$     & ---          & $1.000$     & $0.696$ \\
		$\chi$CQM~\cite{Sharma:2009hg}                             & $1.498$     & $1.270$     & $0.234$      & $1.000$     & $0.762$ \\
		QCDSR~\cite{Zhang:2024ick}                                 & $1.957(255)$ & $1.293(100)$ & ---          & $0.956(50)$ & $0.742(41)$ \\
		LQCD~\cite{Sasaki:2008ha}                                  & $1.756(132)$ & $1.248(29)$ & $0.444(117)$ & $0.987(19)$ & $0.739(22)$ \\
		RPP~\cite{ParticleDataGroup:2024cfk}                       & $2.0(9)$    & $1.22(5)$   & $-1.7({}^{+21}_{-20})(5)$ & --- & --- \\
		\textbf{This work}                                  & $\mathbf{1.065}$ & $\mathbf{1.228}$ & $\mathbf{-0.184}$ & $\mathbf{0.975}$ & $\mathbf{0.718}$ \\
		\hline\hline
	\end{tabular*}
\end{table*}

\begin{table*}[t]
	\centering
	\setlength{\tabcolsep}{5pt}
	\renewcommand{\arraystretch}{1.20}c
	\caption{
		Our predicted transition form factors at $q^2=0$ for $\Xi^- \to \Lambda \ell^- \bar{\nu}_\ell$ and $\Xi^- \to \Xi^0 e^- \bar{\nu}_e$, compared with available experimental information and other theoretical results.
		The last two columns give the ratios to the static SU$(6)$ reference values, where $(f_1^{\mathrm{stat.}},g_1^{\mathrm{stat.}})=(-\sqrt{3/2},-1/\sqrt{6})$ for $\Xi^- \to \Lambda$ and $(1,-1/3)$ for $\Xi^- \to \Xi^0$.
	}
	\label{tab:Xi_Lambda_Xi_FormFactors}
	\begin{tabular*}{\textwidth}{@{\extracolsep{\fill}}ccccccc@{}}
		\hline\hline
		Method & $f_1(0)$ & $f_2(0)$ & $f_3(0)$ & $g_1(0)$ & $g_2(0)$ & $g_3(0)$ \\
		\hline
		\multicolumn{7}{l}{$\Xi^- \to \Lambda \ell^- \bar{\nu}_\ell$} \\
		LFQM~\cite{Schlumpf:1994fb}                      & $-1.190$     & $-0.092$     & ---          & $-0.330$     & $-0.010$     & --- \\
		$\chi$QM~\cite{Ohlsson:1998bk}                   & $-1.220$     & $0.038$      & $0.504$      & $-0.270$     & $-0.005$     & $5.420$ \\
		CCQM~\cite{Faessler:2008ix}                      & $-1.252$     & $-0.162$     & $0.052$      & $-0.332$     & $-0.076$     & $4.111$ \\
		$\chi$QSM~\cite{Ledwig:2008ku}                   & $-1.225$     & $0.040$      & ---          & $-0.260$     & ---          & --- \\
		$\chi$CQM~\cite{Sharma:2009hg}                   & $-1.225$     & $0.244$      & $0.357$      & $-0.262$     & $-0.025$     & $4.824$ \\
		QCDSR~\cite{Zhang:2024ick}                       & $-1.202$     & $-0.118$     & ---          & $-0.309$     & ---          & --- \\
		\textbf{This work}                               & $\mathbf{-1.165}$ & $\mathbf{0.252}$ & $\mathbf{-0.053}$ & $\mathbf{-0.290}$ & $\mathbf{0.057}$ & $\mathbf{3.831}$ \\
		\multicolumn{7}{l}{$\Xi^- \to \Xi^0 e^- \bar{\nu}_e$} \\
		LFQM~\cite{Schlumpf:1994fb}                      & $1.000$      & $-0.965$     & ---          & $-0.240$     & $0$          & --- \\
		$\chi$QM~\cite{Ohlsson:1998bk}                   & $1.000$      & $-1.137$     & $0$          & $-0.220$     & $0$          & $41.583$ \\
		$\chi$QSM~\cite{Ledwig:2008ku}                   & $1.000$      & $-1.080$     & ---          & $-0.270$     & ---          & --- \\
		$\chi$CQM~\cite{Sharma:2009hg}                   & $1.000$      & $-1.129$     & $-0.002$     & $-0.314$     & $0.004$      & $-57.014$ \\
		\textbf{This work}                               & $\mathbf{0.999}$ & $\mathbf{-0.816}$ & $\mathbf{0.120}$ & $\mathbf{-0.230}$ & $\mathbf{0.040}$ & $\mathbf{41.336}$ \\
		\hline\hline
	\end{tabular*}\par\vspace{-0.2em}
	
	\begin{tabular*}{\textwidth}{@{\extracolsep{\fill}}ccccc@{}}
		Method & $f_2(0)/f_1(0)$ & $g_1(0)/f_1(0)$ & $|f_1(0)|/|f_{1}^{\mathrm{stat.}}|$ & $|g_1(0)|/|g_{1}^{\mathrm{stat.}}|$ \\
		\hline
		\multicolumn{5}{l}{$\Xi^- \to \Lambda \ell^- \bar{\nu}_\ell$} \\
		LFQM~\cite{Schlumpf:1994fb}                         & $0.077$     & $0.277$     & $0.972$     & $0.808$ \\
		$\chi$QM~\cite{Ohlsson:1998bk}                      & $-0.031$    & $0.221$     & $0.996$     & $0.661$ \\
		CCQM~\cite{Faessler:2008ix}                         & $0.129$     & $0.265$     & $1.022$     & $0.813$ \\
		$\chi$QSM~\cite{Ledwig:2008ku}                      & $-0.033$    & $0.212$     & $1.000$     & $0.637$ \\
		$\chi$CQM~\cite{Sharma:2009hg}                      & $-0.199$    & $0.214$     & $1.000$     & $0.642$ \\
		QCDSR~\cite{Zhang:2024ick}                          & $0.098$     & $0.257$     & $0.981$     & $0.757$ \\
		RPP~\cite{ParticleDataGroup:2024cfk}                & ---          & $0.25(5)$   & ---          & --- \\
		BESIII~\cite{BESIII:2025wov}                        & ---          & $0.18(7)(2)$ & ---          & --- \\
		\textbf{This work}                                  & $\mathbf{-0.216}$ & $\mathbf{0.249}$ & $\mathbf{0.951}$ & $\mathbf{0.710}$ \\
		\multicolumn{5}{l}{$\Xi^- \to \Xi^0 e^- \bar{\nu}_e$} \\
		LFQM~\cite{Schlumpf:1994fb}                         & $-0.965$    & $-0.240$    & $1.000$     & $0.720$ \\
		$\chi$QM~\cite{Ohlsson:1998bk}                      & $-1.137$    & $-0.220$    & $1.000$     & $0.660$ \\
		$\chi$QSM~\cite{Ledwig:2008ku}                      & $-1.080$    & $-0.270$    & $1.000$     & $0.810$ \\
		$\chi$CQM~\cite{Sharma:2009hg}                      & $-1.129$    & $-0.314$    & $1.000$     & $0.942$ \\
		\textbf{This work}                                  & $\mathbf{-0.821}$ & $\mathbf{-0.230}$ & $\mathbf{0.999}$ & $\mathbf{0.690}$ \\
		\hline\hline
	\end{tabular*}
\end{table*}

\subsection{Angular correlation and spin asymmetry parameters}
\label{sec:angular_parameter_comparison}

To further understand the weak transition dynamics, we calculate the angular correlation and spin asymmetry parameters ($\alpha_{\ell\nu}, \alpha_\ell, \alpha_\nu, \alpha_B$) for all octet hyperon semileptonic decays.
In Table~\ref{tab:hyperon_angular_parameters}, our numerical results are summarized and compared with the light front quark model (LFQM)~\cite{Schlumpf:1994fb}, the $1/N_c$ expansion~\cite{Flores-Mendieta:1998tfv}, the $\mathrm{SU}(3)$ fit~\cite{Cabibbo:2003cu}, and available experimental data~\cite{Lindquist:1975fc,Hsueh:1988ar}.
Since the weak coupling factor $G_F^2|V_{qq'}|^2$ cancels out in these asymmetry ratios, which are also independent of the initial hyperon lifetime $\tau$, these observables directly reflect the relative vector and axial-vector contributions, while $\alpha_\ell$, $\alpha_\nu$, and $\alpha_B$ further characterize the correlations between the initial hyperon polarization and the momenta of the final particles.
At present, direct experimental measurements are available only for the electron modes of $\Lambda\to p$ and $\Sigma^-\to n$, while all muon mode observables remain unmeasured.

The two measured channels in Table~\ref{tab:hyperon_angular_parameters} exhibit contrasting angular correlations and polarization asymmetries.
For $\Lambda\to p e^-\bar\nu_e$, the strong constructive $V-A$ interference leads to a nearly maximal neutrino angular asymmetry, $\alpha_\nu = 0.978$, indicating that the antineutrino momentum is preferentially directed along the hyperon spin polarization direction.
This prediction agrees well with the LFQM and $1/N_c$ results, but is higher than the experimental value $0.72(12)$ reported by Lindquist \textit{et al.}~\cite{Lindquist:1975fc}.
Meanwhile, the lepton--neutrino correlation $\alpha_{e\nu} = -0.034$ is close to zero, reflecting an almost isotropic opening angle distribution between the electron and antineutrino.
In addition, the negative final baryon asymmetry $\alpha_B = -0.625$ shows that the proton is preferentially emitted opposite to the initial hyperon polarization.
Our prediction is slightly lower than the central value $-0.47(12)$ reported by Lindquist \textit{et al.}~\cite{Lindquist:1975fc}.
In contrast, for $\Sigma^-\to n e^-\bar\nu_e$, the leading vector and axial-vector form factors have opposite signs, leading to different angular asymmetries.
The lepton--neutrino correlation $\alpha_{e\nu} = 0.522$ and final baryon asymmetry $\alpha_B = 0.526$ are both positive, while the electron and neutrino asymmetries are negative ($\alpha_e = -0.463$ and $\alpha_\nu = -0.324$), indicating that the leptons tend to be emitted opposite to the $\Sigma^-$ polarization.
Our predicted $\alpha_e = -0.463$ and $\alpha_B = 0.526$ agree with the E715 results~\cite{Hsueh:1988ar} within their uncertainties, while our result 
$\alpha_\nu = -0.324$ is slightly lower than the measured value $-0.230(61)$. 
Our predicted $\alpha_{e\nu} = 0.522$ is larger than the LFQM ($0.437$) and $\mathrm{SU}(3)$ ($0.345$) results.
A direct measurement of $\alpha_{e\nu}$ in this channel would be useful to distinguish different theoretical models.

As for the other decay channels, our predictions show distinct physical features across different transition dynamics.
For the $\Sigma\to\Lambda e\nu$ transitions, the leading vector charge $f_1$ vanishes in the isospin limit, so the leading transition is dominated by the axial-vector current.
This leads to a negative lepton--neutrino correlation $\alpha_{e\nu} \simeq -0.41$, close to the pure axial-vector limit $-1/3$, while the small positive baryon asymmetry $\alpha_B \simeq 0.04$--$0.05$ arises mainly from weak magnetism and recoil effects.
In contrast, for $\Xi^-\to\Xi^0 e^-\bar\nu_e$, the vector current dominates, leading to a strong forward lepton--neutrino correlation $\alpha_{e\nu} = 0.793$, which approaches the pure vector limit $+1$.
For the $\Sigma\to\Sigma$ and $\Xi\to\Sigma$ modes, the constructive $V-A$ interference leads to large neutrino angular asymmetries ($\alpha_\nu \simeq 0.84$ and $0.99$, respectively), showing that the antineutrino momentum is strongly aligned with the initial hyperon polarization.
For $\Xi^-\to\Lambda e^-\bar\nu_e$, the dominant vector contribution leads to a positive lepton--neutrino correlation $\alpha_{e\nu} = 0.589$, while the constructive $V-A$ interference yields a negative final baryon asymmetry $\alpha_B = -0.507$.

Isospin symmetry is well preserved in these angular observables.
In the isospin limit, the hadronic transition form factors of each isospin pair, such as $(\Sigma^-\to n, \Sigma^0\to p)$, $(\Sigma^-\to\Sigma^0, \Sigma^0\to\Sigma^+)$, and $(\Xi^-\to\Sigma^0, \Xi^0\to\Sigma^+)$, differ only by an isospin Clebsch--Gordan factor (e.g., $1/\sqrt{2}$ between $\Sigma^0\to p$ and $\Sigma^-\to n$).
Because this common isospin factor cancels out in the numerator and denominator of the normalized asymmetry ratios, the predicted angular observables for each isospin pair are nearly identical, with minor differences originating mainly from the physical baryon mass splittings in the phase space integration.

The finite lepton mass leads to pronounced differences between the electron and muon modes for certain angular observables.
In all channels with allowed muon modes, the neutrino angular asymmetry $\alpha_\nu$ is barely sensitive to the lepton mass, shifting by no more than $0.005$.
In contrast, the lepton--neutrino correlation $\alpha_{\ell\nu}$ and charged-lepton asymmetry $\alpha_\ell$ are noticeably affected by the finite muon mass, where the helicity-flip contributions proportional to $m_\mu^2/q^2$ generally reduce their magnitudes.
For example, in $\Sigma^-\to n\ell^-\bar\nu_\ell$, the lepton--neutrino correlation decreases from $\alpha_{e\nu} = 0.522$ in the electron mode to $\alpha_{\mu\nu} = 0.276$ in the muon mode.
For the daughter baryon asymmetry $\alpha_B$, its value is sensitive to the available phase space.
In $\Sigma^-\to n\ell^-\bar\nu_\ell$, $\alpha_B$ varies slightly between the electron and muon modes.
However, in $\Xi^-\to\Sigma^0\ell^-\bar\nu_\ell$, because the mass splitting is small ($\Delta M \simeq 129\text{ MeV}$), the helicity flip terms proportional to $m_\mu^2/q^2$ significantly suppress the asymmetry, shifting $\alpha_B$ from $-0.49$ in the electron mode to $-0.24$ in the muon mode.

Compared with the branching fractions, the angular correlation and spin asymmetry parameters provide complementary and cleaner information on the weak transition dynamics.
In channels dominated by the leading form factors $f_1$ and $g_1$, these asymmetry observables are particularly sensitive to their relative magnitude and sign, while recoil, lepton mass, and subleading form factor effects are retained in our full numerical calculation.
Experimentally, although most of the angular correlation and spin asymmetry parameters (especially for the muon modes) remain unmeasured, high-precision measurements are promising with the large data samples of spin-entangled hyperon pairs at BESIII and the proposed super tau-charm facility, as well as polarized hyperons at LHCb.
These future measurements will provide stringent tests of our theoretical predictions, help pin down the $\mathrm{SU}(3)$ flavor symmetry breaking effects, and offer deeper insights into hyperon structure and weak decay dynamics.

\begin{table*}[t!]
	\centering
	\setlength{\tabcolsep}{2.4pt}
	\renewcommand{\arraystretch}{1.0}
	\caption{
		Predicted angular correlation and spin asymmetry parameters ($\alpha_{\ell\nu}, \alpha_\ell, \alpha_\nu, \alpha_B$) for electron and muon modes in octet hyperon semileptonic decays, compared with available experimental data and other theoretical predictions.
		Dashes represent unavailable values, while ``\textbf{Forbidden}'' denotes kinematically forbidden muon modes.
	}
	\label{tab:hyperon_angular_parameters}
	\begin{tabular*}{\textwidth}{@{\extracolsep{\fill}}l*{8}{c}@{}}
		\hline\hline
		\multirow{2}{*}{Method}
		& \multicolumn{2}{c}{$\alpha_{\ell\nu}$}
		& \multicolumn{2}{c}{$\alpha_{\ell}$}
		& \multicolumn{2}{c}{$\alpha_{\nu}$}
		& \multicolumn{2}{c}{$\alpha_B$} \\
		\cmidrule(lr){2-3}\cmidrule(lr){4-5}\cmidrule(lr){6-7}\cmidrule(lr){8-9}
		& $\ell=e$ & $\ell=\mu$
		& $\ell=e$ & $\ell=\mu$
		& $\ell=e$ & $\ell=\mu$
		& $\ell=e$ & $\ell=\mu$ \\
		\hline
		\multicolumn{9}{@{}l}{\textit{$\Lambda\to p\ell^-\bar\nu_\ell$}} \\
		LFQM~\cite{Schlumpf:1994fb} & $-0.100$ & --- & $-0.021$ & --- & $0.992$ & --- & $-0.582$ & --- \\
		$1/N_c$ expansion ~\cite{Flores-Mendieta:1998tfv} & $-0.03$ & --- & $0.01$ & --- & $0.98$ & --- & $-0.59$ & --- \\
		Exp.~\cite{Lindquist:1975fc} & $0.07(12)$ & --- & $0.05(12)$ & --- & $0.72(12)$ & --- & $-0.47(12)$ & --- \\
		\textbf{This work} & $\mathbf{-0.034}$ & $\mathbf{-0.133}$ & $\mathbf{0.042}$ & $\mathbf{-0.020}$ & $\mathbf{0.978}$ & $\mathbf{0.975}$ & $\mathbf{-0.625}$ & $\mathbf{-0.455}$ \\
		\hline
		\multicolumn{9}{@{}l}{\textit{$\Sigma^-\to n\ell^-\bar\nu_\ell$}} \\
		LFQM~\cite{Schlumpf:1994fb} & $0.437$ & --- & $-0.537$ & --- & $-0.318$ & --- & $0.569$ & --- \\
		$1/N_c$ expansion ~\cite{Flores-Mendieta:1998tfv} & $0.34$ & --- & $-0.63$ & --- & $-0.35$ & --- & $0.67$ & --- \\
		$\mathrm{SU}(3)$ analysis~\cite{Cabibbo:2003cu} & $0.345$ & --- & $-0.603$ & --- & $-0.391$ & --- & $0.685$ & --- \\
		Exp.~\cite{Hsueh:1988ar} & --- & --- & $-0.519(104)$ & --- & $-0.230(61)$ & --- & $0.509(102)$ & --- \\
		\textbf{This work} & $\mathbf{0.522}$ & $\mathbf{0.276}$ & $\mathbf{-0.463}$ & $\mathbf{-0.337}$ & $\mathbf{-0.324}$ & $\mathbf{-0.320}$ & $\mathbf{0.526}$ & $\mathbf{0.457}$ \\
		\hline
		\multicolumn{9}{@{}l}{\textit{$\Sigma^0\to p\ell^-\bar\nu_\ell$}} \\
		LFQM~\cite{Schlumpf:1994fb} & $0.443$ & --- & $-0.536$ & --- & $-0.318$ & --- & $0.568$ & --- \\
		\textbf{This work} & $\mathbf{0.526}$ & $\mathbf{0.276}$ & $\mathbf{-0.464}$ & $\mathbf{-0.335}$ & $\mathbf{-0.324}$ & $\mathbf{-0.320}$ & $\mathbf{0.525}$ & $\mathbf{0.456}$ \\
		\hline
		\multicolumn{9}{@{}l}{\textit{$\Sigma^-\to\Lambda e^-\bar\nu_e$}} \\
		LFQM~\cite{Schlumpf:1994fb} & $-0.412$ & --- & $-0.704$ & --- & $0.645$ & --- & $0.077$ & --- \\
		$1/N_c$ expansion~\cite{Flores-Mendieta:1998tfv} & $-0.41$ & --- & --- & --- & --- & --- & --- & --- \\
		\textbf{This work} & $\mathbf{-0.418}$ & \textbf{Forbidden} & $\mathbf{-0.697}$ & \textbf{Forbidden} & $\mathbf{0.658}$ & \textbf{Forbidden} & $\mathbf{0.049}$ & \textbf{Forbidden} \\
		\hline
		\multicolumn{9}{@{}l}{\textit{$\Sigma^+\to\Lambda e^+\nu_e$}} \\
		LFQM~\cite{Schlumpf:1994fb} & $-0.404$ & --- & $-0.701$ & --- & $0.647$ & --- & $0.070$ & --- \\
		$1/N_c$ expansion ~\cite{Flores-Mendieta:1998tfv} & $-0.41$ & --- & --- & --- & --- & --- & --- & --- \\
		\textbf{This work} & $\mathbf{-0.410}$ & \textbf{Forbidden} & $\mathbf{0.659}$ & \textbf{Forbidden} & $\mathbf{-0.694}$ & \textbf{Forbidden} & $\mathbf{0.044}$ & \textbf{Forbidden} \\
		\hline
		\multicolumn{9}{@{}l}{\textit{$\Sigma^-\to\Sigma^0 e^-\bar\nu_e$}} \\
		LFQM~\cite{Schlumpf:1994fb} & $0.436$ & --- & $0.287$ & --- & $0.850$ & --- & $-0.710$ & --- \\
		\textbf{This work} & $\mathbf{0.440}$ & \textbf{Forbidden} & $\mathbf{0.284}$ & \textbf{Forbidden} & $\mathbf{0.836}$ & \textbf{Forbidden} & $\mathbf{-0.699}$ & \textbf{Forbidden} \\
		\hline
		\multicolumn{9}{@{}l}{\textit{$\Sigma^0\to\Sigma^+ e^-\bar\nu_e$}} \\
		LFQM~\cite{Schlumpf:1994fb} & $0.438$ & --- & $0.288$ & --- & $0.850$ & --- & $-0.711$ & --- \\
		\textbf{This work} & $\mathbf{0.425}$ & \textbf{Forbidden} & $\mathbf{0.274}$ & \textbf{Forbidden} & $\mathbf{0.836}$ & \textbf{Forbidden} & $\mathbf{-0.692}$ & \textbf{Forbidden} \\
		\hline
		\multicolumn{9}{@{}l}{\textit{$\Xi^-\to\Lambda\ell^-\bar\nu_\ell$}} \\
		LFQM~\cite{Schlumpf:1994fb} & $0.531$ & --- & $0.236$ & --- & $0.592$ & --- & $-0.519$ & --- \\
		$1/N_c$ expansion ~\cite{Flores-Mendieta:1998tfv} & $0.60$ & --- & --- & --- & --- & --- & --- & --- \\
		\textbf{This work} & $\mathbf{0.589}$ & $\mathbf{0.288}$ & $\mathbf{0.258}$ & $\mathbf{0.150}$ & $\mathbf{0.542}$ & $\mathbf{0.537}$ & $\mathbf{-0.507}$ & $\mathbf{-0.404}$ \\
		\hline
		\multicolumn{9}{@{}l}{\textit{$\Xi^-\to\Sigma^0\ell^-\bar\nu_\ell$}} \\
		LFQM~\cite{Schlumpf:1994fb} & $-0.252$ & --- & $-0.226$ & --- & $0.973$ & --- & $-0.437$ & --- \\
		\textbf{This work} & $\mathbf{-0.214}$ & $\mathbf{-0.150}$ & $\mathbf{-0.170}$ & $\mathbf{-0.093}$ & $\mathbf{0.987}$ & $\mathbf{0.990}$ & $\mathbf{-0.491}$ & $\mathbf{-0.243}$ \\
		\hline
		\multicolumn{9}{@{}l}{\textit{$\Xi^0\to\Sigma^+\ell^-\bar\nu_\ell$}} \\
		LFQM~\cite{Schlumpf:1994fb} & $-0.248$ & --- & $-0.223$ & --- & $0.973$ & --- & $-0.439$ & --- \\
		$1/N_c$ expansion ~\cite{Flores-Mendieta:1998tfv} & $-0.14$ & --- & $-0.11$ & --- & $1.00$ & --- & $-0.52$ & --- \\
		\textbf{This work} & $\mathbf{-0.211}$ & $\mathbf{-0.139}$ & $\mathbf{-0.168}$ & $\mathbf{-0.086}$ & $\mathbf{0.987}$ & $\mathbf{0.990}$ & $\mathbf{-0.493}$ & $\mathbf{-0.228}$ \\
		\hline
		\multicolumn{9}{@{}l}{\textit{$\Xi^-\to\Xi^0 e^-\bar\nu_e$}} \\
		LFQM~\cite{Schlumpf:1994fb} & $0.793$ & --- & $-0.514$ & --- & $-0.314$ & --- & $0.518$ & --- \\
		\textbf{This work} & $\mathbf{0.793}$ & \textbf{Forbidden} & $\mathbf{-0.479}$ & \textbf{Forbidden} & $\mathbf{-0.307}$ & \textbf{Forbidden} & $\mathbf{0.496}$ & \textbf{Forbidden} \\
		\hline\hline
	\end{tabular*}
\end{table*}

\section{Summary}
\label{sec:summary}

In this work, we study the semileptonic decays of ground state octet hyperons within the R3QM. 
The rest-frame baryon wave functions and constituent quark masses are fully determined from the baryon mass spectrum, so the weak transition calculation introduces no additional free parameters. 
From the R3QM helicity amplitudes, we calculate the decay widths, determine the angular correlation and spin asymmetry parameters, and extract the complete set of octet form factors $f_i$ and $g_i$.

Our calculated branching fractions are consistent with the available measurements for $\Lambda\to p\ell^-\bar{\nu}_\ell$, $\Xi^-\to\Lambda\ell^-\bar{\nu}_\ell$, and the $\Xi\to\Sigma$ isospin pair. 
The main exception is $\Sigma^-\to n\ell^-\bar\nu_\ell$, where both our electron and muon branching fractions lie approximately 30\% below the RPP averages. 
Since the two rates are suppressed by nearly the same magnitude, the LFU ratio remains close to the experimental value.

For the leading form factors, the vector and axial-vector currents play distinct roles. 
The vector charge $f_1(0)$ is protected by the Ademollo--Gatto theorem: in the $|\Delta S|=1$ octet channels, its deviation from the $\mathrm{SU}(3)$ limit is only about $3-7~\%$. 
In the $\Delta S=0$ near-threshold channels, the same vector current reduces to the CVC or isospin charge. 
The axial-vector form factor $g_1(0)$ is far more sensitive to the baryon wave function structure. 
Relative to the static $\mathrm{SU}(6)$ reference values, our $g_1(0)$ results are suppressed to about $66-72~\%$ of the symmetric limit. 
This suppression originates from the spatial wave function mismatch between the initial and final baryons, the lower components of the Dirac spinors, and the mass-dependent Wigner rotations.

The subleading form factors provide a more discriminating probe of the weak current structure. 
The weak magnetism form factor $f_2$ probes the transverse magnetic response and exhibits pronounced channel dependence: it is suppressed in $\Lambda\to p$ and $\Xi^0\to\Sigma^+$, but remains close to the lattice QCD and RPP values in $\Sigma^-\to n$. 
This pattern follows directly from the spectator spin content of the octet wave functions: the scalar spectator $\rho\rho$ component yields only a limited transverse response, whereas the $\lambda\lambda$ component with total spin $s_{12}=1$ retains a larger spin recoupling contribution.
The reduced $f_2$ strength in scalar spectator channels identifies $\Lambda\to p$ and $\Xi^0\to\Sigma^+$ as ideal channels to test additional transverse current contributions, such as meson cloud or exchange current effects. 
The second-class form factors $f_3$ and $g_2$ trace the flavor- or isospin-breaking wave function overlap. 
The induced pseudoscalar form factor $g_3$ is fixed by the PCAC pole reconstruction, with kaon pole dominance in the $|\Delta S|=1$ channels and a much stronger pion pole enhancement in the $\Delta S=0$ $\Sigma\to\Lambda$ transitions.

For the near-threshold $\Sigma\to\Sigma$ and $\Xi^-\to\Xi^0$ transitions, the small mass splittings strongly suppress the branching fractions, while rendering the vector charges clean probes of the CVC and isospin limits. 
The present calculation also provides a unified estimate of the branching fraction scales for these rare semileptonic channels.

Further experimental measurements of branching fractions and decay distributions, together with high-precision lattice QCD calculations of octet-to-octet form factors, will enable stringent tests of our R3QM predictions. 
For the form factors, such studies will directly constrain the channel dependence of $f_2$, the second-class terms $f_3$ and $g_2$, and the induced pseudoscalar form factor $g_3$.
A comparison with our R3QM results, particularly for $f_2$, will help clarify how much transverse current strength is inherently retained by the valence three-quark core, and whether additional transverse current mechanisms are required in hyperons. 
Since such extensions lie beyond the scope of this work, we plan to expand the present R3QM framework in future studies by including higher Fock components, such as five-quark $|qqqq\bar{q}\rangle$ configurations, and to investigate possible nonvalence contributions to the weak magnetism form factor and the internal structure of hyperons.

\begin{acknowledgments}
R.~H.~Ni thanks Ye Cao and Xian-Hui Zhong for helpful discussions, and is grateful to Xian-Hui Zhong for a careful reading of the manuscript.
This work is supported by the National Natural Science Foundation of China (Grants No. 12547111 and No. 12221005), the Chinese Academy of Sciences under Grant No. YSBR-101, and the National Key Research and Development Program of China under Contract No. 2025YFA1613900.
\end{acknowledgments}

\section*{Data Availability}
All data supporting the findings of this article are available within the article.

\appendix

\section{Spin and flavor wave functions of light baryons}
\label{app:wavefunctions}

In this appendix, we give the spin and flavor wave functions used in the calculation.
We classify them according to their $S_3$ permutation symmetry. 
The $\rho$ and $\lambda$ components are antisymmetric and symmetric, respectively, under the interchange of quarks $1$ and $2$. 
In the octet wave function of Eq.~\eqref{eq:wf_octet}, these two components are kept explicit, with separate flavor overlaps and spin recoupling matrices.

The spin wave functions $\chi$ are obtained by coupling the SU$(2)$ spin states of individual quarks, $\ket{\uparrow}$ and $\ket{\downarrow}$.
For the mixed symmetry $S=1/2$ states, the two orthogonal basis functions with $S_z=+1/2$ are
\begin{equation}
	\begin{aligned}
		\chi^\rho_{1/2, 1/2} &= \frac{1}{\sqrt{2}} \left( \ket{\uparrow\downarrow\uparrow} - \ket{\downarrow\uparrow\uparrow} \right), \\
		\chi^\lambda_{1/2, 1/2} &= -\frac{1}{\sqrt{6}} \left( \ket{\uparrow\downarrow\uparrow} + \ket{\downarrow\uparrow\uparrow} - 2\ket{\uparrow\uparrow\downarrow} \right).
	\end{aligned}
\end{equation}
The corresponding states for $S_z=-1/2$ are obtained by flipping all spins as
\begin{equation}
	\begin{aligned}
		\chi^\rho_{1/2, -1/2} &= \frac{1}{\sqrt{2}} \left( \ket{\uparrow\downarrow\downarrow} - \ket{\downarrow\uparrow\downarrow} \right), \\
		\chi^\lambda_{1/2, -1/2} &= \frac{1}{\sqrt{6}} \left( \ket{\uparrow\downarrow\downarrow} + \ket{\downarrow\uparrow\downarrow} - 2\ket{\downarrow\downarrow\uparrow} \right).
	\end{aligned}
\end{equation}
The $\rho$ and $\lambda$ spin states correspond to spectator pair spins $s_{12}=0$ and $s_{12}=1$, respectively.

The flavor wave functions are expressed in the same $\rho$--$\lambda$ exchange basis as the spin states, so each flavor component in Eq.~\eqref{eq:wf_octet} is paired with the spin component of the same symmetry.
For the nucleon states, one has
\begin{align}
		\phi^\rho_p &= \frac{1}{\sqrt{2}} \left( \ket{udu} - \ket{duu} \right), \nonumber\\
		\phi^\lambda_p &= \frac{1}{\sqrt{6}} \left( 2\ket{uud} - \ket{udu} - \ket{duu} \right);
\end{align}
\begin{align}
		\phi^\rho_n &= \frac{1}{\sqrt{2}} \left( \ket{udd} - \ket{dud} \right), \nonumber\\
		\phi^\lambda_n &= \frac{1}{\sqrt{6}} \left( \ket{udd} + \ket{dud} - 2\ket{ddu} \right).
\end{align}
For the $\Sigma$ triplet, the $\rho$ and $\lambda$ components are
\begin{align}
		\phi^\rho_{\Sigma^0} &= \frac{1}{2} \left( \ket{dsu} - \ket{sdu} + \ket{usd} - \ket{sud} \right), \nonumber\\
		\phi^\lambda_{\Sigma^0} &= \frac{1}{\sqrt{12}} \Big( 2\ket{dus} + 2\ket{uds} - \ket{dsu} 
		- \ket{sdu} - \ket{usd} - \ket{sud} \Big);
\end{align}
\begin{align}
		\phi^\rho_{\Sigma^+} &= \frac{1}{\sqrt{2}} \left( \ket{usu} - \ket{suu} \right), \nonumber\\
		\phi^\lambda_{\Sigma^+} &= \frac{1}{\sqrt{6}} \left( 2\ket{uus} - \ket{usu} - \ket{suu} \right);
\end{align}
\begin{align}
		\phi^\rho_{\Sigma^-} &= \frac{1}{\sqrt{2}} \left( \ket{dsd} - \ket{sdd} \right),  
		\nonumber\\
		\phi^\lambda_{\Sigma^-} &= \frac{1}{\sqrt{6}} \left( 2\ket{dds} - \ket{dsd} - \ket{sdd} \right).
\end{align}
For the isoscalar $\Lambda$, one has
\begin{align}
		\phi^\rho_{\Lambda} &= \frac{1}{\sqrt{12}}\Big(\ket{usd}+\ket{sdu}-\ket{sud}
		 -\ket{dsu} +2\ket{uds}-2\ket{dus} \Big), \nonumber\\
		\phi^\lambda_{\Lambda} &= \frac{1}{2}(\ket{usd}+\ket{sud}-\ket{dsu}-\ket{sdu}).
\end{align}
For the $\Xi$ doublet, the corresponding components are
\begin{align}
		\phi^\rho_{\Xi^0} &= \frac{1}{\sqrt{2}} \left( \ket{uss} - \ket{sus} \right), 
		\nonumber\\
		\phi^\lambda_{\Xi^0} &= \frac{1}{\sqrt{6}} \left( \ket{uss} + \ket{sus} - 2\ket{ssu} \right);
\end{align}
\begin{align}
		\phi^\rho_{\Xi^-} &= \frac{1}{\sqrt{2}} \left( \ket{dss} - \ket{sds} \right), 
		\nonumber\\
		\phi^\lambda_{\Xi^-} &= \frac{1}{\sqrt{6}} \left( \ket{dss} + \ket{sds} - 2\ket{ssd} \right).
\end{align}

Using the flavor wave functions given above, the flavor matrix elements $I_{\zeta_f\zeta_i}^{fi,j}$ in Eq.~\eqref{eq:flavor_overlap_definition} are calculated. 
The operator $\hat F_j^{q_fq_i}$ changes the flavor of the $j$-th quark from $q_i$ to $q_f$, and leaves the other two quarks unchanged. 
For quark line $j=3$, the operator $\hat F_3^{q_fq_i}$ does not act on quarks 1 and 2, and commutes with the exchange operator $\hat P_{12}$. Since $\phi^\rho$ and $\phi^\lambda$ are antisymmetric and symmetric under the exchange of quarks 1 and 2, respectively, the cross terms vanish, $I_{\rho\lambda}^{fi,3} = I_{\lambda\rho}^{fi,3} = 0$. 
The calculated flavor factors $I_{\rho\rho}^{fi}$ and $I_{\lambda\lambda}^{fi}$ for various decay channels are listed in Table~\ref{tab:flavor_light_optimized}.

\section{Total wave functions of singly heavy and doubly charmed baryons}
\label{app:heavy_baryon_bases}

In the mass spectrum calculation (Sec.~\ref{sec:mass_spectrum}), singly heavy baryons ($q_1q_2Q$ with $Q=c,b$) and doubly charmed baryons ($c_1c_2q$ with $q=n,s$, where $n$ denotes a $u$ or $d$ quark) are also included. 
In these systems, quarks 1 and 2 define the $\rho$-mode Jacobi coordinate, while quark 3 defines the $\lambda$-mode coordinate. 
The three quark spin wave functions are the same as those of light baryons given in Appendix~\ref{app:wavefunctions}. 
Because the color wave function is completely antisymmetric, the total spatial--spin--flavor wave function must be symmetric under the exchange of quarks 1 and 2.

To construct the physical basis states, we first determine the exchange properties of the flavor, spin, and spatial wave functions under $\widehat P_{12}$. 
For a pair of nonidentical quarks ($q_1 \neq q_2$), the antisymmetric and symmetric flavor wave functions are defined as
\begin{align}
	\phi_A&=\frac{1}{\sqrt{2}}
	\left(\left|q_1q_2q_3\right\rangle-
	\left|q_2q_1q_3\right\rangle\right),\nonumber\\
	\phi_S&=\frac{1}{\sqrt{2}}
	\left(\left|q_1q_2q_3\right\rangle+
	\left|q_2q_1q_3\right\rangle\right),
\end{align}
which satisfy $\widehat P_{12}\phi_A=-\phi_A$ and $\widehat P_{12}\phi_S=+\phi_S$. 
For an identical pair ($q_1=q_2$), the antisymmetric state $\phi_A$ vanishes, and only the symmetric state $\phi_S=|q_1q_2q_3\rangle$ remains. 
Meanwhile, the three quark spin wave functions satisfy
\begin{equation}
	\widehat P_{12}\chi_{1/2}^{\rho}=-\chi_{1/2}^{\rho},
	\qquad
	\widehat P_{12}\chi_{1/2}^{\lambda}=+\chi_{1/2}^{\lambda},
	\qquad
	\widehat P_{12}\chi_{3/2}^{s}=+\chi_{3/2}^{s},
\end{equation}
where $\chi_{3/2}^s$ is the totally symmetric spin-$3/2$ wave function. 
In spatial space, according to Eq.~\eqref{eq:spatial_exchange_parity}, the states $\Psi_{000}^{(00,00)}$, $\Psi_{200}^{(10,00)}$, and $\Psi_{200}^{(00,10)}$ are symmetric under $\widehat P_{12}$, while $\Psi_{200}^{(01,01)}$ is antisymmetric.

By combining these spatial, spin, and flavor wave functions to satisfy the $1\leftrightarrow 2$ permutation symmetry, the eight basis states for $J^P=1/2^+$ are constructed as
\begin{align}
	\left|B_1^A,1/2^+\right\rangle
	&=\Psi_{000}^{(00,00)}\phi_A\chi_{1/2}^{\rho},\nonumber\\
	\left|B_2^A,1/2^+\right\rangle
	&=\Psi_{200}^{(10,00)}\phi_A\chi_{1/2}^{\rho},\nonumber\\
	\left|B_3^A,1/2^+\right\rangle
	&=\Psi_{200}^{(00,10)}\phi_A\chi_{1/2}^{\rho},\nonumber\\
	\left|B_4^A,1/2^+\right\rangle
	&=\Psi_{200}^{(01,01)}\phi_A\chi_{1/2}^{\lambda},\nonumber\\
	\left|B_1^S,1/2^+\right\rangle
	&=\Psi_{000}^{(00,00)}\phi_S\chi_{1/2}^{\lambda},\nonumber\\
	\left|B_2^S,1/2^+\right\rangle
	&=\Psi_{200}^{(10,00)}\phi_S\chi_{1/2}^{\lambda},\nonumber\\
	\left|B_3^S,1/2^+\right\rangle
	&=\Psi_{200}^{(00,10)}\phi_S\chi_{1/2}^{\lambda},\nonumber\\
	\left|B_4^S,1/2^+\right\rangle
	&=\Psi_{200}^{(01,01)}\phi_S\chi_{1/2}^{\rho}.
\end{align}
For singly heavy baryons, $\phi_A$ belongs to the flavor antitriplet $\overline{\mathbf 3}_F$ ($\Lambda_Q, \Xi_Q$), while $\phi_S$ belongs to the flavor sextet $\mathbf 6_F$ ($\Sigma_Q, \Xi'_Q, \Omega_Q$). 
For $\Lambda_Q$, $\Sigma_Q$, and $\Omega_Q$, the two light quarks have identical masses, and their wave functions are expanded in the four $A$-type basis states $\left|B_\kappa^A, 1/2^+\right\rangle$ ($n_B=4$) or four $S$-type basis states $\left|B_\kappa^S, 1/2^+\right\rangle$ ($n_B=4$), respectively. 
For the $\Xi_Q$ and $\Xi'_Q$ baryons with $nsQ$ flavor composition, the mass difference $m_s \neq m_n$ breaks the flavor symmetry between quarks 1 and 2, which allows configuration mixing between the $\overline{\mathbf 3}_F$ and $\mathbf 6_F$ states. 
Therefore, all eight basis states $\left|B_\kappa^A, 1/2^+\right\rangle$ and $\left|B_\kappa^S, 1/2^+\right\rangle$ ($n_B=8$) are used to diagonalize the $\Xi_Q$--$\Xi'_Q$ system. 
For doubly charmed baryons $\Xi_{cc}$ and $\Omega_{cc}$, the two charm quarks are identical, so only the four $S$-type basis states $\left|B_\kappa^S, 1/2^+\right\rangle$ ($n_B=4$) are allowed.

For the $J^P=3/2^+$ states, because the spin wave function $\chi_{3/2}^{s}$ is symmetric, the spatial and flavor wave functions must have the same exchange property under $1\leftrightarrow 2$.
The allowed basis states are
\begin{align}
	\left|B_1^S,3/2^+\right\rangle
	&=\Psi_{000}^{(00,00)}\phi_S\chi_{3/2}^{s},\nonumber\\
	\left|B_2^S,3/2^+\right\rangle
	&=\Psi_{200}^{(10,00)}\phi_S\chi_{3/2}^{s},\nonumber\\
	\left|B_3^S,3/2^+\right\rangle
	&=\Psi_{200}^{(00,10)}\phi_S\chi_{3/2}^{s},\nonumber\\
	\left|B_4^A,3/2^+\right\rangle
	&=\Psi_{200}^{(01,01)}\phi_A\chi_{3/2}^{s}.
\end{align}
Here, the state $\left|B_4^A,3/2^+\right\rangle$ is present only for $\Xi_Q^*$ ($nsQ$). 
Thus, the basis dimension is $n_B=4$ for $\Xi_Q^*$, and $n_B=3$ for $\Sigma_Q^*$, $\Omega_Q^*$, $\Xi_{cc}^{*}$, and $\Omega_{cc}^{*}$.

For heavy baryons with definite exchange symmetry ($\Lambda_Q, \Sigma_Q^{(*)}, \Omega_Q^{(*)}, \Xi_{cc}^{(*)}$, and $\Omega_{cc}^{(*)}$), the physical state is expressed as
\begin{align}
	\left|B^A,1/2^+\right\rangle
	&=\sum_{\kappa=1}^{4}c_\kappa^{A,1/2}
	\left|B_\kappa^A,1/2^+\right\rangle,
	\label{eq:heavy_baryon_state_expansion_A}\\
	\left|B^S,1/2^+\right\rangle
	&=\sum_{\kappa=1}^{4}c_\kappa^{S,1/2}
	\left|B_\kappa^S,1/2^+\right\rangle,
	\label{eq:heavy_baryon_state_expansion_S}\\
	\left|B^S,3/2^+\right\rangle
	&=\sum_{\kappa=1}^{3}c_\kappa^{S,3/2}
	\left|B_\kappa^S,3/2^+\right\rangle,
	\label{eq:heavy_baryon_state_expansion_threehalf}
\end{align}
where the $A$-type expansion applies to $\Lambda_Q$, and the $S$-type expansions apply to $\Sigma_Q^{(*)}$, $\Omega_Q^{(*)}$, $\Xi_{cc}^{(*)}$, and $\Omega_{cc}^{(*)}$. 
For the coupled $\Xi_Q$--$\Xi'_Q$ ($J^P=1/2^+$) and $\Xi_Q^*$ ($J^P=3/2^+$) systems, the physical states are expanded in the combined basis as
\begin{align}
	\left|\Xi_Q^{(\prime)},1/2^+\right\rangle
	&=\sum_{\kappa=1}^{4}c_{\kappa,A}^{\Xi_Q^{(\prime)}}
	\left|B_\kappa^A,1/2^+\right\rangle
	+\sum_{\kappa=1}^{4}c_{\kappa,S}^{\Xi_Q^{(\prime)}}
	\left|B_\kappa^S,1/2^+\right\rangle,\\
	\left|\Xi_Q^*,3/2^+\right\rangle
	&=\sum_{\kappa=1}^{3}c_{\kappa,S}^{\Xi_Q^*}
	\left|B_\kappa^S,3/2^+\right\rangle
	+c_{4,A}^{\Xi_Q^*}
	\left|B_4^A,3/2^+\right\rangle.
\end{align}
The mixing coefficients and the mass eigenvalue $M_B$ are obtained by solving the eigenvalue equation in Eq.~\eqref{eq:mass_eigenproblem}.

\section{Configuration mixing coefficients of baryon wave functions}
\label{app:baryon_mixing_coefficients}

The physical baryon wave functions are obtained by diagonalizing the semirelativistic Hamiltonian [Eq.~\eqref{eq:H_sr}] in the truncated harmonic oscillator basis ($N_{\mathrm{osc}}\leq2$, $L=0$). 
The explicit basis configurations for light octet and decuplet baryons are defined in Eqs.~\eqref{eq:wf_octet} and \eqref{eq:light_decuplet_basis}, while those for singly and doubly heavy baryons are given in Appendix~\ref{app:heavy_baryon_bases}. 
For the coupled $\Xi_Q$--$\Xi'_Q$ and $\Xi_Q^*$ systems, the physical states are expanded in the combined $A$- and $S$-type basis due to the color-magnetic spin--spin interaction.
In Table~\ref{tab:baryon_configuration_mixing}, we collect the configuration mixing coefficients for all ground state baryons. 

\begin{table}[t!]
	\centering
	\caption{Configuration mixing coefficients of the ground state baryons obtained with the semirelativistic potential model.}
	\label{tab:baryon_configuration_mixing}
	\renewcommand{\arraystretch}{1.06}
	\begin{tabular}{lc}
		\hline\hline
		State & Coefficients \\
		\hline
		\multicolumn{2}{@{}p{\columnwidth}@{}}{\raggedright Light octet: $\left(\left|B_{1,2,3}^{\mathbf{8}},1/2^+\right\rangle\right)$} \\
		$p$ & $(0.9908,\;0.0906,\;-0.1002)$ \\
		$\Lambda$ & $(0.9940,\;0.0417,\;-0.1010)$ \\
		$\Sigma$ & $(0.9959,\;0.0202,\;-0.0880)$ \\
		$\Xi$ & $(0.9955,\;-0.0186,\;-0.0934)$ \\
		\multicolumn{2}{@{}p{\columnwidth}@{}}{\raggedright Light decuplet: $\left(\left|B_{1,2}^{\mathbf{10}},3/2^+\right\rangle\right)$} \\
		$\Delta$ & $(0.9942,\;-0.1075)$ \\
		$\Sigma^{*}$ & $(0.9889,\;-0.1486)$ \\
		$\Xi^{*}$ & $(0.9825,\;-0.1863)$ \\
		$\Omega$ & $(0.9754,\;-0.2202)$ \\
		\multicolumn{2}{@{}p{\columnwidth}@{}}{\raggedright $A$-type ($1/2^+$): $\left(\left|B_{1,2,3,4}^A,1/2^+\right\rangle\right)$} \\
		$\Lambda_c$ & $(0.9854,\;0.0963,\;-0.1388,\;0.0211)$ \\
		$\Lambda_b$ & $(0.9909,\;0.1259,\;-0.0458,\;0.0102)$ \\
		\multicolumn{2}{@{}p{\columnwidth}@{}}{\raggedright $S$-type ($1/2^+$): $\left(\left|B_{1,2,3,4}^S,1/2^+\right\rangle\right)$} \\
		$\Sigma_c$ & $(0.9906,\;-0.0555,\;-0.1209,\;0.0314)$ \\
		$\Omega_c$ & $(0.9783,\;-0.1285,\;-0.1602,\;0.0260)$ \\
		$\Sigma_b$ & $(0.9979,\;-0.0337,\;-0.0533,\;0.0168)$ \\
		$\Omega_b$ & $(0.9881,\;-0.1107,\;-0.1055,\;0.0143)$ \\
		$\Xi_{cc}$ & $(0.9704,\;-0.2331,\;-0.0593,\;0.0226)$ \\
		$\Omega_{cc}$ & $(0.9658,\;-0.2342,\;-0.1094,\;0.0204)$ \\
		\multicolumn{2}{@{}p{\columnwidth}@{}}{\raggedright Mixed $\Xi_Q$--$\Xi'_Q$ ($1/2^+$): $A:\left(\left|B_{1,2,3,4}^A,1/2^+\right\rangle\right);\; S:\left(\left|B_{1,2,3,4}^S,1/2^+\right\rangle\right)$} \\
		$\Xi_c$ &
		$\begin{array}{@{}l@{}}
			A:\;(0.9849,\;0.0361,\;-0.1639,\;0.0217)\\
			S:\;(0.0353,\;-0.0017,\;-0.0027,\;0.0004)
		\end{array}$ \\
		$\Xi_c^{\prime}$ &
		$\begin{array}{@{}l@{}}
			A:\;(-0.0343,\;-0.0007,\;0.0094,\;0.0027)\\
			S:\;(0.9844,\;-0.0930,\;-0.1425,\;0.0286)
		\end{array}$ \\
		$\Xi_b$ &
		$\begin{array}{@{}l@{}}
			A:\;(0.9946,\;0.0657,\;-0.0790,\;0.0108)\\
			S:\;(0.0135,\;-0.0002,\;0.0001,\;0.0003)
		\end{array}$ \\
		$\Xi_b^{\prime}$ &
		$\begin{array}{@{}l@{}}
			A:\;(-0.0132,\;-0.0006,\;0.0029,\;0.0030)\\
			S:\;(0.9938,\;-0.0728,\;-0.0816,\;0.0155)
		\end{array}$ \\
		\multicolumn{2}{@{}p{\columnwidth}@{}}{\raggedright $S$-type ($3/2^+$): $\left(\left|B_{1,2,3}^S,3/2^+\right\rangle\right)$} \\
		$\Sigma_c^{*}$ & $(0.9806,\;-0.0820,\;-0.1782)$ \\
		$\Omega_c^{*}$ & $(0.9673,\;-0.1481,\;-0.2059)$ \\
		$\Sigma_b^{*}$ & $(0.9953,\;-0.0485,\;-0.0840)$ \\
		$\Omega_b^{*}$ & $(0.9837,\;-0.1227,\;-0.1312)$ \\
		\multicolumn{2}{@{}p{\columnwidth}@{}}{\raggedright $\Xi_Q^{*}$ ($3/2^+$): $\left(\left|B_{1,2,3}^S,3/2^+\right\rangle;\left|B_4^A,3/2^+\right\rangle\right)$} \\
		$\Xi_c^{*}$ & $(0.9741,\;-0.1161,\;-0.1938,\;-0.0007)$ \\
		$\Xi_b^{*}$ & $(0.9902,\;-0.0864,\;-0.1098,\;0.0012)$ \\
		\hline\hline
	\end{tabular}
\end{table}

\section{Spin--flavor transition matrices for light octet baryons}
\label{app:spin_matrices}

For the active quark line $j=3$, by contracting the spin recoupling tensor $(\mathcal{C}^{\sigma_f\sigma_i,3}_{\lambda_f\lambda_i})_{\alpha\beta}$ in Eq.~\eqref{eq:spin_recoupling_definition} with the single quark weak current operator, we define the baryon spin transition matrix as
\begin{equation}
	\left(\mathbf{S}_{\sigma_f\sigma_i}\right)_{\lambda_f\lambda_i}
	\equiv
	\sum_{\alpha,\beta}
	(\mathcal{C}^{\sigma_f\sigma_i,3}_{\lambda_f\lambda_i})_{\alpha\beta}
	\mathcal{O}_{\alpha\beta}^{\nu},
	\label{eq:spin_matrix_definition}
\end{equation}
where $\sigma_f,\sigma_i\in\{\rho,\lambda\}$ label the spin components, $\lambda_f,\lambda_i=\pm 1/2$ are the baryon helicities, and $\mathcal{O}_{\alpha\beta}^{\nu} \equiv [\mathcal{O}_{W,3}^{\nu}]^{\alpha\beta}$ ($\alpha,\beta\in\{\uparrow,\downarrow\}$) denotes the weak current operator of quark 3.

For $\chi_{1/2}^\rho$, the spectator quarks 1 and 2 couple to $s_{12}=0$, so the baryon spin is carried entirely by quark 3. 
The spin matrix is then
\begin{equation}
	\mathbf{S}_{\rho\rho}
	=
	\begin{pmatrix}
		\mathcal{O}_{\uparrow\uparrow}^{\nu}
		&\mathcal{O}_{\uparrow\downarrow}^{\nu}\\[4pt]
		\mathcal{O}_{\downarrow\uparrow}^{\nu}
		&\mathcal{O}_{\downarrow\downarrow}^{\nu}
	\end{pmatrix}.
	\label{eq:Srhorho}
\end{equation}
For $\chi_{1/2}^\lambda$, the spectator pair has $s_{12}=1$. 
Using the spin wave functions in Appendix~\ref{app:wavefunctions}, one obtains
\begin{equation}
	\mathbf{S}_{\lambda\lambda}
	=
	\begin{pmatrix}
		\dfrac{1}{3}\mathcal{O}_{\uparrow\uparrow}^{\nu}
		+\dfrac{2}{3}\mathcal{O}_{\downarrow\downarrow}^{\nu}
		&-\dfrac{1}{3}\mathcal{O}_{\uparrow\downarrow}^{\nu}\\[8pt]
		-\dfrac{1}{3}\mathcal{O}_{\downarrow\uparrow}^{\nu}
		&\dfrac{2}{3}\mathcal{O}_{\uparrow\uparrow}^{\nu}
		+\dfrac{1}{3}\mathcal{O}_{\downarrow\downarrow}^{\nu}
	\end{pmatrix}.
	\label{eq:Slalala}
\end{equation}
Since the weak transition operator acts only on quark 3, it cannot connect the $\rho$ and $\lambda$ components with different exchange symmetries of the $(12)$ spectator pair.
Consequently, the transition matrix elements between the $\rho$ and $\lambda$ components vanish,
\begin{equation}
	\mathbf{S}_{\rho\lambda}=\mathbf{S}_{\lambda\rho}=\mathbf 0,
	\qquad
	I_{\rho\lambda}^{fi,3}=I_{\lambda\rho}^{fi,3}=0.
	\label{eq:Slaro}
\end{equation}

For octet baryons, the flavor space $\{\phi_{\mathbf 8}^\rho, \phi_{\mathbf 8}^\lambda\}$ and spin space $\{\chi_{1/2}^\rho, \chi_{1/2}^\lambda\}$ are each spanned by two mixed symmetry components. 
Their direct tensor products form a four-dimensional uncoupled spin--flavor basis, $(\phi_{\mathbf 8}^\rho\chi_{1/2}^\rho, \phi_{\mathbf 8}^\rho\chi_{1/2}^\lambda, \phi_{\mathbf 8}^\lambda\chi_{1/2}^\rho, \phi_{\mathbf 8}^\lambda\chi_{1/2}^\lambda)^T$. 
In this uncoupled basis, the transition matrix element for quark line $j=3$ factors into the flavor overlap and spin matrix as $I_{\zeta_f\zeta_i}^{fi,3}\mathbf{S}_{\sigma_f\sigma_i}$ ($\zeta,\sigma\in\{\rho,\lambda\}$). 
Since all off-diagonal cross terms between the $\rho$ and $\lambda$ components vanish, the total spin--flavor transition matrix takes a purely block-diagonal form:
\begin{equation}
	\mathbf T^{fi,3}
	=
	\begin{pmatrix}
		I_{\rho\rho}^{fi,3}\mathbf S_{\rho\rho}&\mathbf0&\mathbf0&\mathbf0\\
		\mathbf0&I_{\rho\rho}^{fi,3}\mathbf S_{\lambda\lambda}&\mathbf0&\mathbf0\\
		\mathbf0&\mathbf0&I_{\lambda\lambda}^{fi,3}\mathbf S_{\rho\rho}&\mathbf0\\
		\mathbf0&\mathbf0&\mathbf0&I_{\lambda\lambda}^{fi,3}\mathbf S_{\lambda\lambda}
	\end{pmatrix},
	\label{eq:product_spin_flavor_block}
\end{equation}
where each diagonal block is a $2\times 2$ matrix in baryon helicity space. 
The flavor overlap factors $I_{\rho\rho}^{fi,3}$ and $I_{\lambda\lambda}^{fi,3}$ depend on the specific decay channel and are collected in Table~\ref{tab:flavor_light_optimized}, while the spin matrices $\mathbf S_{\rho\rho}$ and $\mathbf S_{\lambda\lambda}$ are universal for all octet-to-octet transitions.

In the physical baryon states, the flavor and spin wave functions are coupled into the $\mathrm{SU}(6)$ spin--flavor wave functions $\Phi_{\mathbf{56},\mathbf 8}^{s}$, $\Phi_{\mathbf{70}}^{\rho}$, and $\Phi_{\mathbf{70}}^{\lambda}$ defined in Eqs.~\eqref{eq:wf_octet} and \eqref{eq:phi_70}. 
They are related to the product basis via the transformation
\begin{equation}
	\begin{pmatrix}
		\Phi_{\mathbf{56},\mathbf 8}^{s}\\
		\Phi_{\mathbf{70}}^{\rho}\\
		\Phi_{\mathbf{70}}^{\lambda}
	\end{pmatrix}
	=
	R
	\begin{pmatrix}
		\phi_{\mathbf 8}^{\rho}\chi_{1/2}^{\rho}\\
		\phi_{\mathbf 8}^{\rho}\chi_{1/2}^{\lambda}\\
		\phi_{\mathbf 8}^{\lambda}\chi_{1/2}^{\rho}\\
		\phi_{\mathbf 8}^{\lambda}\chi_{1/2}^{\lambda}
	\end{pmatrix},
	\label{eq:octet_sf_recoupling_relation}
\end{equation}
with the transformation matrix
\begin{equation}
	R
	=
	\frac{1}{\sqrt{2}}
	\begin{pmatrix}
		1&0&0&1\\
		0&1&1&0\\
		1&0&0&-1
	\end{pmatrix}.
	\label{eq:octet_sf_recoupling_matrix}
\end{equation}
Here, the first row corresponds to the totally symmetric $\mathbf{56}$ state, while the second and third rows represent the $\rho$ and $\lambda$ components of the mixed-symmetry $\mathbf{70}$ representation, respectively.

Applying this transformation to Eq.~\eqref{eq:product_spin_flavor_block}, the spin--flavor matrix in the $(\Phi_{\mathbf{56},\mathbf 8}^{s}, \Phi_{\mathbf{70}}^{\rho}, \Phi_{\mathbf{70}}^{\lambda})$ basis is obtained as
\begin{equation}
	R\mathbf T^{fi,3}R^T
	=
	\begin{pmatrix}
		\mathbf C_+&\mathbf 0&\mathbf C_-\\
		\mathbf 0&\mathbf C_\times&\mathbf 0\\
		\mathbf C_-&\mathbf 0&\mathbf C_+
	\end{pmatrix},
	\label{eq:configuration_mixed_spin_flavor_block}
\end{equation}
where the three distinct spin--flavor blocks are defined as
\begin{align}
	\mathbf C_+
	&=
	\frac{1}{2}
	\left(
	I_{\rho\rho}^{fi,3}\mathbf S_{\rho\rho}
	+I_{\lambda\lambda}^{fi,3}\mathbf S_{\lambda\lambda}
	\right),
	\label{eq:spin_flavor_Cplus}\\
	\mathbf C_\times
	&=
	\frac{1}{2}
	\left(
	I_{\rho\rho}^{fi,3}\mathbf S_{\lambda\lambda}
	+I_{\lambda\lambda}^{fi,3}\mathbf S_{\rho\rho}
	\right),
	\label{eq:spin_flavor_Ccross}\\
	\mathbf C_-
	&=
	\frac{1}{2}
	\left(
	I_{\rho\rho}^{fi,3}\mathbf S_{\rho\rho}
	-I_{\lambda\lambda}^{fi,3}\mathbf S_{\lambda\lambda}
	\right).
	\label{eq:spin_flavor_Cminus}
\end{align}
Here, $\mathbf C_+$ describes the transitions within the symmetric $\mathbf{56}$ octet states ($\Phi_{\mathbf{56},\mathbf 8}^{s}\to\Phi_{\mathbf{56},\mathbf 8}^{s}$) as well as between the $\lambda$ components of the $\mathbf{70}$ representation ($\Phi_{\mathbf{70}}^{\lambda}\to\Phi_{\mathbf{70}}^{\lambda}$). 
The block $\mathbf C_\times$ gives the transition within the $\rho$ component of the $\mathbf{70}$ representation ($\Phi_{\mathbf{70}}^{\rho}\to\Phi_{\mathbf{70}}^{\rho}$), where the crossed combination of flavor and spin factors arises because $\Phi_{\mathbf{70}}^{\rho}$ pairs flavor and spin functions with crossed exchange symmetries. 
The block $\mathbf C_-$ mediates the configuration mixing transition between $\Phi_{\mathbf{56},\mathbf 8}^{s}$ and $\Phi_{\mathbf{70}}^{\lambda}$. 
The off-diagonal matrix elements between $\Phi_{\mathbf{70}}^{\rho}$ and $(\Phi_{\mathbf{56},\mathbf 8}^{s}, \Phi_{\mathbf{70}}^{\lambda})$ vanish identically because their $(12)$ exchange parities cannot match.

Combining these spin--flavor blocks with the spatial wave functions of the light octet configurations $\left|B_\kappa^{\mathbf 8}, 1/2^+\right\rangle$ ($\kappa=1,2,3$) in Eqs.~\eqref{eq:wf_octet}--\eqref{eq:baryon_state_expansion}, we obtain the spatial--spin--flavor transition factors $\mathcal G_{\kappa_f\kappa_i}^{fi,3}$ for quark line $j=3$ as
\begin{align}
	\mathcal G_{11}^{fi,3}
	&=\widetilde\Psi_{000}^{s*}(\bm{k}'_\rho,\bm{k}'_\lambda)
	\widetilde\Psi_{000}^{s}(\bm{k}_\rho,\bm{k}_\lambda)\mathbf C_+,\nonumber\\
	\mathcal G_{12}^{fi,3}
	&=\widetilde\Psi_{000}^{s*}(\bm{k}'_\rho,\bm{k}'_\lambda)
	\widetilde\Psi_{200}^{s}(\bm{k}_\rho,\bm{k}_\lambda)\mathbf C_+,\nonumber\\
	\mathcal G_{21}^{fi,3}
	&=\widetilde\Psi_{200}^{s*}(\bm{k}'_\rho,\bm{k}'_\lambda)
	\widetilde\Psi_{000}^{s}(\bm{k}_\rho,\bm{k}_\lambda)\mathbf C_+,\nonumber\\
	\mathcal G_{22}^{fi,3}
	&=\widetilde\Psi_{200}^{s*}(\bm{k}'_\rho,\bm{k}'_\lambda)
	\widetilde\Psi_{200}^{s}(\bm{k}_\rho,\bm{k}_\lambda)\mathbf C_+,\nonumber\\
	\mathcal G_{13}^{fi,3}
	&=\frac{1}{\sqrt2}\widetilde\Psi_{000}^{s*}(\bm{k}'_\rho,\bm{k}'_\lambda)
	\widetilde\Psi_{200}^{\lambda}(\bm{k}_\rho,\bm{k}_\lambda)\mathbf C_-,\nonumber\\
	\mathcal G_{23}^{fi,3}
	&=\frac{1}{\sqrt2}\widetilde\Psi_{200}^{s*}(\bm{k}'_\rho,\bm{k}'_\lambda)
	\widetilde\Psi_{200}^{\lambda}(\bm{k}_\rho,\bm{k}_\lambda)\mathbf C_-,\nonumber\\
	\mathcal G_{31}^{fi,3}
	&=\frac{1}{\sqrt2}\widetilde\Psi_{200}^{\lambda*}(\bm{k}'_\rho,\bm{k}'_\lambda)
	\widetilde\Psi_{000}^{s}(\bm{k}_\rho,\bm{k}_\lambda)\mathbf C_-,\nonumber\\
	\mathcal G_{32}^{fi,3}
	&=\frac{1}{\sqrt2}\widetilde\Psi_{200}^{\lambda*}(\bm{k}'_\rho,\bm{k}'_\lambda)
	\widetilde\Psi_{200}^{s}(\bm{k}_\rho,\bm{k}_\lambda)\mathbf C_-,\nonumber\\
	\mathcal G_{33}^{fi,3}
	&=\frac{1}{2}
	\Big[
	\widetilde\Psi_{200}^{\rho*}(\bm{k}'_\rho,\bm{k}'_\lambda)
	\widetilde\Psi_{200}^{\rho}(\bm{k}_\rho,\bm{k}_\lambda)\mathbf C_\times
	\nonumber\\
	&\quad+
	\widetilde\Psi_{200}^{\lambda*}(\bm{k}'_\rho,\bm{k}'_\lambda)
	\widetilde\Psi_{200}^{\lambda}(\bm{k}_\rho,\bm{k}_\lambda)\mathbf C_+
	\Big].
	\label{eq:configuration_space_integrand_matrix}
\end{align}
Here, $\widetilde\Psi_i(\bm{k}_\rho,\bm{k}_\lambda)$ and $\widetilde\Psi_f^{*}(\bm{k}'_\rho,\bm{k}'_\lambda)$ are the momentum space spatial wave functions obtained by Fourier transforming the coordinate space wave functions determined from the baryon mass spectrum in Sec.~\ref{sec:mass_spectrum}.

Substituting $\mathcal G_{\kappa_f\kappa_i}^{fi,3}$ into the master integral of Eq.~\eqref{eq:master_integral} yields the single quark contribution $H_{\kappa_f\kappa_i}^{\nu,3}(\lambda_f,\lambda_i;P_i,P_f)$ for the active quark line $j=3$. 
Because the physical light octet baryon states are totally symmetric under constituent quark permutations, each of the three active quark lines $j=1,2,3$ gives an identical matrix element to the total weak current $J^\nu = \sum_{j=1}^3 J_j^\nu$. 
Thus, the total hadronic transition matrix element is given by
\begin{align}
	&H^\nu(\lambda_f,\lambda_i;P_i,P_f)
	\nonumber\\
	&=
	3
	\sum_{\kappa_f=1}^{3}
	\sum_{\kappa_i=1}^{3}
	\left(c_{\kappa_f}^{\mathbf 8_f,1/2}\right)^*
	c_{\kappa_i}^{\mathbf 8_i,1/2}
	H_{\kappa_f\kappa_i}^{\nu,3}(\lambda_f,\lambda_i;P_i,P_f),
	\label{eq:total_coherent_hadronic_matrix_element}
\end{align}
where the factor of 3 arises from the equal contributions of the three quark lines under the permutation symmetry of the baryon wave functions.

\section{$q^2$ dependence of hyperon semileptonic form factors}
\label{app:hsd_numerical_tables}

\begin{figure*}[t!]
	\centering
	\includegraphics[width=0.98\textwidth]{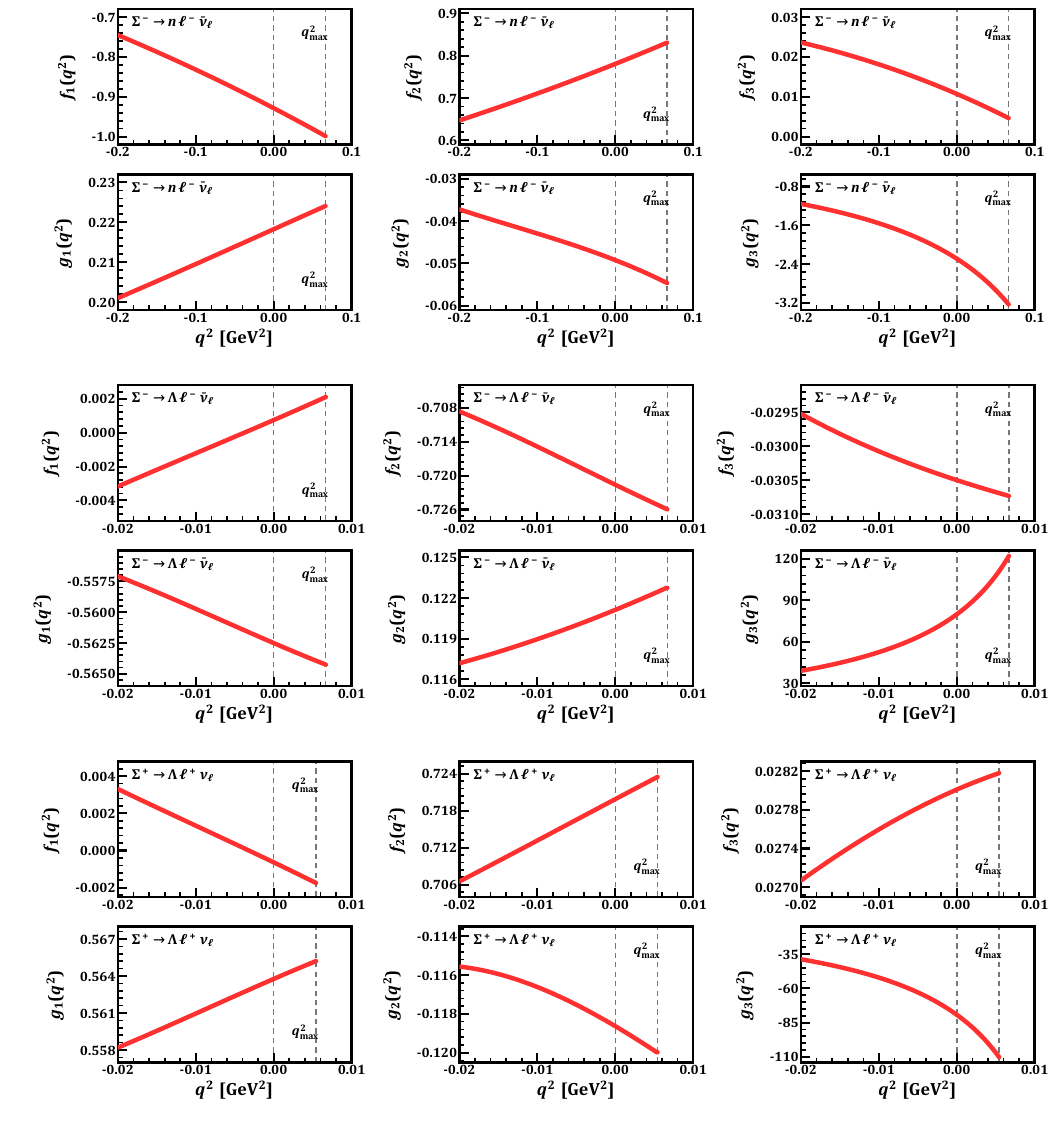}%
	\caption{R3QM form factors $f_i(q^2)$ and $g_i(q^2)$ for the semileptonic $\Sigma$ decays $\Sigma^-\to n\ell^-\bar\nu_\ell$, $\Sigma^-\to\Lambda\ell^-\bar\nu_\ell$, and $\Sigma^+\to\Lambda\ell^+\nu_\ell$, obtained from the corresponding weak transition amplitudes.
        }
	\label{fig:Sigma_form_factors_1}
\end{figure*}

\begin{figure*}[t!]
	\centering
	\includegraphics[width=0.98\textwidth]{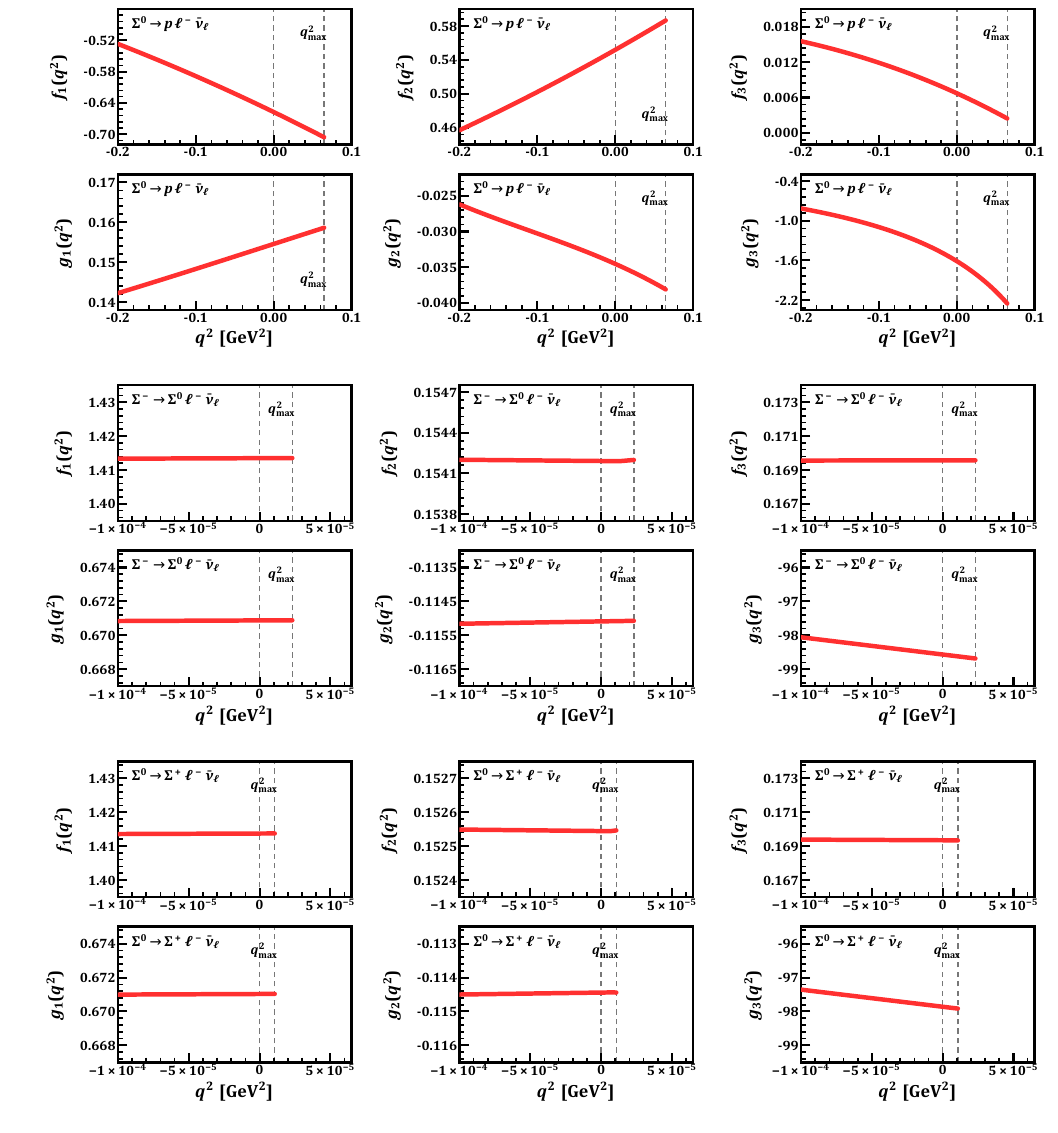}%
	\caption{R3QM form factors $f_i(q^2)$ and $g_i(q^2)$ for the semileptonic $\Sigma$ decays $\Sigma^0\to p\ell^-\bar\nu_\ell$,
		$\Sigma^-\to \Sigma^0\ell^-\bar\nu_\ell$, and
		$\Sigma^0 \to\Sigma^+\ell^-\bar\nu_\ell$, obtained from the corresponding weak transition amplitudes.}
	\label{fig:Sigma_form_factors_2}
\end{figure*}

\begin{figure*}[t!]
	\centering
	\includegraphics[width=0.92\textwidth]{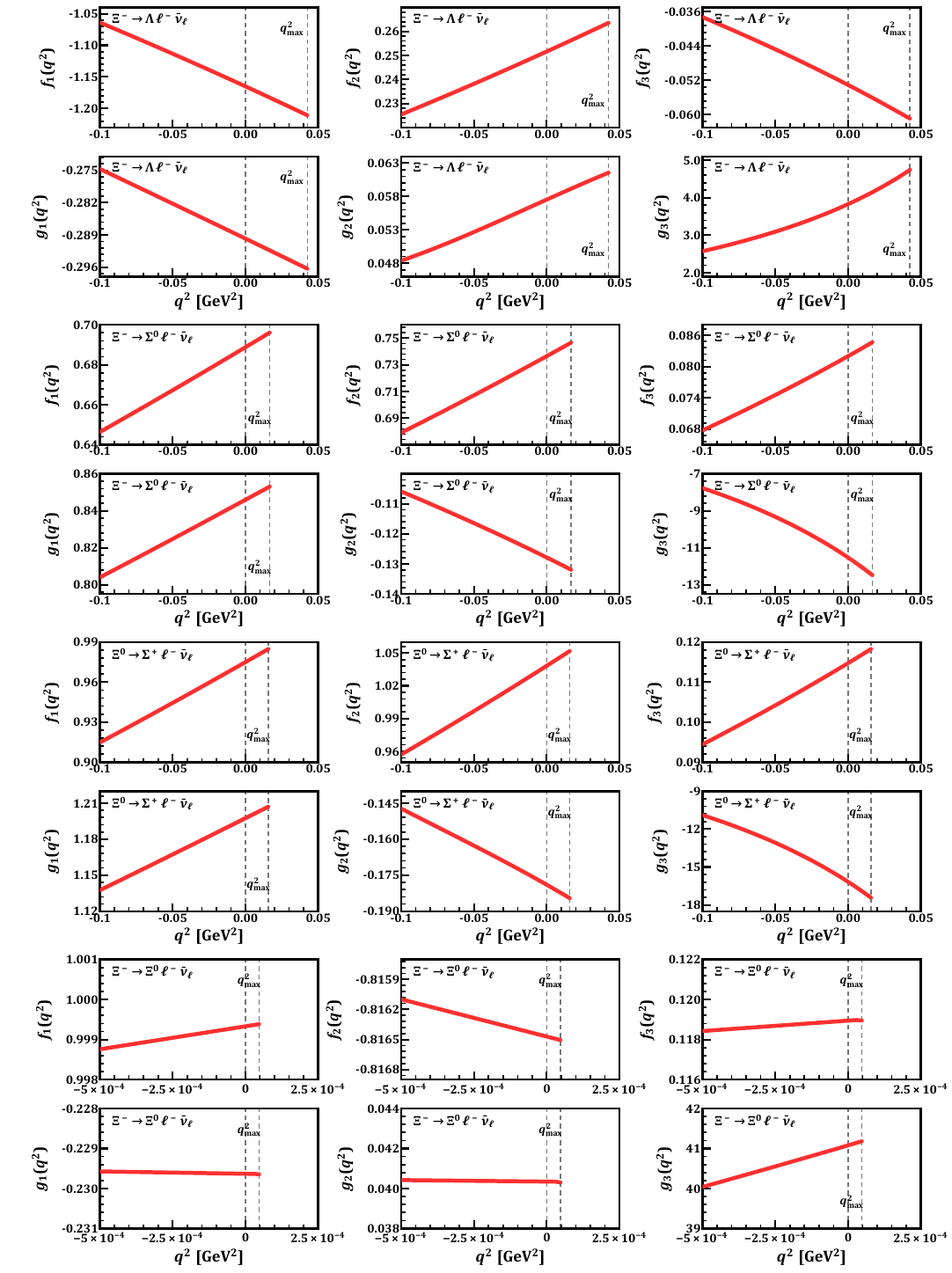}%
	\caption{R3QM form factors $f_i(q^2)$ and $g_i(q^2)$ for the semileptonic $\Xi$ decays $\Xi^-\to \Lambda \ell^-\bar\nu_\ell$,
		$\Xi^-\to \Sigma^0\ell^-\bar\nu_\ell$, 
		$\Xi^0\to\Sigma^+\ell^-\bar\nu_\ell$, and $\Xi^-\to\Xi^0\ell^-\bar\nu_\ell$ obtained from the corresponding weak transition amplitudes.}
	\label{fig:Xi_form_factors}
\end{figure*}

This appendix collects the figures for the $q^2$ dependence of the $\Sigma$ and $\Xi$ hyperon semileptonic form factors. 
Figs.~\ref{fig:Sigma_form_factors_1} and \ref{fig:Sigma_form_factors_2} show the $\Sigma$ hyperon form factors, while Fig.~\ref{fig:Xi_form_factors} shows the $\Xi$ hyperon form factors. 
The $\Lambda\to p$ form factors are compared with the lattice QCD results in Fig.~\ref{fig:Lambda_p_form_factors_vs_LQCD} in the main text.

\bibliography{ref}

@article{Isgur:1978xj,
    author = "Isgur, Nathan and Karl, Gabriel",
    title = "{P Wave Baryons in the Quark Model}",
    reportNumber = "Print-78-1057 (TORONTO)",
    doi = "10.1103/PhysRevD.18.4187",
    journal = "Phys. Rev. D",
    volume = "18",
    pages = "4187",
    year = "1978"
}

@article{Wang:2019alu,
    author = "Wang, Ru-Min and Yang, Mao-Zhi and Li, Hai-Bo and Cheng, Xiao-Dong",
    title = "{Testing SU(3) Flavor Symmetry in Semileptonic and Two-body Nonleptonic Decays of Hyperons}",
    eprint = "1906.08413",
    archivePrefix = "arXiv",
    primaryClass = "hep-ph",
    doi = "10.1103/PhysRevD.100.076008",
    journal = "Phys. Rev. D",
    volume = "100",
    number = "7",
    pages = "076008",
    year = "2019"
}

@article{Schlumpf:1994fb,
    author = "Schlumpf, Felix",
    title = "{Beta decay of hyperons in a relativistic quark model}",
    eprint = "hep-ph/9409272",
    archivePrefix = "arXiv",
    reportNumber = "SLAC-PUB-6661",
    doi = "10.1103/PhysRevD.51.2262",
    journal = "Phys. Rev. D",
    volume = "51",
    pages = "2262--2270",
    year = "1995"
}

@article{Faessler:2008ix,
    author = "Faessler, Amand and Gutsche, Thomas and Holstein, Barry R. and Ivanov, Mikhail A. and Korner, Jurgen G. and Lyubovitskij, Valery E.",
    title = "{Semileptonic decays of the light $J^P = 1/2^+$ ground state baryon octet}",
    eprint = "0809.4159",
    archivePrefix = "arXiv",
    primaryClass = "hep-ph",
    doi = "10.1103/PhysRevD.78.094005",
    journal = "Phys. Rev. D",
    volume = "78",
    pages = "094005",
    year = "2008"
}

@article{Wu:2013kla,
    author = "Wu, Jia-Jun and Zou, Bing-Song",
    title = "{Hyperon Production from Neutrino{\textendash}Nucleon Reaction}",
    eprint = "1307.0574",
    archivePrefix = "arXiv",
    primaryClass = "hep-ph",
    doi = "10.1007/s00601-015-0973-0",
    journal = "Few Body Syst.",
    volume = "56",
    number = "4-5",
    pages = "165--183",
    year = "2015"
}

@article{Zhang:2024ick,
    author = "Zhang, Sheng-Qi and Zhang, Xuan-Heng and Qiao, Cong-Feng",
    title = "{Hyperon semileptonic decays in QCD sum rules}",
    eprint = "2402.15088",
    archivePrefix = "arXiv",
    primaryClass = "hep-ph",
    doi = "10.1007/JHEP06(2024)122",
    journal = "JHEP",
    volume = "06",
    pages = "122",
    year = "2024"
}

@article{Lih:2026dcj,
    author = "Lih, Chong-Chung and Geng, Chao-Qiang",
    title = "{Semileptonic Decays of $\Lambda\to p\ell^{-}\bar{\nu}_{\ell}$ in the Light-Front Dynamics}",
    eprint = "2606.01768",
    archivePrefix = "arXiv",
    primaryClass = "hep-ph",
    month = "6",
    year = "2026"
}

@article{Bacchio:2025auj,
    author = "Bacchio, Simone and Konstantinou, Andreas",
    title = "{Study of the {\ensuremath{\Lambda}}{\textrightarrow}p{\ensuremath{\ell}}{\ensuremath{\nu}}{\textasciimacron}{\ensuremath{\ell}} Semileptonic Decay in Lattice QCD}",
    eprint = "2507.09970",
    archivePrefix = "arXiv",
    primaryClass = "hep-lat",
    doi = "10.1103/bvs9-wcsj",
    journal = "Phys. Rev. Lett.",
    volume = "135",
    number = "23",
    pages = "231901",
    year = "2025"
}

@article{LHCb:2025wld,
    author = "Aaij, Roel and others",
    collaboration = "LHCb",
    title = "{Branching fraction measurement of the $\Lambda \to p\mu^-\bar{\nu}_{\mu}$ decay}",
    eprint = "2511.15681",
    archivePrefix = "arXiv",
    primaryClass = "hep-ex",
    reportNumber = "LHCb-PAPER-2025-030, CERN-EP-2025-223",
    doi = "10.1007/JHEP04(2026)096",
    journal = "JHEP",
    volume = "04",
    pages = "096",
    year = "2026"
}

@article{ParticleDataGroup:2024cfk,
    author = "Navas, S. and others",
    collaboration = "Particle Data Group",
    title = "{Review of particle physics}",
    doi = "10.1103/PhysRevD.110.030001",
    journal = "Phys. Rev. D",
    volume = "110",
    number = "3",
    pages = "030001",
    year = "2024"
}

@article{BESIII:2025wov,
    author = "Ablikim, Medina and others",
    collaboration = "BESIII",
    title = "{Measurements of the absolute branching fraction of the semileptonic decay $\Xi^- \to \Lambda e^- \bar{\nu}_{e}$ and the axial charge of the $\Xi^-$}",
    eprint = "2512.15273",
    archivePrefix = "arXiv",
    primaryClass = "hep-ex",
    month = "12",
    year = "2025"
}

@article{Geng:2020gjh,
    author = "Geng, C. Q. and Liu, Chia-Wei and Tsai, Tien-Hsueh",
    title = "{Semileptonic weak decays of antitriplet charmed baryons in the light-front formalism}",
    eprint = "2012.04147",
    archivePrefix = "arXiv",
    primaryClass = "hep-ph",
    doi = "10.1103/PhysRevD.103.054018",
    journal = "Phys. Rev. D",
    volume = "103",
    number = "5",
    pages = "054018",
    year = "2021"
}

@article{Ebert:2006rp,
    author = "Ebert, D. and Faustov, R. N. and Galkin, V. O.",
    title = "{Semileptonic decays of heavy baryons in the relativistic quark model}",
    eprint = "hep-ph/0604017",
    archivePrefix = "arXiv",
    reportNumber = "HU-EP-06-11",
    doi = "10.1103/PhysRevD.73.094002",
    journal = "Phys. Rev. D",
    volume = "73",
    pages = "094002",
    year = "2006"
}

@article{BESIII:2021ynj,
    author = "Ablikim, M. and others",
    collaboration = "BESIII",
    title = "{First Measurement of the Absolute Branching Fraction of $\Lambda \to p \mu^- \bar{\nu}_{\mu}$}",
    eprint = "2107.06704",
    archivePrefix = "arXiv",
    primaryClass = "hep-ex",
    doi = "10.1103/PhysRevLett.127.121802",
    journal = "Phys. Rev. Lett.",
    volume = "127",
    number = "12",
    pages = "121802",
    year = "2021"
}

@article{Bakamjian:1953kh,
    author = "Bakamjian, B. and Thomas, L. H.",
    title = "{Relativistic particle dynamics. 2}",
    doi = "10.1103/PhysRev.92.1300",
    journal = "Phys. Rev.",
    volume = "92",
    pages = "1300--1310",
    year = "1953"
}

@article{Andreadis:1974ey,
    author = "Andreadis, P. and Baltas, A. and Le Yaouanc, A. and Oliver, L. and Pene, O. and Raynal, J. C.",
    title = "{A Study of Neutrino-Production of Nucleon Isobars in the Quark Model}",
    reportNumber = "LPTHE-TH-74/5",
    doi = "10.1016/0003-4916(74)90403-5",
    journal = "Annals Phys.",
    volume = "88",
    pages = "242",
    year = "1974"
}

@article{Faustov:1970af,
    author = "Faustov, R.",
    title = "{Magnetic moment of the relativistic composite system}",
    doi = "10.1007/BF02728769",
    journal = "Nuovo Cim. A",
    volume = "69",
    pages = "37--46",
    year = "1970"
}

@article{Faustov:1972rp,
    author = "Faustov, R. N.",
    title = "{Relativistic wave function and form-factors of the bound system}",
    doi = "10.1016/0003-4916(73)90007-9",
    journal = "Annals Phys.",
    volume = "78",
    pages = "176--189",
    year = "1973"
}

@article{Donoghue:1981uk,
    author = "Donoghue, John F. and Holstein, Barry R.",
    title = "{Quark Model Calculation of the Weak Electric Coupling in Semileptonic Baryon Decay}",
    reportNumber = "UMHEP-152",
    doi = "10.1103/PhysRevD.25.206",
    journal = "Phys. Rev. D",
    volume = "25",
    pages = "206",
    year = "1982"
}

@article{Barik:1985in,
    author = "Barik, N. and Dash, B. K. and Das, M.",
    title = "{WEAK ELECTRIC AND MAGNETIC FORM-FACTORS FOR SEMILEPTONIC BARYON DECAYS IN AN INDEPENDENT QUARK MODEL}",
    doi = "10.1103/PhysRevD.32.1725",
    journal = "Phys. Rev. D",
    volume = "32",
    pages = "1725--1732",
    year = "1985"
}

@article{Beyer:1986cf,
    author = "Beyer, M. and Singh, S. K.",
    title = "{Quark Model Analysis of Hyperon Semileptonic Decays}",
    doi = "10.1007/BF01588039",
    journal = "Z. Phys. C",
    volume = "31",
    pages = "421--428",
    year = "1986"
}

@article{Cabibbo:2003cu,
    author = "Cabibbo, Nicola and Swallow, Earl C. and Winston, Roland",
    title = "{Semileptonic hyperon decays}",
    eprint = "hep-ph/0307298",
    archivePrefix = "arXiv",
    doi = "10.1146/annurev.nucl.53.013103.155258",
    journal = "Ann. Rev. Nucl. Part. Sci.",
    volume = "53",
    pages = "39--75",
    year = "2003"
}

@article{Kadeer:2005aq,
    author = "Kadeer, A. and Korner, J. G. and Moosbrugger, U.",
    title = "{Helicity analysis of semileptonic hyperon decays including lepton mass effects}",
    eprint = "hep-ph/0511019",
    archivePrefix = "arXiv",
    reportNumber = "MZ-TH-05-24",
    doi = "10.1140/epjc/s10052-008-0801-5",
    journal = "Eur. Phys. J. C",
    volume = "59",
    pages = "27--47",
    year = "2009"
}

@article{Flores-Mendieta:2001fcy,
    author = "Flores-Mendieta, Ruben and Garcia, A. and Martinez, A. and Torres, J. J.",
    title = "{Radiative corrections to the semileptonic Dalitz plot with angular correlation between polarized decaying and emitted hyperons: Effects of the four body region}",
    eprint = "hep-ph/0111117",
    archivePrefix = "arXiv",
    doi = "10.1103/PhysRevD.65.074002",
    journal = "Phys. Rev. D",
    volume = "65",
    pages = "074002",
    year = "2002"
}

@article{Singleton:1990ye,
    author = "Singleton, Robert L.",
    title = "{Semileptonic baryon decays with a heavy quark}",
    reportNumber = "SLAC-PUB-5335",
    doi = "10.1103/PhysRevD.43.2939",
    journal = "Phys. Rev. D",
    volume = "43",
    pages = "2939--2950",
    year = "1991"
}

@article{Korner:1991ph,
    author = "Korner, J. G. and Kramer, M.",
    title = "{Polarization effects in exclusive semileptonic $\Lambda_c$ and $\Lambda_b$ charm and bottom baryon decays}",
    reportNumber = "DESY-91-123, MZ-TH-91-05",
    doi = "10.1016/0370-2693(92)91623-H",
    journal = "Phys. Lett. B",
    volume = "275",
    pages = "495--505",
    year = "1992"
}

@inproceedings{Korner:2014bca,
    author = {K{\"o}rner, J. G.},
    title = "{Helicity Amplitudes and Angular Decay Distributions}",
    booktitle = "{Helmholtz International Summer School on Physics of Heavy Quarks and Hadrons}",
    eprint = "1402.2787",
    archivePrefix = "arXiv",
    primaryClass = "hep-ph",
    reportNumber = "MITP-14-004",
    pages = "169--184",
    year = "2014"
}

@article{Julia-Diaz:2004yqv,
    author = "Julia-Diaz, B. and Riska, D. O.",
    title = "{Baryon magnetic moments in relativistic quark models}",
    eprint = "hep-ph/0401096",
    archivePrefix = "arXiv",
    doi = "10.1016/j.nuclphysa.2004.03.078",
    journal = "Nucl. Phys. A",
    volume = "739",
    pages = "69--88",
    year = "2004"
}

@article{Gaillard:1984ny,
    author = "Gaillard, Jean Marc and Sauvage, Gilles",
    title = "{HYPERON BETA DECAYS}",
    reportNumber = "CERN-EP-84-62",
    doi = "10.1146/annurev.ns.34.120184.002031",
    journal = "Ann. Rev. Nucl. Part. Sci.",
    volume = "34",
    pages = "351--402",
    year = "1984"
}

@article{Gursey:1964keh,
    author = {G{\"u}rsey, F. and Pais, A. and Radicati, L. A.},
    title = "{Spin and Unitary Spin Independence of Strong Interactions}",
    doi = "10.1103/PhysRevLett.13.299",
    journal = "Phys. Rev. Lett.",
    volume = "13",
    number = "8",
    pages = "299--301",
    year = "1964"
}

@book{LeYaouanc:1988fx,
    author = "Le Yaouanc, A. and Oliver, L. and Pene, O. and Raynal, J. C.",
    title = "{HADRON TRANSITIONS IN THE QUARK MODEL}",
    publisher = "Gordon and Breach",
    address = "New York",
    year = "1988"
}

@article{Garcia:1971wc,
    author = "Garcia, Augusto",
    title = "{Analysis of spin correlations in the beta decay of the lambda hyperon}",
    reportNumber = "C00-264-569, EFI-70-75",
    doi = "10.1103/PhysRevD.3.2638",
    journal = "Phys. Rev. D",
    volume = "3",
    pages = "2638--2648",
    year = "1971"
}

@book{Garcia:1985xz,
    author = "Garcia, A. and Kielanowski, P.",
    title = "{THE BETA DECAY OF HYPERONS}",
    doi = "10.1007/3-540-15184-2",
    series = "Lect. Notes Phys.",
    volume = "222",
    publisher = "Springer",
    year = "1985"
}

@article{Lu:2024ajt,
    author = "Lu, Yu and Jing, Hao-Jie and Wu, Jia-Jun",
    title = "{Phase Conventions in Hadron Physics from the Perspective of the Quark Model}",
    eprint = "2407.17131",
    archivePrefix = "arXiv",
    primaryClass = "hep-ph",
    doi = "10.3390/sym16081061",
    journal = "Symmetry",
    volume = "16",
    number = "8",
    pages = "1061",
    year = "2024"
}

@article{Rabl:1975zy,
    author = "Rabl, Veronika and Campbell, Jr., George and Wali, Kameshwar C.",
    title = "{SU(4) Clebsch-Gordan Coefficients}",
    reportNumber = "SU-4206-50, COO-3533-50",
    doi = "10.1063/1.522491",
    journal = "J. Math. Phys.",
    volume = "16",
    pages = "2494",
    year = "1975"
}

@article{Chen:1979qz,
    author = "Chen, Jin-Quan and Gao, Mei-Juan and Wang, Fan",
    title = "{ON THE PHASE AND THE REPRESENTATION TRANSFORMATIONS OF SU(N) BARYON AND MESON WAVE FUNCTIONS. (IN CHINESE)}",
    journal = "HEPNP",
    volume = "3",
    pages = "408--417",
    year = "1979"
}

@article{Guadagnoli:2006gj,
    author = "Guadagnoli, D. and Lubicz, V. and Papinutto, M. and Simula, S.",
    title = "{First Lattice QCD Study of the $\Sigma \to n$ Axial and Vector Form Factors with $\mathrm{SU}(3)$ Breaking Corrections}",
    eprint = "hep-ph/0606181",
    archivePrefix = "arXiv",
    reportNumber = "RM3-TH-06-7, TUM-HEP-632-06",
    doi = "10.1016/j.nuclphysb.2006.10.022",
    journal = "Nucl. Phys. B",
    volume = "761",
    pages = "63--91",
    year = "2007"
}

@article{Sasaki:2008ha,
    author = "Sasaki, Shoichi and Yamazaki, Takeshi",
    title = "{Lattice study of flavor SU(3) breaking in hyperon beta decay}",
    eprint = "0811.1406",
    archivePrefix = "arXiv",
    primaryClass = "hep-ph",
    reportNumber = "TKYNT-08-27, YITP-08-86",
    doi = "10.1103/PhysRevD.79.074508",
    journal = "Phys. Rev. D",
    volume = "79",
    pages = "074508",
    year = "2009"
}

@article{Ademollo:1964sr,
    author = "Ademollo, M. and Gatto, R.",
    title = "{Nonrenormalization Theorem for the Strangeness Violating Vector Currents}",
    doi = "10.1103/PhysRevLett.13.264",
    journal = "Phys. Rev. Lett.",
    volume = "13",
    pages = "264--265",
    year = "1964"
}

@article{Flores-Mendieta:1998tfv,
    author = "Flores-Mendieta, Ruben and Jenkins, Elizabeth E. and Manohar, Aneesh V.",
    title = "{SU(3) symmetry breaking in hyperon semileptonic decays}",
    eprint = "hep-ph/9805416",
    archivePrefix = "arXiv",
    reportNumber = "UCSD-PTH-98-17",
    doi = "10.1103/PhysRevD.58.094028",
    journal = "Phys. Rev. D",
    volume = "58",
    pages = "094028",
    year = "1998"
}

@article{Mateu:2005wi,
    author = "Mateu, V. and Pich, A.",
    title = "{$V_{us}$ determination from hyperon semileptonic decays}",
    eprint = "hep-ph/0509045",
    archivePrefix = "arXiv",
    reportNumber = "IFIC-05-34, FTUV-05-0907",
    doi = "10.1088/1126-6708/2005/10/041",
    journal = "JHEP",
    volume = "10",
    pages = "041",
    year = "2005"
}

@article{Bristol-Geneva-Heidelberg-Orsay-Rutherford-Strasbourg:1983rna,
    author = "Bourquin, M. and others",
    title = "{Measurements of Hyperon Semileptonic Decays at the CERN Super Proton Synchrotron. 4. Tests of the Cabibbo Model}",
    doi = "10.1007/BF01648773",
    journal = "Z. Phys. C",
    volume = "21",
    pages = "27--44",
    year = "1983"
}

@article{Lacour:2007wm,
    author = "Lacour, A. and Kubis, B. and Meissner, Ulf-G.",
    title = "{Hyperon decay form factors in chiral perturbation theory}",
    eprint = "0708.3957",
    archivePrefix = "arXiv",
    primaryClass = "hep-ph",
    reportNumber = "HISKP-TH-07-21, FZJ-IKP-TH-2007-23",
    doi = "10.1088/1126-6708/2007/10/083",
    journal = "JHEP",
    volume = "10",
    pages = "083",
    year = "2007"
}

@article{Ledwig:2008ku,
    author = "Ledwig, Tim and Silva, Antonio and Kim, Hyun-Chul and Goeke, Klaus",
    title = "{Semileptonic hyperon decays in the self-consistent SU(3) chiral quark-soliton model}",
    eprint = "0806.4072",
    archivePrefix = "arXiv",
    primaryClass = "hep-ph",
    reportNumber = "INHA-NTG-10-2008",
    doi = "10.1088/1126-6708/2008/07/132",
    journal = "JHEP",
    volume = "07",
    pages = "132",
    year = "2008"
}

@article{Ahmadi:2025oal,
    author = "Ahmadi, M. and Najjar, Z. Rajabi and Azizi, K.",
    title = "{Study of the semileptonic decay {\ensuremath{\Lambda}}{\textrightarrow}p{\ensuremath{\ell}}{\ensuremath{\nu}}{\textasciimacron}{\ensuremath{\ell}} in QCD}",
    eprint = "2509.23421",
    archivePrefix = "arXiv",
    primaryClass = "hep-ph",
    doi = "10.1103/wdt2-stt2",
    journal = "Phys. Rev. D",
    volume = "112",
    number = "9",
    pages = "094035",
    year = "2025"
}

@article{Geng:2009ik,
    author = "Geng, L. S. and Martin Camalich, J. and Vicente Vacas, M. J.",
    title = "{$\mathrm{SU}(3)$-breaking corrections to the hyperon vector coupling $f_1(0)$ in covariant baryon chiral perturbation theory}",
    eprint = "0903.4869",
    archivePrefix = "arXiv",
    primaryClass = "hep-ph",
    doi = "10.1103/PhysRevD.79.094022",
    journal = "Phys. Rev. D",
    volume = "79",
    pages = "094022",
    year = "2009"
}

@article{Luty:1993gi,
    author = "Luty, M. A. and White, Martin J.",
    title = "{Decouplet contributions to hyperon axial vector form-factors}",
    eprint = "hep-ph/9305203",
    archivePrefix = "arXiv",
    doi = "10.1016/0370-2693(93)90812-V",
    journal = "Phys. Lett. B",
    volume = "319",
    pages = "261--268",
    year = "1993"
}

@article{Sasaki:2012ne,
    author = "Sasaki, Shoichi",
    title = "{Hyperon vector form factor from 2+1 flavor lattice QCD}",
    eprint = "1209.6115",
    archivePrefix = "arXiv",
    primaryClass = "hep-lat",
    doi = "10.1103/PhysRevD.86.114502",
    journal = "Phys. Rev. D",
    volume = "86",
    pages = "114502",
    year = "2012"
}

@article{Ramalho:2015jem,
    author = "Ramalho, G. and Tsushima, K.",
    title = "{Axial form factors of the octet baryons in a covariant quark model}",
    eprint = "1512.01167",
    archivePrefix = "arXiv",
    primaryClass = "hep-ph",
    doi = "10.1103/PhysRevD.94.014001",
    journal = "Phys. Rev. D",
    volume = "94",
    number = "1",
    pages = "014001",
    year = "2016"
}

@article{Ohlsson:1998bk,
author = "Ohlsson, Tommy and Snellman, Hakan",
title = "{Weak form-factors for semileptonic octet baryon decays in the chiral quark model}",
eprint = "hep-ph/9803490",
archivePrefix = "arXiv",
doi = "10.1007/s100529800908",
journal = "Eur. Phys. J. C",
volume = "6",
pages = "285--296",
year = "1999"
}

@article{Sharma:2009hg,
author = "Sharma, Neetika and Dahiya, Harleen and Chatley, P. K. and Gupta, Manmohan",
title = "{Weak vector and axial-vector form factors in the chiral constituent quark model with configuration mixing}",
eprint = "0904.2246",
archivePrefix = "arXiv",
primaryClass = "hep-ph",
doi = "10.1103/PhysRevD.79.077503",
journal = "Phys. Rev. D",
volume = "79",
pages = "077503",
year = "2009"
}

@article{Ledwig:2014rfa,
    author = "Ledwig, T. and Martin Camalich, J. and Geng, L. S. and Vicente Vacas, M. J.",
    title = "{Octet-baryon axial-vector charges and SU(3)-breaking effects in the semileptonic hyperon decays}",
    eprint = "1405.5456",
    archivePrefix = "arXiv",
    primaryClass = "hep-ph",
    doi = "10.1103/PhysRevD.90.054502",
    journal = "Phys. Rev. D",
    volume = "90",
    pages = "054502",
    year = "2014"
}

@article{Sasaki:2017jue,
    author = "Sasaki, Shoichi",
    title = "{Continuum limit of hyperon vector coupling $f_1(0)$ from 2+1 flavor domain wall QCD}",
    eprint = "1708.04008",
    archivePrefix = "arXiv",
    primaryClass = "hep-lat",
    doi = "10.1103/PhysRevD.96.074509",
    journal = "Phys. Rev. D",
    volume = "96",
    number = "7",
    pages = "074509",
    year = "2017"
}

@article{KTeV:2001djr,
    author = "Alavi-Harati, A. and others",
    collaboration = "KTeV",
    title = "{First Measurement of Form-Factors of the Decay $\Xi^0 \to \Sigma^+ e^- \bar{\nu}_e$}",
    eprint = "hep-ex/0105016",
    archivePrefix = "arXiv",
    reportNumber = "EFI-01-11, FERMILAB-PUB-01-492-E",
    doi = "10.1103/PhysRevLett.87.132001",
    journal = "Phys. Rev. Lett.",
    volume = "87",
    pages = "132001",
    year = "2001"
}

@article{KTeVE832E799:1999tte,
    author = "Affolder, Anthony A. and others",
    collaboration = "KTeV E832/E799",
    title = "{Observation of the Decay $\Xi^0 \to \Sigma^+ e^- \bar{\nu_e}$}",
    reportNumber = "FERMILAB-PUB-98-403-E",
    doi = "10.1103/PhysRevLett.82.3751",
    journal = "Phys. Rev. Lett.",
    volume = "82",
    pages = "3751--3754",
    year = "1999"
}

@article{NA48I:2006yat,
    author = "Batley, J. R. and others",
    collaboration = "NA48/I",
    title = "{Measurement of the branching ratios of the decays $\Xi^0 \to \Sigma^+ e^- \bar{\nu}_e$ and $\bar{\Xi}^0 \to \bar{\Sigma}^+ e^+ \nu_e$}",
    eprint = "hep-ex/0612043",
    archivePrefix = "arXiv",
    reportNumber = "CERN-PH-EP-2006-032",
    doi = "10.1016/j.physletb.2006.12.028",
    journal = "Phys. Lett. B",
    volume = "645",
    pages = "36--46",
    year = "2007"
}

@article{NA481:2012dtx,
    author = "Batley, J. R. and others",
    collaboration = "NA48/1",
    title = "{Measurement of the branching ratio of the decay $\Xi^{0}\rightarrow \Sigma^{+} \mu^{-} \bar{\nu}_{\mu}$}",
    eprint = "1212.3131",
    archivePrefix = "arXiv",
    primaryClass = "hep-ex",
    reportNumber = "CERN-PH-EP-2012-288",
    doi = "10.1016/j.physletb.2013.01.023",
    journal = "Phys. Lett. B",
    volume = "720",
    pages = "105--110",
    year = "2013"
}

@article{Donoghue:1986th,
    author = "Donoghue, John F. and Holstein, Barry R. and Klimt, Stefan W.",
    title = "{$K_M$ Angles and SU(3) Breaking in Hyperon Beta Decay}",
    reportNumber = "UMHEP-254",
    doi = "10.1103/PhysRevD.35.934",
    journal = "Phys. Rev. D",
    volume = "35",
    pages = "934",
    year = "1987"
}

@article{Villadoro:2006nj,
    author = "Villadoro, Giovanni",
    title = "{Chiral corrections to the hyperon vector form factors}",
    eprint = "hep-ph/0603226",
    archivePrefix = "arXiv",
    reportNumber = "HUTP-06-A0010",
    doi = "10.1103/PhysRevD.74.014018",
    journal = "Phys. Rev. D",
    volume = "74",
    pages = "014018",
    year = "2006"
}

@article{Bristol-Geneva-Heidelberg-Orsay-Rutherford-Strasbourg:1981uuv,
    author = "Bourquin, M. and others",
    collaboration = "Bristol-Geneva-Heidelberg-Orsay-Rutherford-Strasbourg",
    title = "{Measurements of Hyperon Semileptonic Decays at the CERN Super Proton Synchrotron. 1. The $\Sigma^- \to \Lambda e^-\bar\nu$ Decay Mode}",
    reportNumber = "CERN-EP/81-165",
    doi = "10.1007/BF01557576",
    journal = "Z. Phys. C",
    volume = "12",
    pages = "307",
    year = "1982"
}

@article{Tanenbaum:1975bp,
    author = "Tanenbaum, William M. and others",
    title = "{Leptonic Decays of the $\Sigma^-$ and $\Xi^-$ Hyperons}",
    doi = "10.1103/PhysRevD.12.1871",
    journal = "Phys. Rev. D",
    volume = "12",
    pages = "1871--1883",
    year = "1975"
}

@article{BESIII:2021emv,
    author = "Ablikim, M. and others",
    collaboration = "BESIII",
    title = "{Search for the hyperon semileptonic decay $\Xi^- \to \Xi^0 e^- \bar{\nu}_e$}",
    eprint = "2108.09948",
    archivePrefix = "arXiv",
    primaryClass = "hep-ex",
    doi = "10.1103/PhysRevD.104.072007",
    journal = "Phys. Rev. D",
    volume = "104",
    number = "7",
    pages = "072007",
    year = "2021"
}

@article{Bristol-Geneva-Heidelberg-Orsay-Rutherford-Strasbourg:1983jzt,
    author = "Bourquin, M. and others",
    collaboration = "Bristol-Geneva-Heidelberg-Orsay-Rutherford-Strasbourg",
    title = "{Measurements of Hyperon Semileptonic Decays at the CERN Super Proton Synchrotron. 2. The $\Lambda \to pe\bar\nu$, $\Xi^- \to \Lambda e\bar\nu$, and $\Xi^- \to \Sigma^0 e\bar\nu$ Decay Modes}",
    reportNumber = "CERN-EP/83-78",
    doi = "10.1007/BF01648771",
    journal = "Z. Phys. C",
    volume = "21",
    pages = "1",
    year = "1983"
}

@article{Bristol-Geneva-Heidelberg-Orsay-Rutherford-Strasbourg:1983jpz,
    author = "Bourquin, M. and others",
    collaboration = "Bristol-Geneva-Heidelberg-Orsay-Rutherford-Strasbourg",
    title = "{Measurements of Hyperon Semileptonic Decays at the {CERN} Super Proton Synchrotron. 3. The $\Sigma^- \to n e^-$ Anti-neutrino Decay Mode}",
    reportNumber = "CERN-EP/83-79",
    doi = "10.1007/BF01648772",
    journal = "Z. Phys. C",
    volume = "21",
    pages = "17",
    year = "1983"
}

@article{Wise:1980xx,
    author = "Wise, J. and Jensen, D. A. and Kreisler, M. N. and Lomanno, F. and Poster, R. and Rabin, M. S. Z. and Raychaudhuri, K. and Way, M. and Humphrey, J.",
    title = "{Precise Measurement of the Ratio of the Axial Vector Coupling to Vector Coupling in $\Lambda^0 \to p e^- \bar{\nu}$}",
    reportNumber = "UMASS HEP-142",
    doi = "10.1016/0370-2693(81)90381-6",
    journal = "Phys. Lett. B",
    volume = "98",
    pages = "123",
    year = "1981",
    note = "[Erratum: Phys. Lett. B 100, 519 (1981)]"
}

@article{Dworkin:1990dd,
    author = "Dworkin, J. and others",
    title = "{High statistics measurement of $g_A/g_V$ in $\Lambda \to p e^- \bar{\nu}_e$}",
    doi = "10.1103/PhysRevD.41.780",
    journal = "Phys. Rev. D",
    volume = "41",
    pages = "780--800",
    year = "1990"
}

@article{Hsueh:1988ar,
    author = "Hsueh, S. Y. and others",
    title = "{A High Precision Measurement of Polarized $\Sigma^-$ Beta Decay}",
    reportNumber = "FERMILAB-PUB-88-017-E",
    doi = "10.1103/PhysRevD.38.2056",
    journal = "Phys. Rev. D",
    volume = "38",
    pages = "2056",
    year = "1988"
}

@article{Lindquist:1975fc,
    author = "Lindquist, J. and others",
    title = "{Measurement of the Angular Correlation Parameters in the beta Decay of Polarized Lambda Hyperons}",
    reportNumber = "EFI-75-54-CHICAGO, COO-1545-167",
    doi = "10.1103/PhysRevD.16.2104",
    journal = "Phys. Rev. D",
    volume = "16",
    pages = "2104",
    year = "1977"
}

@article{BESIII:2025hgj,
    author = "Ablikim, Medina and others",
    collaboration = "BESIII",
    title = "{Determination of CKM matrix element and axial vector form factors from weak decays of quantum-entangled strange baryons}",
    eprint = "2509.09266",
    archivePrefix = "arXiv",
    primaryClass = "hep-ex",
    month = "9",
    year = "2025"
}

@article{Wigner:1939cj,
    author  = "Wigner, E. P.",
    title   = "{On unitary representations of the inhomogeneous {L}orentz group}",
    doi     = "10.2307/1968551",
    journal = "Ann. Math.",
    volume  = "40",
    pages   = "149--204",
    year    = "1939"
}

@article{Guadagnoli:2004qw,
    author = "Guadagnoli, D. and Martinelli, G. and Papinutto, M. and Simula, S.",
    title = "{Semileptonic hyperon decays on the Lattice: An Exploratory study}",
    eprint = "hep-lat/0409048",
    archivePrefix = "arXiv",
    doi = "10.1016/j.nuclphysbps.2004.11.316",
    journal = "Nucl. Phys. B Proc. Suppl.",
    volume = "140",
    pages = "390--392",
    year = "2005"
}

@article{Ping:2004sh,
    author = "Ping, R. G. and Zou, B. S. and Chiang, H. C.",
    title = "{P-wave charmonium decays into baryon and antibaryon pairs in quark pair creation model}",
    doi = "10.1140/epja/i2004-10069-9",
    journal = "Eur. Phys. J. A",
    volume = "23",
    pages = "129--133",
    year = "2004"
}

@article{Ping:2004wz,
    author = "Ping, R. G. and Chiang, H. C. and Zou, B. S.",
    title = "{A Study of the Roper resonance as a hybrid state from $J/\psi$ decays}",
    eprint = "nucl-th/0408007",
    archivePrefix = "arXiv",
    doi = "10.1016/j.nuclphysa.2004.07.004",
    journal = "Nucl. Phys. A",
    volume = "743",
    pages = "149--169",
    year = "2004"
}

@article{Ping:2002uj,
    author = "Ping, R. G. and Chiang, H. C. and Zou, B. S.",
    title = "{Study of the structure of baryons from $J/\psi \to B\bar{B}$ decays in the quark model}",
    doi = "10.1103/PhysRevD.66.054020",
    journal = "Phys. Rev. D",
    volume = "66",
    pages = "054020",
    year = "2002"
}

@article{Ebert:2004ck,
    author = "Ebert, D. and Faustov, R. N. and Galkin, V. O. and Martynenko, A. P.",
    title = "{Semileptonic decays of doubly heavy baryons in the relativistic quark model}",
    eprint = "hep-ph/0404280",
    archivePrefix = "arXiv",
    reportNumber = "HU-EP-04-24",
    doi = "10.1103/PhysRevD.70.014018",
    journal = "Phys. Rev. D",
    volume = "70",
    pages = "014018",
    year = "2004",
    note = "[Erratum: Phys.Rev.D 77, 079903 (2008)]"
}

@article{Becirevic:2004bb,
    author = "Becirevic, D. and Guadagnoli, D. and Isidori, G. and Lubicz, V. and Martinelli, G. and Mescia, F. and Papinutto, M. and Simula, S. and Tarantino, C. and Villadoro, G.",
    editor = "Chen, He-Sheng and Du, Dong-Sheng and Li, Wei-Guo and Lu, Cai-Dian",
    title = "{SU(3)-breaking effects in kaon and hyperon semileptonic decays from lattice QCD}",
    eprint = "hep-lat/0411016",
    archivePrefix = "arXiv",
    reportNumber = "LPT-ORSAY-04-73, RM3-TH-04-22, ROME1-1388-2004",
    doi = "10.1140/epjad/s2005-05-012-0",
    journal = "Eur. Phys. J. A",
    volume = "24S1",
    pages = "69--73",
    year = "2005"
}

@article{Sasaki:2006jp,
    author = "Sasaki, Shoichi and Yamazaki, Takeshi",
    editor = "Blum, Tom and Creutz, Michael and DeTar, Carleton and Karsch, Frithjof and Kronfeld, Andreas and Morningstar, Colin and Richards, David and Shigemitsu, Junko and Toussaint, Doug",
    title = "{SU(3) breaking effects in hyperon beta decay from lattice QCD}",
    eprint = "hep-lat/0610082",
    archivePrefix = "arXiv",
    reportNumber = "RBRC-616",
    doi = "10.22323/1.032.0092",
    journal = "PoS",
    volume = "LAT2006",
    pages = "092",
    year = "2006"
}

@article{Sasaki:2011hu,
    author = "Sasaki, Shoichi",
    editor = "Hosaka, Atsushi and Khemchandani, Kanchan and Nagahiro, Hideko and Nawa, Kanabu",
    title = "{Hyperon vector coupling {$f_1(0)$} from 2+1 flavor lattice QCD}",
    eprint = "1102.4934",
    archivePrefix = "arXiv",
    primaryClass = "hep-lat",
    reportNumber = "TKYNT-11-04",
    doi = "10.1063/1.3647427",
    journal = "AIP Conf. Proc.",
    volume = "1388",
    number = "1",
    pages = "443--446",
    year = "2011"
}

@article{Shanahan:2015dka,
    author = "Shanahan, P. E. and Cooke, A. N. and Horsley, R. and Nakamura, Y. and Rakow, P. E. L. and Schierholz, G. and Thomas, A. W. and Young, R. D. and Zanotti, J. M.",
    title = "{SU(3) breaking in hyperon transition vector form factors}",
    eprint = "1508.06923",
    archivePrefix = "arXiv",
    primaryClass = "nucl-th",
    reportNumber = "ADP-15-27-T929, EDINBURGH-2015-17, DESY-15-150",
    doi = "10.1103/PhysRevD.92.074029",
    journal = "Phys. Rev. D",
    volume = "92",
    number = "7",
    pages = "074029",
    year = "2015"
}

@article{Wang:2006yz,
    author = "Wang, Z. G.",
    title = "{Analysis of the Sigma-n form factors with light-cone QCD sum rules}",
    eprint = "hep-ph/0609155",
    archivePrefix = "arXiv",
    doi = "10.1088/0954-3899/34/3/007",
    journal = "J. Phys. G",
    volume = "34",
    pages = "493--504",
    year = "2007"
}

@article{Krause:1990xc,
    author = "Krause, A.",
    title = "{Baryon Matrix Elements of the Vector Current in Chiral Perturbation Theory}",
    doi = "10.5169/seals-116214",
    journal = "Helv. Phys. Acta",
    volume = "63",
    pages = "3--70",
    year = "1990"
}

@article{Anderson:1993as,
    author = "Anderson, Jeffrey and Luty, Markus A.",
    title = "{Chiral corrections to hyperon vector form-factors}",
    eprint = "hep-ph/9301219",
    archivePrefix = "arXiv",
    reportNumber = "LBL-33435",
    doi = "10.1103/PhysRevD.47.4975",
    journal = "Phys. Rev. D",
    volume = "47",
    pages = "4975--4980",
    year = "1993"
}

@article{Jiang:2008aqa,
    author = "Jiang, Fu-Jiun and Tiburzi, Brian C.",
    title = "{Chiral corrections to hyperon axial form factors}",
    eprint = "0801.2535",
    archivePrefix = "arXiv",
    primaryClass = "hep-lat",
    reportNumber = "UMD-40762-405",
    doi = "10.1103/PhysRevD.77.094506",
    journal = "Phys. Rev. D",
    volume = "77",
    pages = "094506",
    year = "2008"
}

@article{Jiang:2009sf,
    author = "Jiang, Fu-Jiun and Tiburzi, Brian C.",
    title = "{Hyperon Axial Charges in Two-Flavor Chiral Perturbation Theory}",
    eprint = "0905.0857",
    archivePrefix = "arXiv",
    primaryClass = "nucl-th",
    reportNumber = "UMD-40762-450",
    doi = "10.1103/PhysRevD.80.077501",
    journal = "Phys. Rev. D",
    volume = "80",
    pages = "077501",
    year = "2009"
}

@article{Sauerwein:2021jxb,
    author = "Sauerwein, Ulrich and Lutz, Matthias F. M. and Timmermans, Rob G. E.",
    title = "{Axial-vector form factors of the baryon octet and chiral symmetry}",
    eprint = "2105.06755",
    archivePrefix = "arXiv",
    primaryClass = "hep-ph",
    doi = "10.1103/PhysRevD.105.054005",
    journal = "Phys. Rev. D",
    volume = "105",
    number = "5",
    pages = "054005",
    year = "2022"
}

@article{Cabibbo:2003ea,
    author = "Cabibbo, Nicola and Swallow, Earl C. and Winston, Roland",
    title = "{Semileptonic hyperon decays and CKM unitarity}",
    eprint = "hep-ph/0307214",
    archivePrefix = "arXiv",
    doi = "10.1103/PhysRevLett.92.251803",
    journal = "Phys. Rev. Lett.",
    volume = "92",
    pages = "251803",
    year = "2004"
}

@article{Ratcliffe:1998wn,
    author = "Ratcliffe, Philip G.",
    editor = "Kalman, Calvin S. and Bozzo, M. and Caso, C. and Pallavicini, M. and Morettini, P. and McKenna, J. and Sanchis-Lozano, M. A.",
    title = "{$\mathrm{SU}(3)$ breaking in hyperon beta decays: A prediction for $\Xi^0 \to \Sigma^+ e^- \bar{\nu}_e$}",
    eprint = "hep-ph/9807394",
    archivePrefix = "arXiv",
    reportNumber = "EPTCO-98-003",
    doi = "10.1016/S0920-5632(99)00325-4",
    journal = "Nucl. Phys. B Proc. Suppl.",
    volume = "75",
    pages = "60--62",
    year = "1999"
}

@article{Pham:2012db,
    author = "Pham, T. N.",
    title = "{Test of SU(3) Symmetry in Hyperon Semileptonic Decays}",
    eprint = "1210.3981",
    archivePrefix = "arXiv",
    primaryClass = "hep-ph",
    doi = "10.1103/PhysRevD.87.016002",
    journal = "Phys. Rev. D",
    volume = "87",
    number = "1",
    pages = "016002",
    year = "2013"
}

@article{Dahiya:2024ekj,
    author = "Dahiya, Harleen and Girdhar, Aarti and Randhawa, Monika",
    title = "{Study of vector and axial-vector form factors and the decay parameters for the semileptonic hyperon decays}",
    eprint = "2405.00444",
    archivePrefix = "arXiv",
    primaryClass = "hep-ph",
    doi = "10.1007/s12648-024-03224-1",
    journal = "Indian J. Phys.",
    volume = "98",
    number = "14",
    pages = "4961--4971",
    year = "2024"
}

@article{Capstick:1986ter,
    author = "Capstick, S. and Isgur, Nathan",
    title = "{Baryons in a Relativized Quark Model with Chromodynamics}",
    doi = "10.1103/PhysRevD.34.2809",
    journal = "Phys. Rev. D",
    volume = "34",
    pages = "2809--2835",
    year = "1986"
}

@article{Eichten:1978tg,
    author = "Eichten, E. and Gottfried, K. and Kinoshita, T. and Lane, K. D. and Yan, Tung-Mow",
    title = "{Charmonium: The Model}",
    doi = "10.1103/PhysRevD.17.3090",
    journal = "Phys. Rev. D",
    volume = "17",
    pages = "3090",
    year = "1978"
}

@article{Zhong:2024mnt,
    author = "Zhong, Hui-Hua and Liu, Ming-Sheng and Ni, Ru-Hui and Chen, Mu-Yang and Zhong, Xian-Hui and Zhao, Qiang",
    title = "{Unified study of nucleon and $\Delta$ baryon spectra and their strong decays with chiral dynamics}",
    eprint = "2409.07998",
    archivePrefix = "arXiv",
    primaryClass = "hep-ph",
    doi = "10.1103/PhysRevD.110.116034",
    journal = "Phys. Rev. D",
    volume = "110",
    pages = "116034",
    year = "2024"
}

@article{Huang:2015nja,
    author = "Huang, Fei and Shen, Peng Nian and Dong, Yu Bing and Zhang, Zong Ye",
    title = "{Understanding the structure of $d^*(2380)$ in chiral quark model}",
    eprint = "1505.05395",
    archivePrefix = "arXiv",
    primaryClass = "nucl-th",
    doi = "10.1007/s11433-015-5767-3",
    journal = "Sci. China Phys. Mech. Astron.",
    volume = "59",
    number = "2",
    pages = "622002",
    year = "2016"
}

@article{Silvestre-Brac:1996myf,
    author = "Silvestre-Brac, B.",
    title = "{Spectrum and static properties of heavy baryons}",
    doi = "10.1007/s006010050028",
    journal = "Few Body Syst.",
    volume = "20",
    pages = "1--25",
    year = "1996"
}

@article{Lu:1997sd,
    author = "Lu, Ding-Hui and Thomas, Anthony William and Williams, Anthony Gordon",
    title = "{Electromagnetic form-factors of the nucleon in an improved quark model}",
    eprint = "nucl-th/9706019",
    archivePrefix = "arXiv",
    reportNumber = "ADP-97-16-T-253",
    doi = "10.1103/PhysRevC.57.2628",
    journal = "Phys. Rev. C",
    volume = "57",
    pages = "2628--2637",
    year = "1998"
}

@article{Lu:2001yc,
    author = "Lu, Ding-Hui and Yang, S. N. and Thomas, Anthony William",
    editor = "Cheung, Chi-Yee and Ho, Y. K. and Lee, T. S. H. and Yang, Shin-Nan",
    title = "{On the role of the pion cloud in nucleon electromagnetic form-factors}",
    doi = "10.1016/S0375-9474(01)00483-3",
    journal = "Nucl. Phys. A",
    volume = "684",
    pages = "296--298",
    year = "2001"
}

@article{Ramalho:2011pp,
    author = "Ramalho, G. and Tsushima, K.",
    title = "{Octet baryon electromagnetic form factors in a relativistic quark model}",
    eprint = "1107.1791",
    archivePrefix = "arXiv",
    primaryClass = "hep-ph",
    reportNumber = "ADP-11-24-T746",
    doi = "10.1103/PhysRevD.84.054014",
    journal = "Phys. Rev. D",
    volume = "84",
    pages = "054014",
    year = "2011"
}

@article{an:2006zf,
    author = "An, C. S. and Li, Q. B. and Riska, D. O. and Zou, B. S.",
    title = "{The qqqq anti-q components and hidden flavor contributions to the baryon magnetic moments}",
    eprint = "nucl-th/0610009",
    archivePrefix = "arXiv",
    doi = "10.1103/PhysRevC.75.069901",
    journal = "Phys. Rev. C",
    volume = "74",
    pages = "055205",
    year = "2006",
    note = "[Erratum: Phys.Rev.C 75, 069901 (2007)]"
}

@article{Flores-Mendieta:2004cyh,
    author = "Flores-Mendieta, Ruben",
    title = "{V(us) from hyperon semileptonic decays}",
    eprint = "hep-ph/0410171",
    archivePrefix = "arXiv",
    doi = "10.1103/PhysRevD.70.114036",
    journal = "Phys. Rev. D",
    volume = "70",
    pages = "114036",
    year = "2004"
}

@article{Bijnens:1985kj,
    author = "Bijnens, J. and Sonoda, H. and Wise, Mark B.",
    title = "{On the Validity of Chiral Perturbation Theory for Weak Hyperon Decays}",
    reportNumber = "CALT-68-1221",
    doi = "10.1016/0550-3213(85)90569-3",
    journal = "Nucl. Phys. B",
    volume = "261",
    pages = "185--198",
    year = "1985"
}

@article{Jenkins:1990jv,
    author = "Jenkins, Elizabeth Ellen and Manohar, Aneesh V.",
    title = "{Baryon chiral perturbation theory using a heavy fermion Lagrangian}",
    reportNumber = "UCSD-PTH-90-23",
    doi = "10.1016/0370-2693(91)90266-S",
    journal = "Phys. Lett. B",
    volume = "255",
    pages = "558--562",
    year = "1991"
}

@article{Jenkins:1991es,
    author = "Jenkins, Elizabeth Ellen and Manohar, Aneesh V.",
    title = "{Chiral corrections to the baryon axial currents}",
    reportNumber = "UCSD-PTH-91-05",
    doi = "10.1016/0370-2693(91)90840-M",
    journal = "Phys. Lett. B",
    volume = "259",
    pages = "353--358",
    year = "1991"
}

@article{Zhu:2000zf,
    author = "Zhu, Shi-Lin and Puglia, S. and Ramsey-Musolf, M. J.",
    title = "{Recoil order chiral corrections to baryon octet axial currents}",
    eprint = "hep-ph/0009159",
    archivePrefix = "arXiv",
    doi = "10.1103/PhysRevD.63.034002",
    journal = "Phys. Rev. D",
    volume = "63",
    pages = "034002",
    year = "2001"
}

@article{Kaiser:2001yc,
    author = "Kaiser, Norbert",
    title = "{Isospin breaking in neutron beta decay and SU(3) violation in semileptonic hyperon decays}",
    eprint = "nucl-th/0105043",
    archivePrefix = "arXiv",
    doi = "10.1103/PhysRevC.64.028201",
    journal = "Phys. Rev. C",
    volume = "64",
    pages = "028201",
    year = "2001"
}

@article{Alexandrou:2026noh,
    author = "Alexandrou, Constantia and Bacchio, Simone and Konstantinou, Andreas and Vakana, Eleni",
    title = "{Scalar and tensor form factors for $\Lambda \to p\ell\bar{\nu}_{\ell}$ from lattice QCD}",
    eprint = "2604.16025",
    archivePrefix = "arXiv",
    primaryClass = "hep-lat",
    month = "4",
    year = "2026"
}

@article{Kim:1999uf,
    author = "Kim, Hyun-Chul and Praszalowicz, Michal and Goeke, Klaus",
    title = "{Hyperon semileptonic decays and quark spin content of the proton}",
    eprint = "hep-ph/9910282",
    archivePrefix = "arXiv",
    reportNumber = "PNU-NTG-04-99, RUB-TPII-15-99",
    doi = "10.1103/PhysRevD.61.114006",
    journal = "Phys. Rev. D",
    volume = "61",
    pages = "114006",
    year = "2000"
}

@article{Goity:1999by,
    author = "Goity, Jose L. and Lewis, Randy and Schvellinger, Martin and Zhang, Long-Zhe",
    title = "{The Goldberger-Treiman discrepancy in SU(3)}",
    eprint = "hep-ph/9901374",
    archivePrefix = "arXiv",
    reportNumber = "JLAB-THY-98-51",
    doi = "10.1016/S0370-2693(99)00217-8",
    journal = "Phys. Lett. B",
    volume = "454",
    pages = "115--122",
    year = "1999"
}

@article{Goldberger:1958vp,
    author = "Goldberger, M. L. and Treiman, S. B.",
    title = "{Form-factors in Beta decay and muon capture}",
    doi = "10.1103/PhysRev.111.354",
    journal = "Phys. Rev.",
    volume = "111",
    pages = "354--361",
    year = "1958"
}

@article{Brown:2014ena,
    author = "Brown, Zachary S. and Detmold, William and Meinel, Stefan and Orginos, Kostas",
    title = "{Charmed bottom baryon spectroscopy from lattice QCD}",
    eprint = "1409.0497",
    archivePrefix = "arXiv",
    primaryClass = "hep-lat",
    reportNumber = "JLAB-THY-14-1950",
    doi = "10.1103/PhysRevD.90.094507",
    journal = "Phys. Rev. D",
    volume = "90",
    number = "9",
    pages = "094507",
    year = "2014"
}

@article{LHCb:2020XiCCMass,
    author = "Aaij, Roel and others",
    collaboration = "LHCb",
    title = "{Precision measurement of the $\Xi_{cc}^{++}$ mass}",
    eprint = "1911.08594",
    archivePrefix = "arXiv",
    primaryClass = "hep-ex",
    reportNumber = "CERN-EP-2019-258, LHCb-PAPER-2019-037",
    doi = "10.1007/JHEP02(2020)049",
    journal = "JHEP",
    volume = "02",
    pages = "049",
    year = "2020"
}

@misc{Xu:2026OmegaCC,
    author = "Xu, Ao",
    title = "{Observation of the strange doubly charmed baryon $\Omega_{cc}^{+}$}",
    howpublished = "Hadron Physics Online Forum, No. 134",
    note = "{Preliminary LHCb result; LHCb-PAPER-2026-022, in preparation}",
    month = "6",
    year = "2026",
    url = "https://indico.itp.ac.cn/event/451/"
}

\end{document}